\documentclass[%
 reprint,onecolumn,
superscriptaddress,
nofootinbib,
 amsmath,amssymb,
]{revtex4-2}

\usepackage{graphicx}
\usepackage{dcolumn}
\usepackage{bm}

\usepackage{cancel}
\usepackage{amsfonts}
\usepackage{xcolor}

\newcommand{\beqa}{\begin{eqnarray}}
\newcommand{\eeqa}{\end{eqnarray}}
\newcommand{\nn}{\nonumber}
\newcommand{\commutator}[3][1.2em]{[ \makebox[#1]{$#2$} , \makebox[#1]{$#3$} ]}

\DeclareMathOperator{\arccoth}{arcCoth} 
\begin{document}

\title{Vortex Flows and Streamline Topology in Curved Biological Membranes}

\date{\today}
\author{R. Samanta}
\affiliation{Indian Statistical Institute, 203 B. T. Road, Kolkata 700108, India}
 \affiliation{Raymond and Beverly Sackler School of Physics and Astronomy, Tel Aviv University, Tel Aviv 69978, Israel}
 \affiliation{The Center for Physics and Chemistry of Living Systems, Tel Aviv University}
\author{N. Oppenheimer}
 \affiliation{Raymond and Beverly Sackler School of Physics and Astronomy, Tel Aviv University, Tel Aviv 69978, Israel}
\affiliation{The Center for Physics and Chemistry of Living Systems, Tel Aviv University}

\begin{abstract}

When considering flows in biological membranes, they are usually treated as flat, though more often than not, they are curved surfaces, even extremely curved, as in the case of the endoplasmic reticulum. Here, we study the topological effects of curvature on flows in membranes.  
Focusing on a system of many point vortical defects, we are able to cast the viscous dynamics of the defects in terms of a geometric Hamiltonian. In contrast to the planar situation, the flows generate additional defects of positive index. For the  simpler situation of two vortices, we analytically predict the location of these stagnation points. At the low curvature limit, the dynamics resemble that of vortices in an ideal fluid, but considerable deviations occur at high curvatures. The geometric formulation allows us to construct the spatio-temporal evolution of streamline topology of the flows resulting from hydrodynamic interactions between the vortices. The streamlines reveal novel dynamical bifurcations leading to spontaneous defect-pair creation and fusion. Further, we find that membrane curvature mediates defect binding and imparts a global rotation to the many-vortex system, with the individual vortices still interacting locally. 

\end{abstract}

\maketitle
\section {Introduction}

We study two dimensional (2D) flows in curved biological membranes arising from the dynamics of rotating embedded particles.  In particular, we explore the spatio-temporal evolution of topological features of such 2D flows. The study is relevant in the context of biological systems featuring vortical defects. In this work we are primarily motivated by proteins embedded in lipid membranes, in particular the rotating ATP synthase proteins \cite{Ueno2005}, abundant in the  endoplasmic reticulum. A detailed knowledge of such flows will shed light on possible mechanisms of mixing in biological viscous fluids \cite{Duprat2015}, as well as serve as a guiding principle to engineer molecular rotors \cite{rotorimg}, artificially controlled microswimmers and nano-carriers in targetted drug delivery \cite{optnvg}, or in wound detection and healing \cite{mrotor}. Moreover, the rotating inclusions that we consider in much of our analysis can be realized in experiments e.g. by paramagnetic microscopic particles in a rotating magnetic field \cite{grzy}, birefringent particles rotated by laser tweezers \cite{terr} and biological swimmers such as bacteria, Volvox algae and diatoms \cite{Lauga2006, vlvx}. \\

The dynamics of physical systems in the presence of topological defects and curvature is currently an active area of research. Topological defects can play a major role in key macroscopic properties of the system --- be it in driving phase transitions, creating fluid flow patterns, or the emergence of turbulence. Mostly, defect dynamics are investigated in planar 2D systems, a few prime example are vortices in  superfluids \cite{mermin,nelson}, Abrikosov vortex lattices in superconductors \cite{abrik,tink}, and vortex driven Berezinskii Kosterlitz Thouless (BKT) transition \cite{bz,bkt}. The dynamics of vortices in ideal fluids, including integrability, chaos, and stability analysis is also a topic of intense research \cite{bgm,km,arefbrons,newton2000, boatto1, boatto2}. The natural world often features motion of defects on curved surfaces. Point vortex dynamics in a spherical geometry may be used as an approximation for air flow in the earth's atmosphere and oceans \cite{Held1995}. In recent years, experimental advances in condensed matter (Bose-Einstein condensates in particular) have also opened up the possibility to explore superfluid vortices in a curved substrate or in optical traps \cite{vitelli2004, nelson2010}. These systems thus allow a rich interplay between curvature and dynamics of topological defects. \\

In the biological world, curvature and defects feature in the vast majority of living systems e.g. cell membranes, tumor growth, and morphogenesis \cite{Maroudas-Sacks2020, tamreview, atam19}. For planar systems, topological defects play a vital role in BKT-like phase transitions in active nematics \cite{srcb2018}. Such defects lead to turbulent flow patterns even in such highly viscous fluids \cite{Sanchez2012, gbmsc,giomi2015, giomi2017, dgturb, dg3dnm, dunkel}.  Recently, motivated by biological examples, defect dynamics in nematic films have been explored on curved surfaces as well \cite{bw_spnm, hcs2017}. Polar active fluids confined to curved surfaces show flocking and topologically protected sound modes  \cite{sbc_topsound}. 
 \\ 


In this work we perform a detailed study of vortical defects in curved biological membranes. The investigation of in-plane and out-of-plane dynamics of membranes separating two viscous fluids has been a subject of much interest \cite{saff1,saff2,hughes,evans}.  A detailed analysis of different modes in such membranes, including force response, mobility calculations, and many-body interactions was performed \cite{ levinemck1,levinemck2, Stone1998, oppdiamant1, oppdiamant2, oppdiamant3, Seki2014}. In particular, a study of 2D flows and dynamics of rotors was carried out in detail in \cite{lenz1, oppshelley}.
The quasi-2D nature of membranes leads to a new length-scale (here termed the Saffman length) given by the ratio of membrane and solvent viscosities. This length scale acts as a natural cutoff for the logarithmic divergence of two dimensional flows. Beyond the Saffman length, the external solvent contributes to the in-plane dynamics, regulating the divergence.\\

A detailed study of biophysical transport applicable to curved membranes was carried out in a few recent works \cite{henlev2008,henlev2010,wood, woodhouse, atzbergershape,atz2016, atz2018,atz2019}. In particular \cite{henlev2008,henlev2010} generalized the pioneering works of Saffman and Delbr\"uck \cite{saff1,saff2} for curved surfaces of static geometry. For a spherical membrane, particle mobility was computed. In the limit of high curvature (small radius), one finds a reduced mobility, with the sphere radius playing the role of the Saffman length. On the other hand, at low curvature, the Saffman length still continues to regulate the logarithmic divergence. Further, the study reveals the existence of a zero mode due to curvature, that imparts a global motion to the system. The initial part of our analysis here is a direct follow-up of these works \cite{henlev2008,henlev2010}. \\

The work presented here outlines the following aspects of 2D flows in curved membranes: 
\begin{enumerate}
\item At the single particle level, we extend the works of Henle and Levine \cite{henlev2008,henlev2010} to account for rotational flows, such as those generated by point vortices and torque dipoles, as applicable to ATP synthase proteins.  
\item We compute the relevant Green's function in real-space in closed form using Appell Hypergeometric Functions which prove extremely useful in analyzing many-particle dynamics. We analytically predict the location of singularities in the flow field using this approach. 
\item For low curvatures, we find a surprising structural similarity between the equations of viscous membrane hydrodynamics sourced by rotating inclusions and the equations of point vortices in ideal fluids on curved surfaces. This analogy suggests that ideal point vortex models may be a useful tool to gain a basic understanding of defect mediated biological turbulent flows observed in the viscous low Reynolds regime, see \cite{dunkel}.  
\item At high curvatures, we find that the dynamics in membranes deviate from the ideal fluid case. There is a soft mode due to curvature which imparts a global rotation to the many-body system. 	
\item We  provide explicit formulas for the dynamical equations and flows (Eqs.~\ref{dynmeq}, \ref{field}) and the rotation rates (Eq.~\ref{omlc} and Eq.~\ref{glbomega}) in the full parameter space of the biological model.
\item We construct a geometric Hamiltonian describing the dynamics, with associated conservation laws. We use the Hamiltonian to construct the spatio-temporal evolution of the streamlines resulting from the hydrodynamic interactions between point rotors. 
\item For curved membranes, we find there are new vortical defects of positive index (centers), this is in contrast to the planar situation. The number of such new stagnation points  is strictly governed by the Euler Characteristic of the surface, consistent with Poincare Index Theorem. 
\item For many point rotors with varying circulations, we find novel dynamical bifurcations leading to defect-pair fission and fusion. We are able to demonstrate all these effects with a relatively small number of rotors.  Further, we observe that the global rotation imparted by the membrane curvature can drive the binding of defects with opposite index, similar to activity driven defect binding and unbinding phenomena observed in 2D nematic fluids \cite{srcb2018}. 
\item From an experimental point of view, one may expect to achieve the transition from low to high curvature regime in a more controlled fashion by tuning the solvent viscosities, keeping the radius of the membrane fixed. Viewed this way, the high curvature regime may be achieved by reducing the external solvent viscosity compared to that of the internal solvent.
\end{enumerate}
While our emphasis in this work has been on 2D viscous flows on curved membranes and associated streamline topology, it is worth mentioning parallel efforts on 3D viscous Navier Stokes equations \cite{trkal94,ers17,puk20}. Analytic approaches have been used to explore chaotic streamlines, complicated Lagrangian structures \cite{dombre86}, stationary points \cite{ershkov2016} in many interesting flows, for example the ABC (Arnold, Beltrami and Childress) flows.\\\\
The paper is organized as follows : In Section \ref{scsetup} we present a short review of the basic equations for viscous hydrodynamics on curved membranes. In Section \ref{sbf}, \ref{sbtq} and \ref{sbctq}, we provide typical examples of the 2D fluid flows in spherical membranes due to 3 types of sources: a point force, a point torque and a torque dipole. The detailed calculations are presented in the Appendix \ref{apf}, \ref{aptq}, \ref{apctq}. In particular, Section \ref{sbtq} explores the  connections to equations arising in vortex dynamics in ideal fluids on curved surfaces. The rotating solutions allow us to construct a Hamiltonian description for a system of rotating inclusions embedded in the membrane. This Hamiltonian description is presented in Section \ref{sbhm} along with basic equations \ref{sbgeq} to explore the streamline topology of the hydrodynamic flow fields. These equations are used to construct the spatio-temporal evolution of streamline topology of the in-plane flow fields. We explore the streamline flows for different vortex circulations along with an analytic understanding of the associated stagnation points. Next in Section \ref{scmany} we present some interesting scenarios of spontaneous creation of defect pairs and defect fusion  that arise in such systems in the chaotic regime of many interacting point rotors. Finally in section \ref{scgz} and \ref{sccl} we conclude with possible generalizations.\\
The Appendix contains many details of the calculations and formulas used in the main text. Appendix \ref{apf}, \ref{aptq} \ref{apctq} describe the full structure of the real space Green's functions used in our study, while Appendix \ref{aproot} discusses the pole structure of the Green's function in Legendre basis. Appendix \ref{apstg} supplements an analytic investigation of stagnation points and streamline topology carried out in the main text for the situation of two vortices. 

\section { Setup : Viscous hydrodynamics in curved membranes coupled to external solvents. }
\label{scsetup}

Let us start by describing the hydrodynamic equations for curved membranes. We use the pioneering works of Saffman and Delbr\"uck \cite{saff1,saff2} as adapted to a spherical membrane \cite{henlev2008,henlev2010}. We approximate the membrane as a two-dimensional viscous fluid surrounded above and below by three-dimensional viscous fluids. We also assume strictly tangential flows within the membrane. In such situations, the appropriate generalization of the Stokes equations describing 2D flows is
\beqa
\boxed {D^\alpha v_\alpha =0, ~~~
\sigma^{ext}_{\alpha} =-\eta_{2d} \left( K(\vec{x}) v_\alpha  + D^\mu D_\mu v_\alpha \right) + D_\alpha p +  \left( \sigma^{3d}_{\alpha z}|_{z\rightarrow 0^-} - \sigma^{3d}_{\alpha z}|_{z\rightarrow 0^+}\right) },
\label{curvedstokes2d}
\eeqa
 where $x$ represents a general coordinate on the surface, $v_{\alpha}$ is the in-plane 2D fluid velocity ($\alpha$ runs over surface coordinates) and $\eta_{2D}$ denotes the viscosity of the 2D membrane fluid. $D$ is the two dimensional covariant derivative which generalizes the partial derivative of flat space,  $K(\vec{x})$ is the local Gaussian curvature, $p$ is the local membrane pressure, $\sigma^{3D}$ denotes the bulk fluid stress tensor while $z$ denotes a generalized co-ordinate in the normal direction to the surface. The first of the two equations in Eq.~\ref{curvedstokes2d} ensures incompressibility of the membrane fluid while the second equation is a stress balance condition on the membrane surface.  The point source embedded in the membrane provides  $\sigma^{ext}$.  The external source term is balanced by the stress provided by the 2D membrane fluid and  the external solvents above and below the membrane. In the limit of vanishing curvature, $K=0$, one recovers the usual Stokes equations. In curved surfaces, the covariant derivatives fail to commute, originating the curvature term $K(\vec{x})$. For more details, see Appendix \ref{apf}.\\
These equations need to be supplemented by the appropriate Stokes equations for the 3D outer fluid, 
 
  \beqa 
  \boxed {\eta _{\pm}\nabla^2 {\bf v}_{\pm} = \nabla_{\pm} p^{\pm},  \nabla \cdot {\bf v}_{\pm} = 0},
  \label{stokes3d}
  \eeqa
  where ${\bf v}^{+}$ (${\bf v}^{-}$) is the fluid velocity above (below) the membrane, with similar notation for pressure $p_{\pm}$ and viscosities $\eta_{\pm}$. One can define two length scales given by the ratio of membrane and solvent viscosities
  \beqa
  \lambda_{\pm}= \frac{\eta_{2D}}{\eta_{\pm}}.
  \label{saffl}
  \eeqa
The curvature introduces a new scale in the problem. In spherical membranes for example, this will be the radius $R$.  The coupling between the 2D membrane flows and the 3D external solvents is mediated via the no-slip boundary condition and by the stress balance on the membrane surface (the last term in Eq.~\ref{curvedstokes2d}). For membranes of arbitrary shape, these equations can be solved numerically, e.g. see \cite{atz2018}. However, in the simpler situation where the curvature is constant, one can analytically extract flows, which we describe next.\\
  
The incompressibility requirement $D^\alpha v_\alpha =0$ allows us to express the flow field in terms of a stream function  as follows:

\beqa
 v_\alpha(x) = \epsilon_{\alpha \gamma} D^{\gamma}   \phi(\vec{x}),
\eeqa
where $\epsilon_{\alpha \beta}$ is the antisymmetric Levi-Civita symbol. One thus needs to solve for $\phi(\vec{x})$, given a point source  $\sigma^{ext}$, taking into account the membrane curvature and boundary conditions. As shown in detail in Refs.~\cite{henlev2008,henlev2010} (see also our appendix \ref{apf}, \ref{aptq} and \ref{apctq}), such a response calculation is conceptually simple. One needs to invert the curved surface Laplace operator in the presence of the curvature and traction terms \footnote { Without the traction contribution, the curvature term gives rise to a zero mode. However, in the presence of traction, the zero mode is removed and the operator can be inverted without issues.}. For a non-trivial spatially varying curvature $K(x)$, a Fourier decomposition can be done numerically but requires knowledge of the spectrum. For surfaces of constant curvature, the spectrum is often known. One can use the known eigenfunctions to perform the inversion in Fourier Space. For example, for a sphere, one decomposes the above equations in the basis of spherical harmonics, taking into account the stick and stress boundary conditions.  Henle and Levine \cite{henlev2008,henlev2010} express the final solution of the stream function in terms of such eigenmodes. Using  Appell Hypergeometric Functions, we are able to perform the inverse Fourier transform and find closed-form expressions for the stream function in real space. We present a detailed description in Appendix \ref{apf}, \ref{aptq} and \ref{apctq} for each of the sources: point force, point torque and a torque dipole respectively.\\

Before proceeding, let us briefly mention some general topological constraints that the flow fields on the spherical membrane must satisfy. First, the hairy ball theorem implies that flow fields on the spherical membrane must feature stagnation points where the velocity field vanishes. Second,  each of the singular points of the flow can be assigned an index which keeps track of the winding of the flow field around the core of the singularity. The sum of these indices \footnote{ We will present a more general criterion for the situation of many embedded particles, see Eq.~\ref{indxth} in Section \ref{sbhm}} must equate to 2, the Euler Characteristic of the sphere (Poincare Index Theorem).  In all the examples involving single inclusions that we are about to study, we will observe these topological features in the flow fields. \\
We now describe the flow fields resulting from a point force, point torque, and torque-dipole, one by one:
  
\subsection {Velocity field due to a point force on the spherical membrane.}
\label{sbf}
The main body of this work concerns vortices in a membrane, but for completeness and consistency, in this section we reproduce the results of Henle and Levine for a point force acting on a spherical membrane.
The velocity field at an arbitrary point $ (\theta, \phi)$ on the sphere due to a point force localized on the membrane surface at $ (\theta_0, \phi_0)$ is summarized by an Oseen tensor on ${\mathbb S^2}$ given by
${\bf v} = {\bf G}(\theta,\theta_0,\phi,\phi_0) {\bf F}$,
%
\begin{figure}
   \centering
\begin{tabular}{lcc}
\includegraphics[width=5cm]{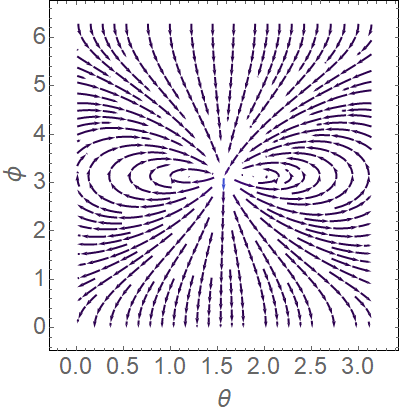}&
\raisebox{.1\height}{\includegraphics[width=4.5cm]{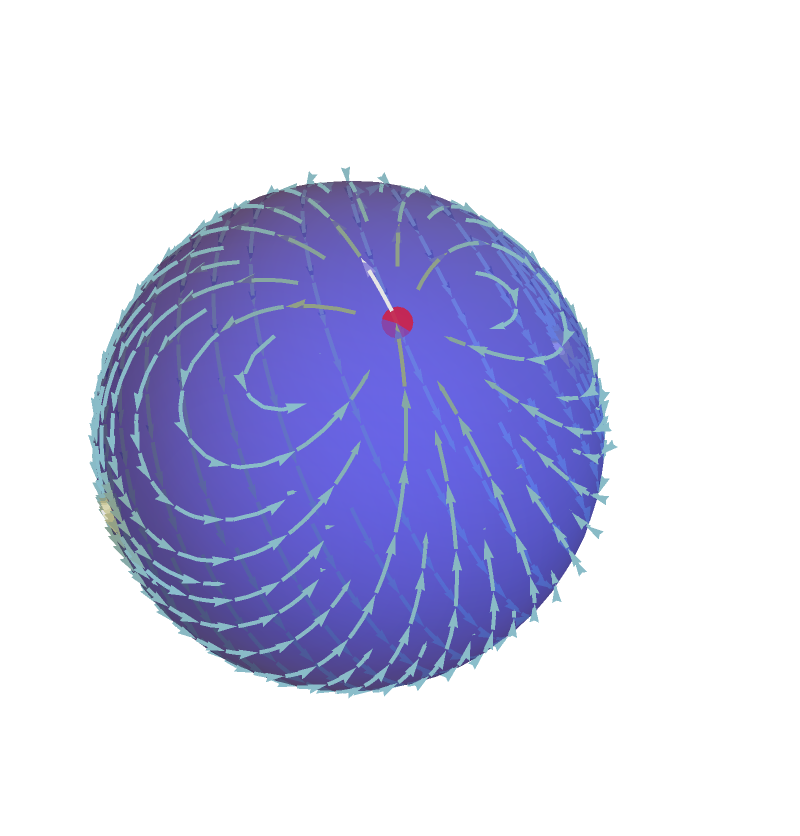}}\\
\includegraphics[width=5cm]{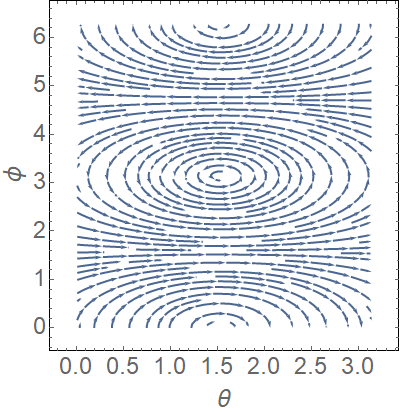}&
\raisebox{.4\height}{\includegraphics[width=3cm]{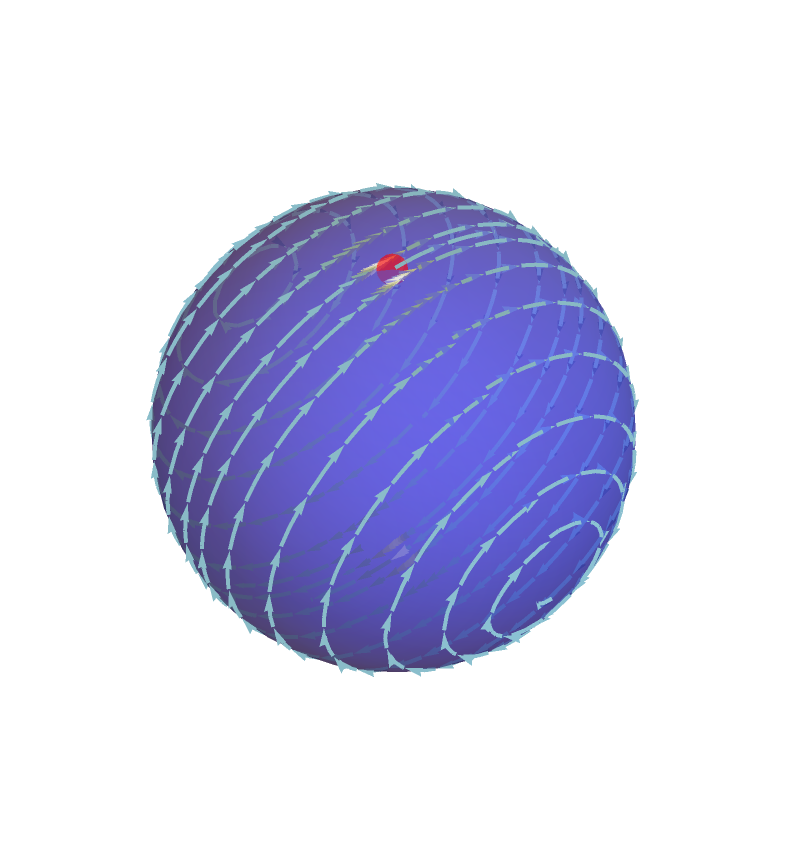}}\\

\end{tabular}
 \caption{Streamline plot of the velocity field in the low (top) and high (bottom) curvature regimes, in response to a force localized on the spherical membrane. On the left, the flow field is shown in a $\theta, \phi$ chart while on the right, the flow field is wrapped on a spherical membrane. Location of the force is marked in red. Note the creation of two vortical defects around the force. In the top row (low curvature regime) we show the flow field for a point force localized at $\theta=1.5$. In the bottom row (high curvature regime) for a point force localized at $\theta=0$, the vortices migrate to the equatorial regions. }
     \label{figptf} 
\end{figure}
where the different components of the Green's function can be expressed in terms of double derivatives of a function $S$
\beqa
& G_{\theta {\theta_0}} = \frac{\csc \theta \csc \theta_0}{ 4 \pi \eta_{2D}} \partial_\phi \partial_{\phi_0} S, \ \ \ 
G_{\theta {\phi_0}} = -\frac{\csc \theta }{ 4 \pi \eta_{2D}} \partial_\phi \partial_{\theta_0} S, \ \ \ 
G_{\phi {\theta_0}} =- \frac{\csc \theta_0}{ 4 \pi \eta_{2D}} \partial_\theta \partial_{\phi_0} S, \ \ \ 
G_{\phi {\phi_0}} = \frac{1}{ 4 \pi \eta_{2D}} \partial_\theta \partial_{\theta_0} S,
\eeqa
and the function $S$ is defined  in the basis of Legendre Polynomials,
$S := \sum_{l=1}^\infty \frac{ (2l+1)}{  s_l~ l (l+1)}  P_l (\cos \gamma)$,
where
$s_l=l(l+1) -2 +\frac{R}{\lambda_-}(l-1)+\frac{R}{\lambda_+}(l+2)$,
and $\cos \gamma$ is the cosine of the geodesic angle between the source at $ (\theta_0, \phi_0)$ and response at $ (\theta, \phi)$ 
 \beqa
\cos \gamma = \sin \theta \sin \theta_0 \cos(\phi -\phi_0) + \cos \theta \cos \theta_0.
\label{gdsc}
\eeqa
The function $S$ thus varies with the geodesic angle and the physical parameters, namely the sphere radius, and the membrane and solvent viscosities. The full structure of the function $S$ in real space is presented in Appendix \ref{apf} in terms of Appell Hypergeometric Functions. 
With the knowledge of the real space Green's function in the full parameter space at hand, we now plot the resulting flows due to a Stokeslet (point force) localized on the spherical membrane. In these plots we have chosen $\eta_+= \eta_- = \eta_{3d}$ in Eq.~\ref{saffl}. Thus, one can compare the radius of the sphere $R$ with respect to the unique Saffman Length  $\lambda = \frac{\eta_{2d}}{2 \eta_{3d}}$. There are two distinct regimes $R > \lambda$ (low curvature) or $R < \lambda$ (high curvature)\footnote{The ratio $\lambda/R$ is often quoted as the Boussinesq number in the surfactant dynamics literature.}.

In the low curvature regime, the velocity field  exhibits a dipole-like structure around the point of application of the force. The dipole has a topological index $+2$ which agrees with the Euler Characteristic of the sphere.  As the curvature is increased, the dipole structure breaks into two $+1$ vortices which migrate away to diametrically opposite points. These features were predicted first in \cite{henlev2008,henlev2010} and generalized to lipid bilayers with slip velocity in \cite{atz2016}. We observe that our real-space Green's function (Appendix \ref{apf}) also reproduces these effects. This provides a consistency check of our summation procedure explained in Appendix \ref{apf}. \\
Similarly for a force dipole, one expects the flow field to be characterized by 4 vortical defects surrounding a saddle of negative index at the core of the dipole. Besides there exists an additional saddle of negative index such that the net index is $+2$, the Euler Characteristic of the sphere. This additional saddle will be absent in the plane. Such force dipoles are used as models for a wide class of active inclusions, colloids, and 'swimmers' \cite{mik,mnk,chis}.

\subsection {Velocity field due to a point torque on the spherical membrane}
\label{sbtq}
\begin{figure}[h]
\begin{tabular}{ccc}
\includegraphics[width=5cm]{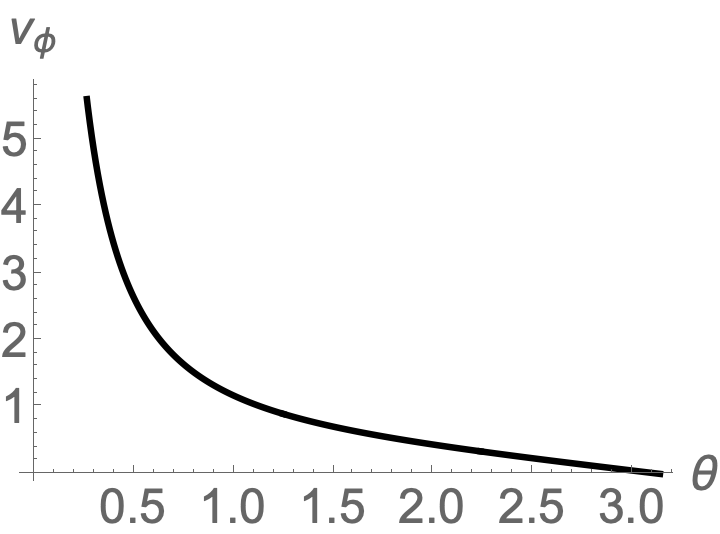}&
\includegraphics[width=4cm]{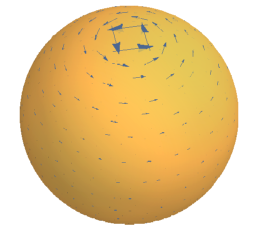}\\
\includegraphics[width=5cm]{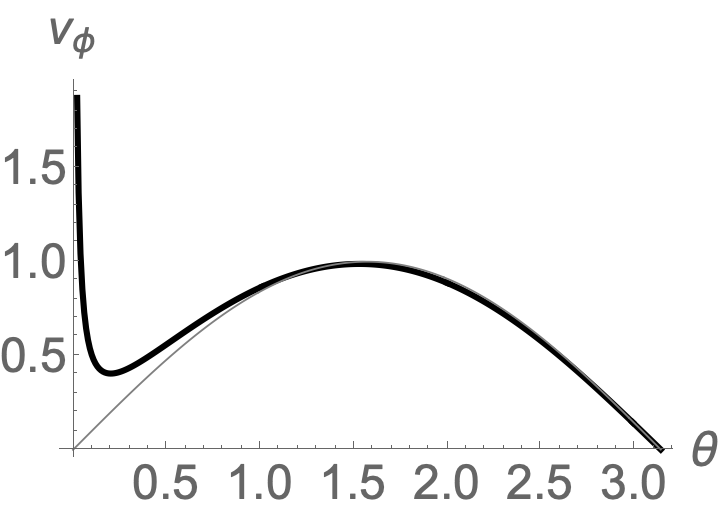}&
\includegraphics[width=2.7cm]{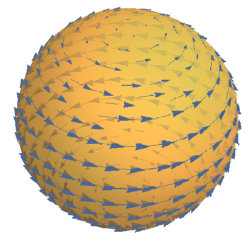}\\

\end{tabular}
\caption{Flow fields due to a vortex localized at north pole. Top row shows the low-curvature case (Eq.~\ref{reptq1}) and bottom row shows the high-curvature one (Eq.~\ref{reptq2}). On the left, the magnitude of the azimuthal velocity $v_{\phi}$ is shown in the $\theta$ direction. In the high curvature case the flow develops a local maximum due to the effect of the global rotation, shown as a gray line (see further discussion in Sec.~\ref{sbglb}). Right panels show the flow field wrapped on a spherical membrane.}
    \label{figptq} 
\end{figure}

The velocity field at a point  $ (\theta, \phi) $  produced by a point torque of circulation $\tau$ localized at $(\theta_0,\phi_0)$ on the sphere can be expressed as 
\beqa 
\boxed{{\bf v} =\frac{\tau}{\eta_{2D}} [\nabla_\perp ^{\mathbb S^2}] ~\bm{\psi}},
\label{strmtq}
\eeqa
where~ $[ \nabla_\perp ^{\mathbb S^2}] =\left(\hat {\theta} \frac{1}{R \sin \theta}\partial_\phi - \hat{\phi} \frac{1}{R} \partial_\theta\right)$ and $\bm{\psi}$ represents the dimensionless stream function. 
In terms of Legendre modes, $\bm{\psi}$ is given by
\beqa
\bm{\psi}[\theta, \phi, \theta_0,\phi_0]=\sum_l \frac{ (2l+1)}{4 \pi ~  s_l}  P_l (\cos \gamma). 
\label{appsum}
\eeqa
where
$\cos \gamma = \sin \theta \sin \theta_0 \cos(\phi -\phi_0) + \cos \theta \cos \theta_0$ is the cosine of the geodesic angle between the source and response locations. The real space representation of $\bm{\psi}$  in the full parameter space is presented in Appendix \ref{aptq}. There are two different representations of the stream function, one valid at low curvatures (Eq.~\ref{reptq1}) and the other valid at high curvature (Eq.~\ref{reptq2}). \\
We now focus on the associated topology of the flow-field due to the rotating inclusion localized at the north pole, see Fig. \ref{figptq}. We find that a new vortical defect of positive index (center) develops at the south pole to make the total index +2 as required by the topology of the sphere.\\
Let us briefly comment here on an interesting connection between the point vortex flows we study in this viscous set-up and the ideal point vortex problem on curved surfaces. For simplicity, let us consider the equation of stress balance for viscous hydrodynamics of the membrane fluid (with no external solvent \footnote{Let us note that the solvent contributions are important at low curvature, at distance scales beyond the Saffman length, similar to the planar situation. However they do not affect the flow topology and number of defects, when compared with ideal vortices on a sphere. Hence, in order to illustrate the connection to ideal vortex equations in the literature, we prefer to include only the curvature and membrane contributions. However, the solvent contributions will be important to determine the precise location of the defects. One also needs to keep the solvent contributions for studying the vortex dynamics eg. rotation rate of a two vortex configuration. For all such computations which appear in the later parts of the paper, we use the full solution for a point torque that includes solvent contributions.}) in the presence of a point rotor of unit strength. \footnote {Please see Appendix \ref{aptq} where we show that the membrane pressure vanishes in this situation.}
\beqa
 \eta_{2D} \left(K(x) + \Delta \right) v_\alpha = \epsilon_{\alpha \beta} D^\beta \underbrace{ \frac{\delta ( \theta- \theta_0) \delta ( \phi -\phi_0)}{R^2 \sin \theta}}_{\omega},
\eeqa
where we have defined $\omega$ as suggested in the above equation.\\
In terms of the scalar stream function defined as $ v _\alpha = \epsilon_{\alpha \beta} D^\beta \phi$, the above equation reads

\beqa
\eta_{2D}( 2 K(x) \epsilon_{\alpha \gamma} D^\gamma \phi + \epsilon_{\alpha \gamma} D^\gamma \Delta \phi) = \epsilon_{\alpha \gamma} D^\gamma \omega \nn\\
\eeqa
In the limit of low curvature where $K(x)$ can be ignored, we are left with\footnote {For the spherical membrane of constant curvature, the complete equation reads $w= \eta_{2D} (\frac{2}{R^2}  \phi +  \Delta \phi)$}
\beqa
\eta_{2D}   \Delta \phi = \omega.
\label{idv}
\eeqa
This equation is identical to the that of a point vortex in an ideal fluid on a curved surface, where $\Delta$ is the surface Laplace Beltrami Operator. Due to this equivalence, we expect that at low curvatures (and hence on the plane in particular) the response to rotating inclusions in a viscous fluid is similar to point vortex flows in an ideal fluid. The situation in the biological model we consider here departs from the ideal vortex problem once the curvature term becomes important. Indeed, we find that at low curvatures where the radius of the sphere is much larger compared to the Saffman length, the flows resemble those of ideal point vortex problem on the sphere. However, as the radius decreases, the curvature contributes to a zero mode that imparts a global rotation to the many body system, while the rotating inclusions individually still continue to interact in a manner similar to local point-vortex like interactions. At high curvatures, due to the effect of the global rotation, the flow is no longer monotonically decreasing, see Fig.~\ref{figptq}.\\

\subsection {Velocity field due to a torque dipole on the spherical membrane}
\label{sbctq}
Rotating inclusions that arise in biological examples feature no external torque e.g. rotor proteins such as ATP synthase. To account for this additional structure, we also construct a model of counter-rotating torque-dipole \cite{oppshelley, lenz1}. On a spherical membrane the solution is constructed in Appendix \ref{apctq}. The dynamics are very similar, mainly giving rise to a faster spatial decay. The solutions are identical in terms of their topology, and thus, in the rest of this paper we will focus on the solutions due to point rotors, but the Appendix outlines the results for torque-dipoles as well.

\subsection{ Emergence of Global Rotation at high curvature}
\label{sbglb}
Having understood the flow fields due to simple source terms on the spherical membrane, we now illustrate the emergence of the global rotation in the high curvature regime for all three situations. This global rotation was first reported in \cite{henlev2008,henlev2010} for a Stokeslet. In all the situations we studied, the dimensionless stream function on the spherical membrane has the generic structure in the basis of Legendre polynomials:
\beqa
\bm{\psi}  [\theta, \phi, \theta_0, \phi_0] =  \sum_l \frac{ f_l}{4 \pi ~  s_l~ g_l}  P_l (\cos \gamma), 
\label{stgen}
\eeqa
where $f_l$ and $g_l$ are polynomials in Legendre modes denoted by $l$ and $s_l=l(l+1) -2 +\frac{R}{\lambda_-}(l-1)+\frac{R}{\lambda_+}(l+2)$. The geodesic angle between the source and response locations is denoted by $\gamma$.
One can understand the emergence of the global rotation from the common denominator $s_l$ arising in all three situations. For simplicity, let us assume the Saffman lengths associated with external and internal solvents to be the same and denote it by $\lambda$. The classification of high and low curvature regimes is then simply determined by the ratio $\lambda/R$. \\
In the high curvature limit of $\lambda/R \gg 1$ 
\beqa
&s_l \sim l(l+1)-2.
\eeqa
The zero mode $l=1$ dominates the Legendre sum in the stream function of Eq.~\ref{stgen} and generates the global rotation. To see this, let us consider the situation where we place a rotating inclusion at the north pole.  The stream function in Eq.~\ref{stgen} leads to a velocity field with flows only in the azimuthal direction given by 
\beqa
v_{\phi}=\frac{\tau}{4 \pi \eta_{2d} R} \sum_l \frac{(2l+1) P_l^1 (\cos \theta)}{s_l}, 
\eeqa
where $ P_l^1 (\cos \theta)$ denotes the associated Legendre function of first order. In the limit of high curvature, the zero mode $l=1$ dominates the sum and we get
\beqa
v_{\phi}=\frac{\tau \sin \theta}{4 \pi R^2 \eta_{+} },
\label{vph}
\eeqa
which corresponds to a global rotation of the flow with 
\beqa
\Omega= \frac{\tau }{4 \pi R^3 \eta_{+} }.
\label{omega_one}
\eeqa
Let us note that the global rotation rate is purely regulated by the external solvent. This can be physically argued as follows : In the limit of high curvature, the zero mode causes the entire spherical membrane along with the internal solvent to rotate like a rigid body. The zero mode leads to zero dissipation in the membrane and internal fluid. Hence the primary shear in this situation is provided by the external solvent. In general, one expects the global rotation to be present for all closed compact membrane surfaces with a finite volume of fluid inside (internal solvent). However, for non-compact surfaces such global rotation is not possible because the fluid velocity field has to decay rapidly towards asymptotic infinity.  \\
One can also use the asymmetry of external and internal solvents at a fixed radius R to generate this global rotation. This happens when $\lambda_+ \gg \lambda_-$.
\beqa
&s_l=l(l+1) -2 +\frac{R}{\lambda_-}(l-1)+\frac{R}{\lambda_+}(l+2) \nn\\ 
& \sim l(l+1) -2 +\frac{R}{\lambda_-}(l-1).
\eeqa
Let us note that $l=1$ continues to be the zero mode in this situation leading to the global rotation. This is unlike the opposite limit when $\lambda_+ \ll \lambda_-$.  Viewed this way, the high curvature regime corresponds to reducing the external solvent viscosity such that $\eta_+ \ll \eta_-$.\\

It is interesting to note how our real space representation of the stream function Eq.~\ref{strmtq} in terms of Appell Hypergeometric functions capture the global rotation. As shown in Appendix\ref{aptq}, \ref{apctq}, depending on the roots of $s_l=0$ (the roots are analyzed in Appendix \ref{aproot}) the Legendre sum defined in Eq.~\ref{appsum} leads to two different representations of the stream function in real space.  One of these representations (Eq.~\ref{reptq1}) is valid in the regime of low curvature, while the other (Eq.~\ref{reptq2}) is valid in the high curvature regime. The appropriate stream function in the high curvature regime indeed shows a dominance of the global rotation term, see Fig. \ref{figptq}.\\

\section{Streamline Topology for Membrane Vortices}
\label{scst}
In this section we set up the equations needed to explore the streamline topology of flows on the biological membrane due to rotating inclusions. 
\subsection{Dynamic Equations for an ensemble of membrane vortices}
\label{sbgeq}
Let us consider $N$ rotating inclusions embedded in a spherical membrane with viscosity $\eta_{2d}$, surrounded by external solvents with viscosities $\eta_\pm$. The evolution equations for purely hydrodynamic interactions between the vortices are given by
\beqa
\dot{\theta}_i= \frac{1}{\eta_{2d} R^2} \sum_{j \neq i}^N \frac { \tau_j}{\sin \theta_i}~~ \bm{\psi}^{\prime} [\gamma _{ij}]~~ \partial_{\phi_i} [ \gamma_{ij}],\nn\\
\dot{\phi}_i=- \frac{1}{\eta_{2d} R^2} \sum_{j \neq i}^N \frac { \tau_j}{\sin \theta_i}~~ \bm{\psi}^{\prime} [\gamma _{ij}]~~ \partial_{\theta_i} [ \gamma_{ij}].
\label{dynmeq}
\eeqa
where $\bm{\psi}^{\prime}$ is the derivative of the stream function defined in Eq.~\ref{strmtq} for point rotors, with the explicit structure in terms of Appell Hypergeometric Functions given in Appendix \ref{aptq} by Eqs.~\ref{reptq1} and \ref{reptq2} for the low and high curvatures respectively (see Eq.~\ref{repctq1}, Eq.~\ref{repctq2} in Appendix \ref{apctq} for the corresponding expressions for torque dipoles).  Let us note that the stream function has two different representations that are dictated by curvature. While performing the dynamical simulations, one needs to insert the appropriate representation of $\bm{\psi}$ into Eq.~\ref{dynmeq}. Finally, $\gamma_{ij}$ is defined in Eq.~\ref{gdsc}.\\
Let us add some comments on the absence of a self-drive term in Eq.~\ref{dynmeq}. This can be argued from symmetry considerations. Due to spherical symmetry, there is no preferred direction and a single vortex does not move. This argument works for the planar situation as well. However, in generic surfaces with no (or restricted) symmetry, one needs to treat the self interaction term with a proper regularization procedure. This will in general lead to a self drive term. For the spherical membrane however, such a regularization leads to a constant (due to symmetry) and has no effect on the dynamics.\\
We will be interested in the flow fields resulting from the above dynamics as well. For this purpose, one constructs the hydrodynamic velocity field at any given point $p$ via superposition (taking into account the contributions from all point rotors).
\beqa
v_{\theta_p}=   \frac{1}{\eta_{2d} R} \sum_{j }^N \frac { \tau_j}{\sin \theta_p}~~ \bm{\psi}^{\prime} [\gamma _{pj}]~~ \partial_{\phi_p} [ \gamma_{pj}],\nn\\
v_{\phi_p}=  - \frac{1}{\eta_{2d} R} \sum_{j }^N  \tau_j~~ \bm{\psi}^{\prime} [\gamma _{pj}]~~ \partial_{\theta_p} [ \gamma_{pj}].
\label{field}
\eeqa
where $\gamma_{pj}$ denotes the geodesic angle between  $ (\theta_p, \phi_p)$ and  $ (\theta_j, \phi_j)$ and now the sum runs over all vortices. 

\subsection { Hamiltonian Formulation}
\label{sbhm}
The dynamical equations (Eq. \ref{dynmeq}) can be cast in terms  of a geometric Hamiltonian. In terms of canonical coordinates, $Q_i= \sqrt{|\tau_i|} \phi_i$,  $P_i= \sqrt{|\tau_i|} \cos \theta_i$, Eq. \ref{dynmeq} can be re-written as \footnote { Let us mention that the so defined canonical momentum $ P $ in our Hamilton description is thus a function of the coordinate $\theta$. The Poisson bracket relation $[Q_i, P_j] =\frac{\delta_{ij}}{ \tau_i}$ thus implies that position coordinates $\theta, \phi$ do not commute in this geometric formulation. This may seem surprising at first sight, however this is not uncommon. For example, consider motion of electrons in a magnetic field (say in the $\hat{z}$ direction),  where the canonical momentum is given by $p_x = m \dot{x} -\frac{e B}{m c} y$. The canonical Poisson brackets ---  $[x, p_x] =1 $  in the large $B$ limit reads $[x,y] \sim -\frac{m c}{e B} \neq 0$. Indeed this has been a major motivation to explore if vortices in superfluids can form quantum Hall states \cite{viefers, beeler, tim}. }
\beqa
\dot{Q}_i = \partial_{P_i} H, ~~ \dot{P}_i =- \partial_{Q_i} H,\nn\\
H= \frac{1}{R^2 \eta_{2D}}\sum _{i<j} \tau_i \tau_j \bm{\psi}[ \gamma_{ij}].
\label{hm}
\eeqa
In the torque dipole case the stream function is given by Eq.~\ref{strmctrq}. \\

In general, it is expected that the dynamics of the rotating inclusions on a surface of a non-trivial topology and curvature will be different from the planar problem. We expect significant changes in the streamlines of the flow fields as well. Let us  recall that in the low curvature limit, there exists a structural similarity between the equations of viscous hydrodynamics (sourced by point rotors) with vortices in an ideal fluid (Eq.~\ref{idv}). This motivates us to borrow some terminology and concepts from vortex literature \cite{bgm,km,arefbrons,newton2000, boatto1, boatto2} that will prove useful: 
\begin{enumerate}
\item {\bf Hairy Ball Theorem.} The theorem forbids the existence of a nowhere vanishing vector field on the sphere, i.e. there is at least one point where the flow field vanishes on the sphere.\\

\item {\bf Poincare Index Theorem.} The topological defects in the flow field on a closed, compact surface can be assigned an index corresponding to the winding of the field around the singular core. The theorem implies that the sum of the indices over all singularities is equal to the Euler Characteristic of the surface. For a spherical membrane (or membranes deformable to a sphere), the Euler characteristic is 2. As we will see later,  the topology of the spherical membrane leads to creation of $N_c$ new centers (vortex defects of index +1 where the velocity field vanishes, i.e. a stagnation point of the flow-field), a phenomenon not observed on the plane \cite{arefbrons}. Such centers have unit positive index. In addition, one has $N_S$ saddles of unit negative index (an anti-vortex). Further, each of the $N$ rotating inclusions that we consider contribute a positive index of +1. The index counting thus demands
\beqa
N + N_c - N_s =2.
\label{indxth}
\eeqa

The circulation $\tau$ of the vortices is independent of the index. In particular, both positive and negative circulations have the same index +1. \\

\item {\bf Integrability and the Liouville-Arnold theorem.} A Hamiltonian system  with $2N$ dimensional phase-space is  integrable  if there exist $N$ independent
integrals of motion which are all mutually Poisson
commuting, i.e. they are in mutual involution. The Hamiltonian we constructed in Eq.~\ref{hm} has the same set of symmetries as ideal point vortices on a sphere which have 3 mutually commuting conserved quantities. Thus, the system loses integrability for $N \geq 4$ vortices and the $N = 4$ situation is integrable if the total circulation of vortices is zero \cite{newton2000}.\\

\end{enumerate}

We are now ready to discuss the dynamics of the vortices and spatio-temporal evolution of vortical defects in the flows within the biological membrane. The basic methodology we adopt here is very simple: we first simulate the dynamics of the vortices as given in Eq.~\ref{dynmeq} and feed the dynamic locations of the vortices into Eq.~\ref{field} to get the hydrodynamic vector field $(v_\theta, v_\phi)$. Using Mathematica \cite{wolfram}, we next plot the streamlines associated with this flow field in the $(\theta, \phi)$ chart. \footnote { Drawing the streamlines on the ($\theta, \phi$) chart is convenient to keep track of the evolution of vortical defects over the entire spherical domain. However one can equally well wrap the flow field on the spherical membrane.} In order to understand the dynamics and flows better, we restrict our discussions in this section to two vortices and discuss the many vortex situation in the next section.  We separate the discussion of two vortices into regimes of low and high curvature. In each regime, we first consider the simple situation of two vortices with same (opposite) circulations $\tau$, discuss the dynamics and the associated streamline topology of the flow field. In appropriate places, we comment on important distinctions from the planar dynamics and flows. From our discussions near Eq.~\ref{idv}, we expect the dynamics and flows to be similar to ideal vortices in the low curvature regime. The flow departs from ideal vortices once curvature becomes important.

\subsection{ Low Curvature Regime }
\label{sblc}

\begin{figure}[h]
 \centering
\begin{tabular}{lcc}
\includegraphics[width=5cm]{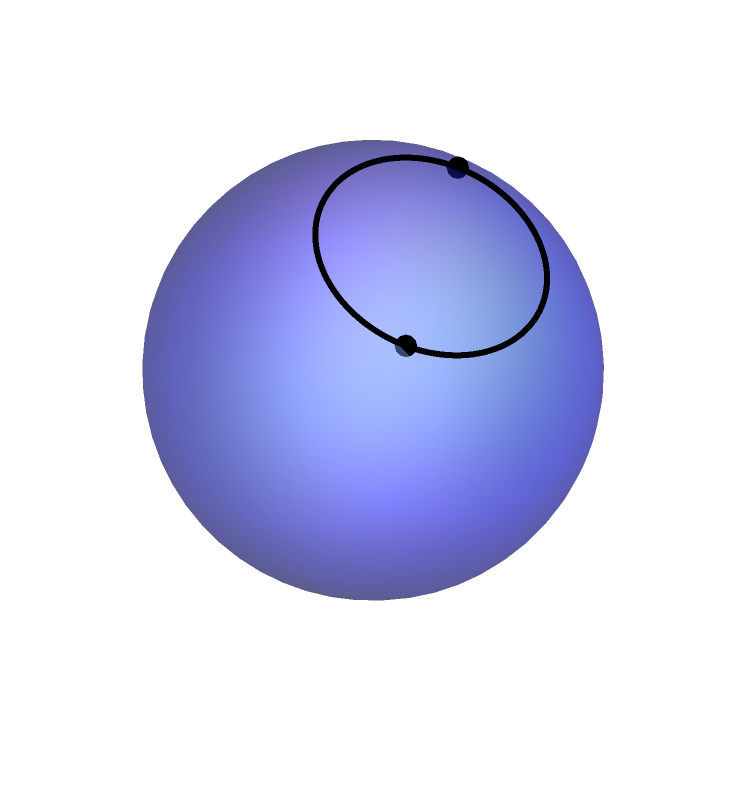}&
\includegraphics[width=5cm]{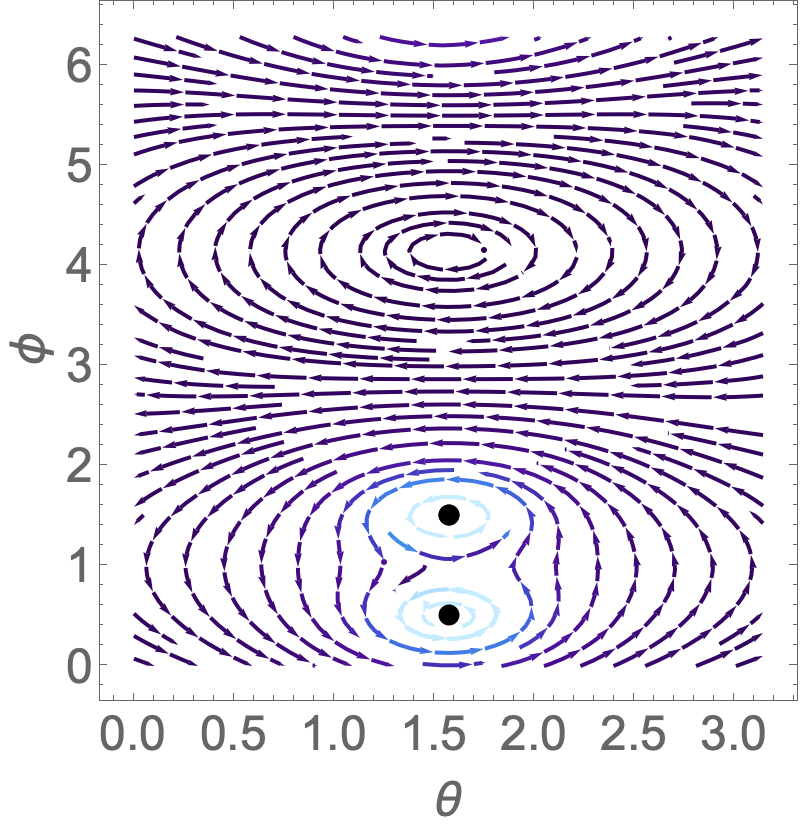}&
\includegraphics[width=5cm]{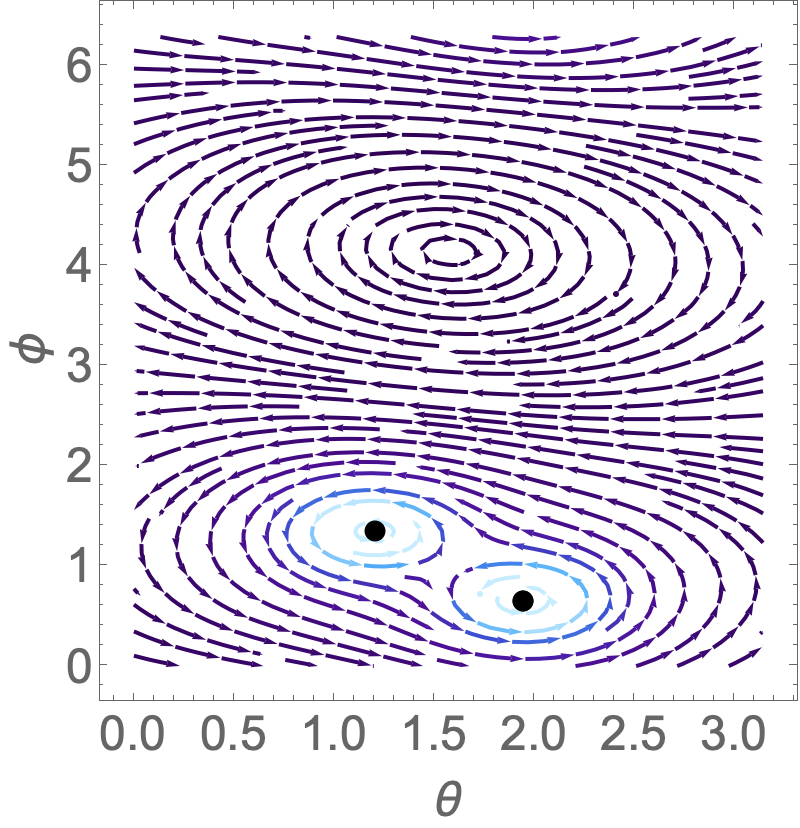}\\
\includegraphics[width=5cm]{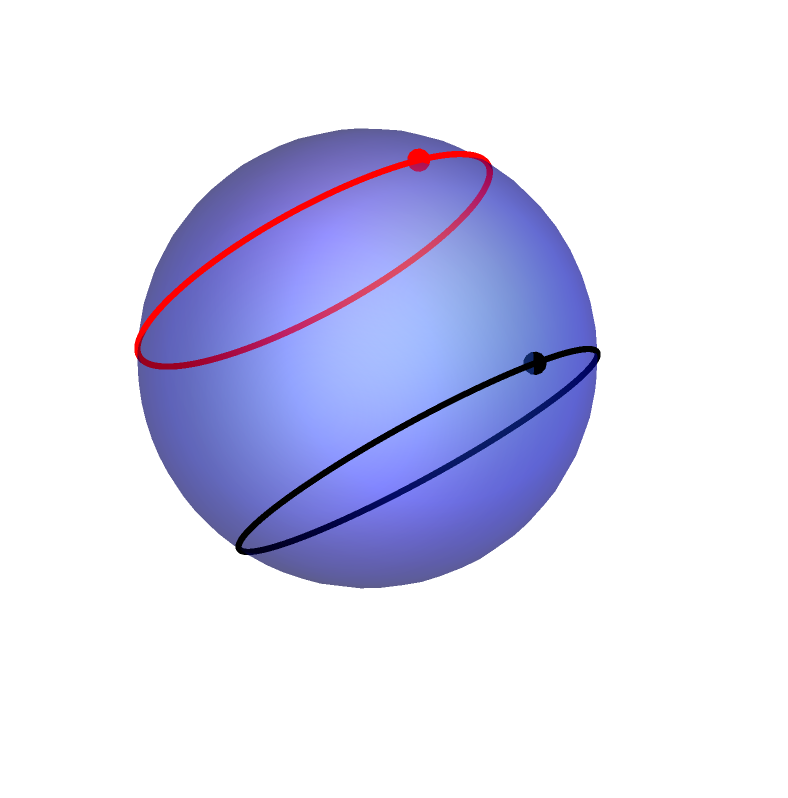}&
\includegraphics[width=5cm]{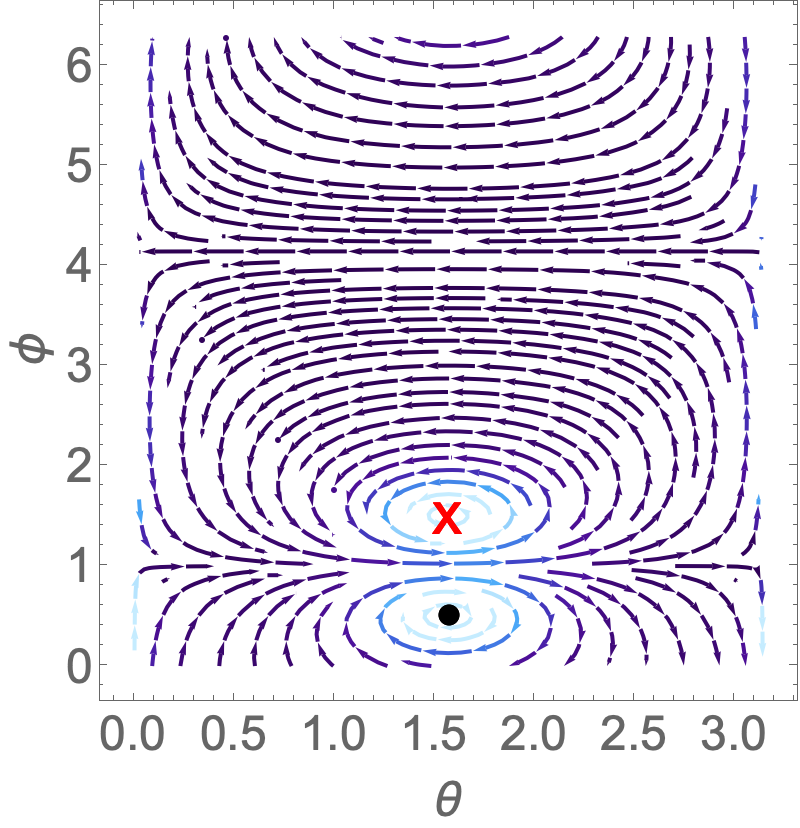}&
\includegraphics[width=5cm]{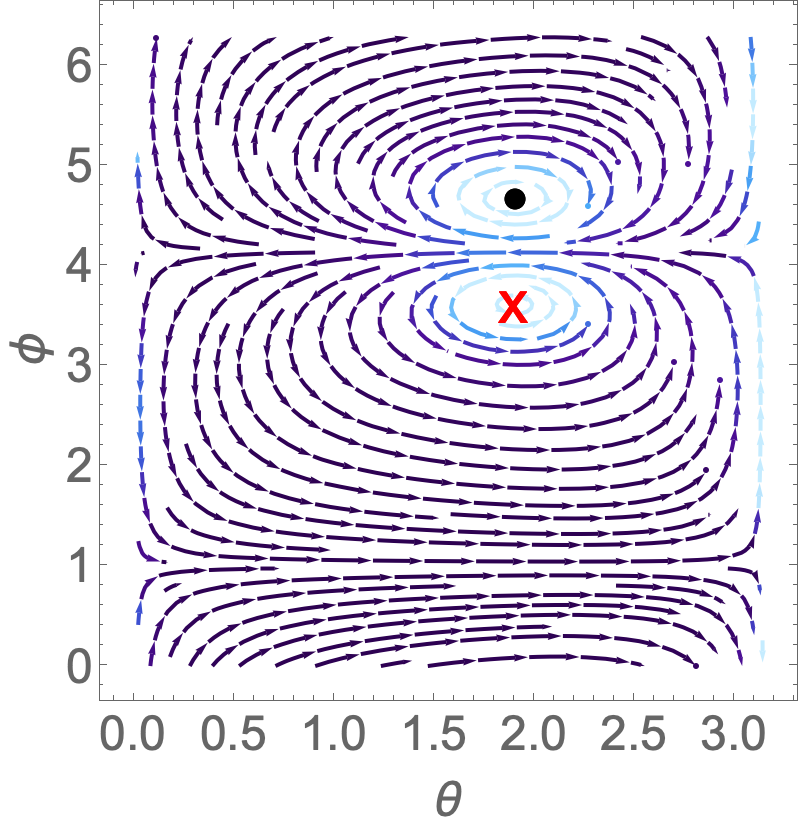}\\
\end{tabular}

    \caption{Streamline plot for two vortices at the low curvature regime, both released at the equator with one at $\phi =0.5$ and the other at $\phi =1.5$ in the low curvature regime, top row with same strength and bottom row with opposite strength, red marks the vortex with the negative circulation. Color code throughout the text signifies the magnitude of the velocity, going from dark blue to white with increasing magnitude. }
        \label{figlc2} 
\end{figure}

\begin{figure}[h]
 \centering
\includegraphics[width=4cm]{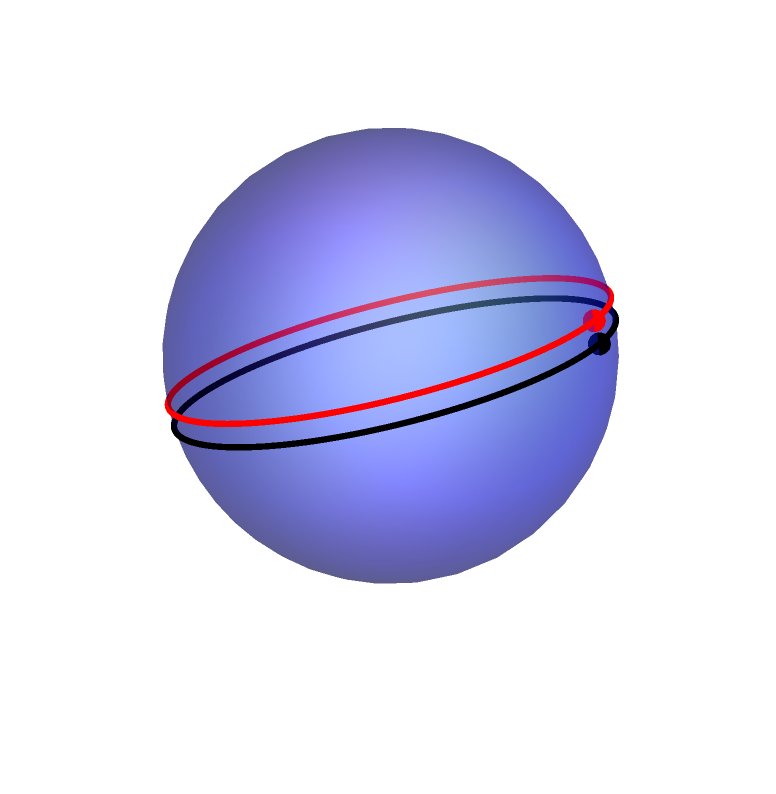}\\
 \caption{Vortex dipole at low curvature traces a geodesic.}
        \label{figkmlc} 
\end{figure}

\begin{figure}[h]
\begin{tabular}{lcc}
\includegraphics[width=5cm]{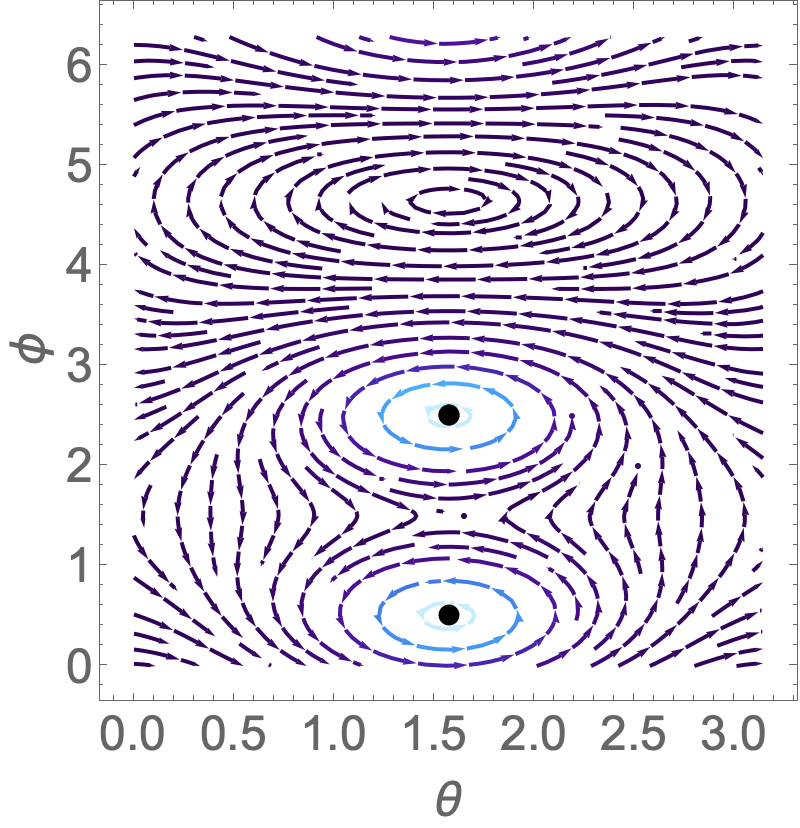}&
\includegraphics[width=5cm]{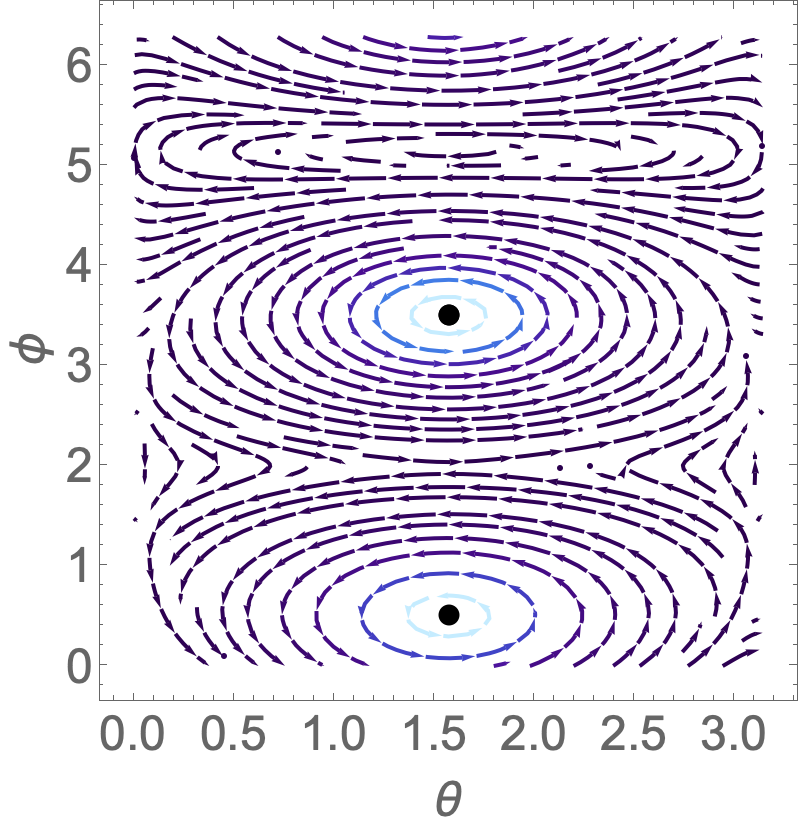}&
\includegraphics[width=5cm]{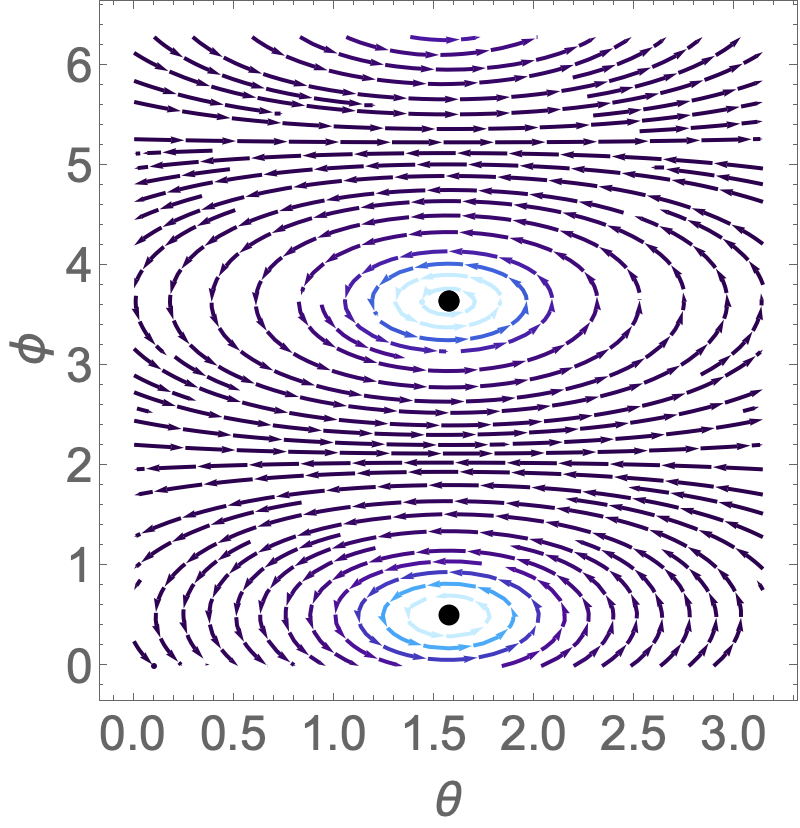}

\end{tabular}
   
    \caption{Location of staganation points for various relative distances between two vortices of the same strength. Plots are shown for $\Delta \phi = 2,3, \pi$ in the low curvature regime in the $\theta-\phi$ plane. A saddle stagnation point (anti-vortex) is shown between the two positive vortex defects. A continuum of stagnation points occur at $\phi=\pi$ shown in the rightmost figure.}
     \label{figlc2_sp} 
\end{figure}

A single vortex does not move due to spherical symmetry. The dynamics becomes interesting once we have two vortices or more. To understand the dynamics of two vortices better, we first note that the Hamiltonian Eq.~\ref{hm} preserves the chord distance $C_{12} =|\vec{X}_2 -\vec{X}_1|$ between the vortices and they orbit around each other. The rotation rate can be easily estimated by converting the dynamical equations for two vortices (Eq.~\ref{dynmeq}) into Cartesian form:
\beqa
\frac{d}{dt}\vec{X}_1 =\frac{\tau_2}{R \eta_{2D}}  \frac{\bm{\psi}^\prime[\gamma[C_{12}]]}{\sin[\gamma[C_{12}]]} \frac{\vec{X}_1 \times \vec{X}_2}{R^2}, \nn\\
\frac{d}{dt}\vec{X}_2 =\frac{\tau_1}{R \eta_{2D}}  \frac{\bm{\psi}^\prime[\gamma[C_{12}]]}{\sin[\gamma[C_{12}]]} \frac{\vec{X}_2 \times \vec{X}_1}{R^2}.
\label{2rotlc}
\eeqa
where $\gamma[C_{12}]$ is related to the chord distance $C_{12} =|\vec{X}_2 -\vec{X}_1|$ via $ \gamma[C_{12}] = 2  \arcsin \left[\frac{C_{12}}{2R}\right]$.\\
The center of vorticity vector $\vec{M}$ in this situation is given by
\beqa
\vec{M} = \frac{\tau_1 \vec{X}_1 + \tau_2 \vec{X}_2}{\tau},
\label{cov2}
\eeqa
where $\tau = \tau_1+\tau_2$ is the total circulation. It is easy to see that Eq.~\ref{2rotlc} can be written using the above vector $\vec{M}$   as
\beqa
\frac{d}{dt}\vec{X}_1 =\frac{1}{R \eta_{2D}}  \frac{\bm{\psi}^\prime[\gamma[C_{12}]]}{\sin[\gamma[C_{12}]]} \frac{\vec{X}_1 \times \tau \vec{M}}{R^2}, \nn\\
\frac{d}{dt}\vec{X}_2 =\frac{1}{R \eta_{2D}}  \frac{\bm{\psi}^\prime[\gamma[C_{12}]]}{\sin[\gamma[C_{12}]]} \frac{\vec{X}_2 \times \tau \vec{M}}{R^2}. 
\label{2rotlcM}
\eeqa
\\ From Eq.~\ref{2rotlcM}, we can read the rotation rate at low curvature $\omega^{2 vortices}_{LC}$ as
\beqa
\boxed{\Omega^{2 vortices}_{LC} =\frac{ \tau |\vec{M}|}{ R^3 \eta_{2D}} \frac{\bm{\psi}^\prime[\gamma[C_{12}]]}{\sin[\gamma[C_{12}]]}.}
\label{omlc}
\eeqa
Let us note that the same formula also holds for the torque-dipole case with appropriate $ \bm{\psi}$ given in Appendix \ref{apctq}. We now elaborate on the dynamics for two vortices in Fig.~\ref{figlc2}.\\
\textbf{Same Circulation}: In this situation, the vortices orbit each other with an angular frequency $\omega^{2 vortices}_{LC}$ given by Eq.~\ref{omlc}. In Fig.~\ref{figlc2} we show an example of an orbit (the trajectory of the vortices) and the associated flow fields at two instants of time. One observes that in addition to the original centers created by the vortices themselves, the flow field exhibits a new center and a saddle where the velocity vanishes. Overall the index adds up to 2 , consistent with Poincare Index Theorem, as expected from the topology of the spherical membrane. In Fig.~\ref{figlc2_sp} , we show how the location of stagnation points change as we change the distance between the vortices (of same circulation). In particular, let us note that when the distance is $\pi$ then there is a continuum of stagnation points formed  along the mid-line between them (left most figure in lower panel of Fig.~\ref{figlc2_sp}). The saddle is always formed between the vortices, as can be seen also from continuity of the flow.\\
\textbf{Opposite Circulation}: In this case the vortices move together, such that the perpendicular bisector of the line joining them follows a geodesic. Interestingly, the flow fields exhibit no new center or saddle. The Poincare Index Theorem is still satisfied, since the index contribution from the two vortices of opposite circulation is 2. 
Let us mention that Kimura \cite{km} predicted that a vortex dipole (vortices of equal and opposite strength placed close to each other) traces a geodesic in all closed Riemann Surfaces deformable to the sphere. For our spherical membrane, we indeed find this property holds true, see Fig.~\ref{figkmlc}.\\

\begin{figure}[h]
   \centering
\begin{tabular}{ccc}

\raisebox{.5\height}{\includegraphics[width=5cm]{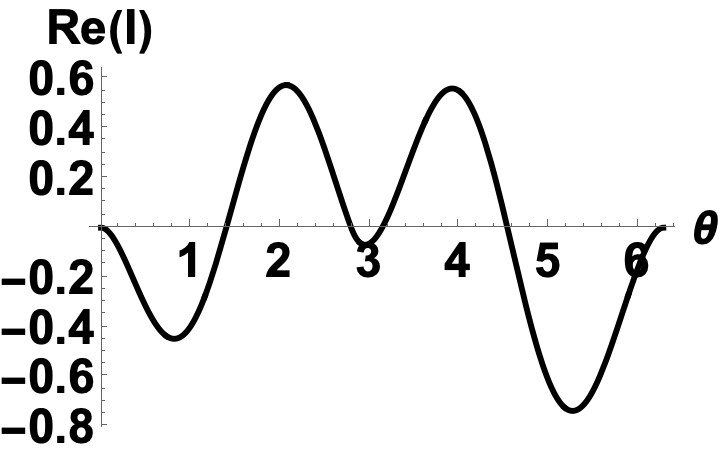}}&
\raisebox{.5\height}{\includegraphics[width=5cm]{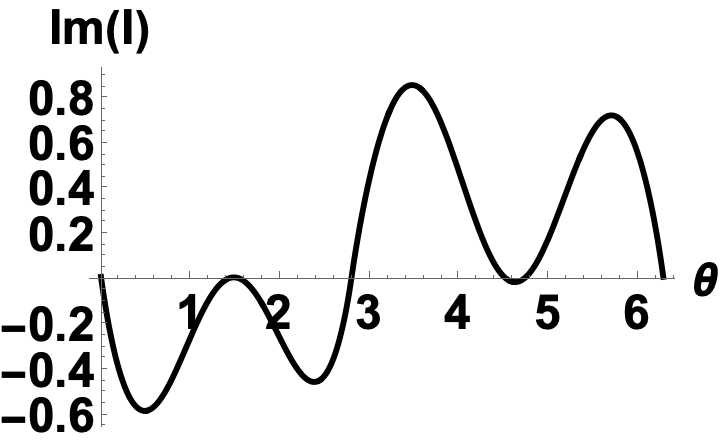}}&
\includegraphics[width=5cm]{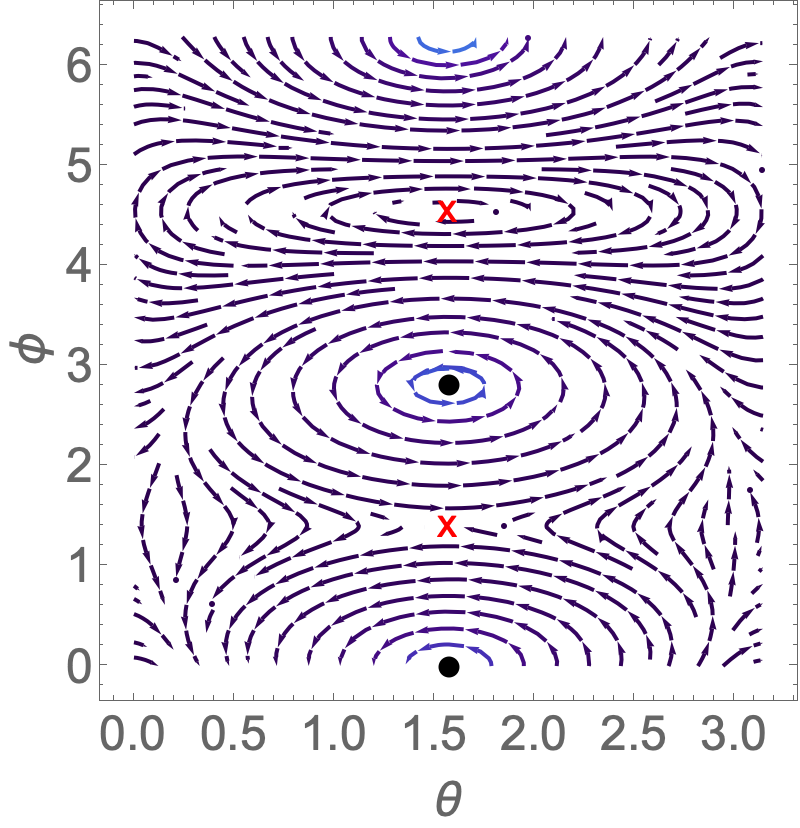}

\end{tabular}
    
    \caption{Zeros of the Real and Imaginary parts of the L.H.S. of Eq.~\ref{rootlc_txt} as a function of $\theta_p$. The new stagnation points are located at around the common zeroes i.e.  $\theta_p= 1.4, ~ 4.5$, marked with a red cross in the streamline plot on the right.}
    \label{figrootlc} 
\end{figure}

\textbf { Calculation of Stagnation Points.} We now proceed to a calculation of the locations of the stagnation points. As shown in Appendix \ref{apstg}, one can project the dynamical equations, Eq.~\ref{dynmeq}, via stereographic projection on the plane. 
Using complex coordinates to denote the locations of the vortices, the equation of a tracer particle in the presence of the vortices can be cast in complex notation
  \beqa
 \frac{d}{dt} \bar{z}_p = \frac{i}{\eta_{2D} R^2} \frac{(1+ |z_p|^2)^2}{2} ~\partial_{z_p} H_p,
 \label{hmcp_txt}
 \eeqa
 where $z_p$ denotes the location of the tracer particle,   $H_p$  is  defined using the streamfunction $\bm{\psi}$ (with the structure presented in appendix \ref{aptq} , see appendix \ref{apctq} for the torque dipole case)
  \beqa
H_p= \sum_{j}^N \tau_j ~\bm{\psi}[\gamma_{pj}],
\label{hmp_txt}
\eeqa
where the geodesic distance in complex notation is given by
 \beqa
\gamma_{pj}= \arccos \left(\frac{(1-|z_p|^2)(1-|z_j|^2)+4 ~ Re[z_p \bar{z}_j ]}{(1+|z_p|^2)(1+|z_j|^2)}\right).
\eeqa
It follows that solving for the stagnation points amounts to finding solutions to
  \beqa
\frac{d}{dt} \bar{z}_p= 0.
  \label{stgeq_txt}
 \eeqa
 As shown in Appendix~\ref{apstg}, this amounts to solving an equation of the general form 
 \beqa
 \sum_{j}^N \tau_j ~F[z_p,z_j]~ G [ z_p, z_j] =0,
\label{fgeq_txt}
\eeqa
where the factors $F$ and $G$ arise from the derivative of the stream function i.e. $\partial_{z_p} \bm{\psi} = \frac{\partial \bm{\psi}}{\partial \cos \gamma}~~ \partial_{z_p} \cos \gamma : = F \times G$. Let us note that although $F$ is dependent on the choice of parameters, the factor $G$ is purely geometric. To proceed further, we need to compute $F$ from the appropriate stream function $\bm{\psi}$. The full structure of the stream function in Appendix \ref{aptq} makes the analysis somewhat complicated, however one can choose a set of parameters for the model to simplify the stream function. As explained in Appendix \ref{apstg}, for a particular choice $\eta_{2D} = 3/2, \eta_- =1, \eta_+=2, R=1$ for which $\lambda/R =1/2$, the stream function is given a relatively simple expression (Eq.~\ref{strmsimple}). Specializing to the case of two vortices on the spherical membrane, let their positions in the complex plane be denoted by $z_1$ and $z_2$.  Since the stagnation points are always constrained to lie on the great circle joining the two locations \cite{newton2000}, we can essentially map the dynamics to the unit circle on the complex plane \footnote {Via stereographic projection, the  azimuthal angle on the spherical membrane maps to the polar angle on the plane.}. Using polar representation $z= r e^{i \tilde{\theta}}$, we choose without loss of generality
\beqa
z_1 = 1, z_2 = e^{i \phi}, \tau_1=1, \tau_2 = \tau.
\label{loc_txt}
\eeqa
Plugging Eq.~\ref{loc_txt} into Eq~.\ref{fgeq_txt} , we convert it to an effective two parameter problem where the stagnation point $ z_p = e^{i \theta_p}$ has to be solved as a function of the relative circulation $\tau$ and the location of the second vortex parametrized by $\phi$ , from the equation (see Appendix \ref{apstg} for a complete derivation)

\beqa
I = f[\cos \theta_p] \frac{1}{4} (1- e^{-2 i \theta_p}) + \tau f [\cos(\theta_p -\phi)] \frac{1}{2} \left(e^{-i \phi} - \cos(\theta_p -\phi) e^{-i \theta_p} \right) = 0,
\label{rootlc_txt}
\eeqa
where
\beqa
f= \frac{10 - 8 \sqrt{2-2 x} + x \left( -1 +5 \sqrt{2-2 x} +15 (-1+\sqrt{2-2 x}) x + 30 x~ (x^2-1) \arccoth(1+\sqrt{2-2 x})\right)}{12 \pi (x^2-1)}. 
\eeqa
We can now systematically search for the location of the new stagnation points $\theta_p$  as a function of the relative vortex circulation $\tau$ and relative distance between the two original vortices $\phi$.  As an example we plot the results for $\phi=2.8$, and $\tau=1$. We plot the real and imaginary parts of $I$ in Eq.~\ref {rootlc_txt} as a function of $\theta$. The common zeros of the plots on the left of Fig.~\ref{figrootlc} are in good agreement with the streamline plot on the right.

\subsection{ High Curvature Regime}
\label{sbhc}
As we saw in Sec.~\ref{scsetup}, the curvature term in Eq.~\ref{curvedstokes2d}  imparts a global rotation to the system. For a single vortex, we predicted this rate in Eq.~\ref{omega_one}. In the generic situation of more than one interacting vortices  with varied circulations, one can proceed as follows:\\
In the high curvature regime, we saw in section \ref{sbglb} that the  $l=1$ term dominates the Legendre sum in Eq.~\ref{stgen}. This term, as we saw in earlier sections, leads to a global rotation. The global rotation rate for a system of many rotating inclusions can be easily extracted by noting that for $R \ll\lambda$, as far as global effects are concerned, one can ignore the local hydrodynamic interactions and keep only the global term in the stream function appearing in the dynamical equation, Eq.~\ref{dynmeq}, i.e.
\beqa
\frac{\bm{\psi}^\prime[C_{ij}]}{\sin[\gamma[C_{ij}]]} \sim \frac{\eta_{2D}}{4 \pi R \eta_+} ~~~~~\forall ~(i,j).
\eeqa
Using this approximation in the high curvature regime gives \beqa
\frac{d}{dt}\vec{X}_i= \sum_{j \neq i} \frac{\tau_j}{R \eta_{2D}} \frac{\bm{\psi}^\prime[C_{ij}]}{\sin[\gamma [C_{ij}]]} \frac{\vec{X}_i \times \vec{X}_j}{R^2} \sim \sum_{j } \frac{\tau_j}{R \eta_{2D}} \frac{\eta_{2D}}{4 \pi R \eta_+} \frac{\vec{X}_i \times \vec{X}_j}{R^2} = \vec{X}_i  \times  \frac{\sum_{j } \tau_j \vec{X}_j}{4 \pi R^4 \eta^+}.
\eeqa

Thus the global rotation rate is
\beqa
\boxed{\omega^{N vortices}_{HC} \sim \frac{\tau |\vec{M}|}{4 \pi R^4 \eta_+},}
\label{glbomega}
\eeqa
where $\vec{M} = \frac{\sum_{j }^N \tau_j \vec{X}_j}{\tau}$
denotes the conserved center of vorticity vector, $\tau$ is the total circulation $\tau = \sum_j \tau_j$. Let us note that the above rate agrees with the one found for a single vortex Eq.~\ref{omega_one}. The same formula holds for the case of torque-dipoles, with the replacement $ \tau \rightarrow \frac{\tau d}{R}$  where $d$ is the finite distance between the counter-rotating inclusions, separated along the sphere radial direction
(see Eq.~\ref{strmctrq}).\\ 
 \begin{figure}[h]
 \centering
\begin{tabular}{lcc}
\includegraphics[width=5cm]{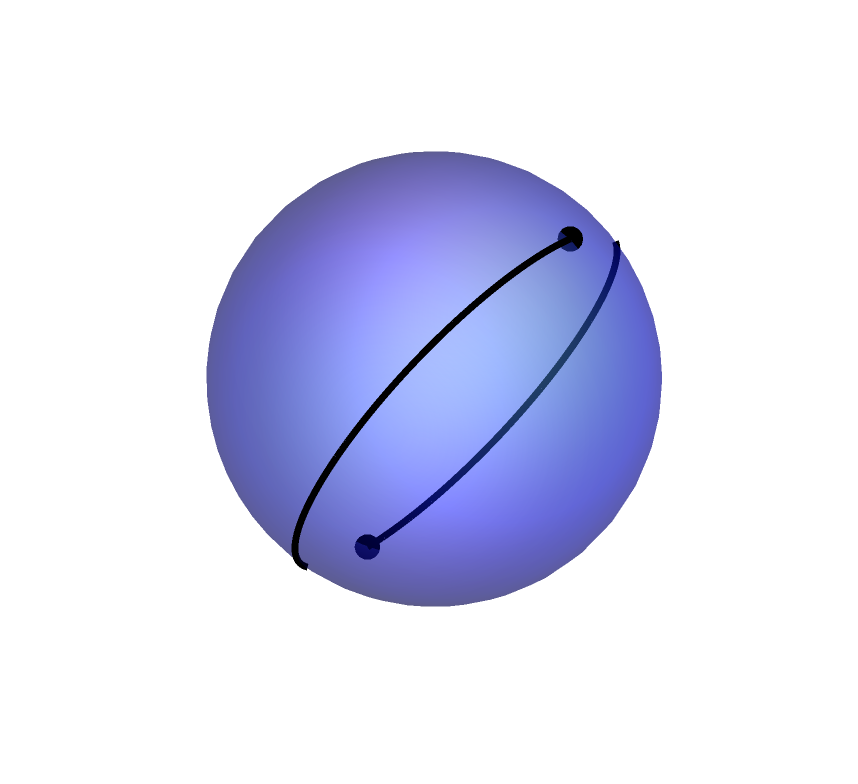}&
\includegraphics[width=5cm]{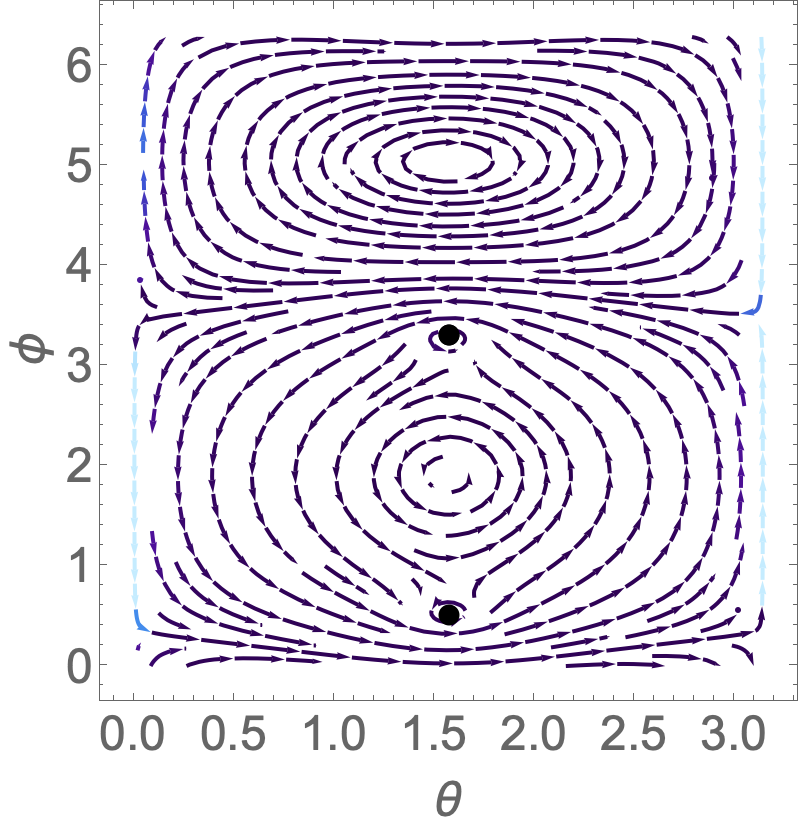}&
\includegraphics[width=5cm]{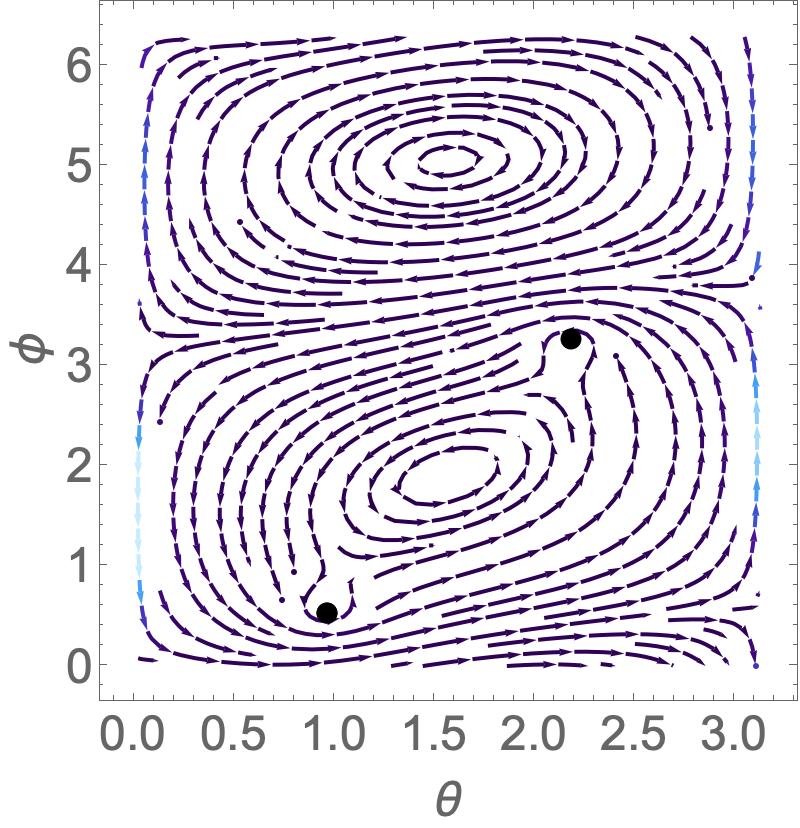}\\
\includegraphics[width=5cm]{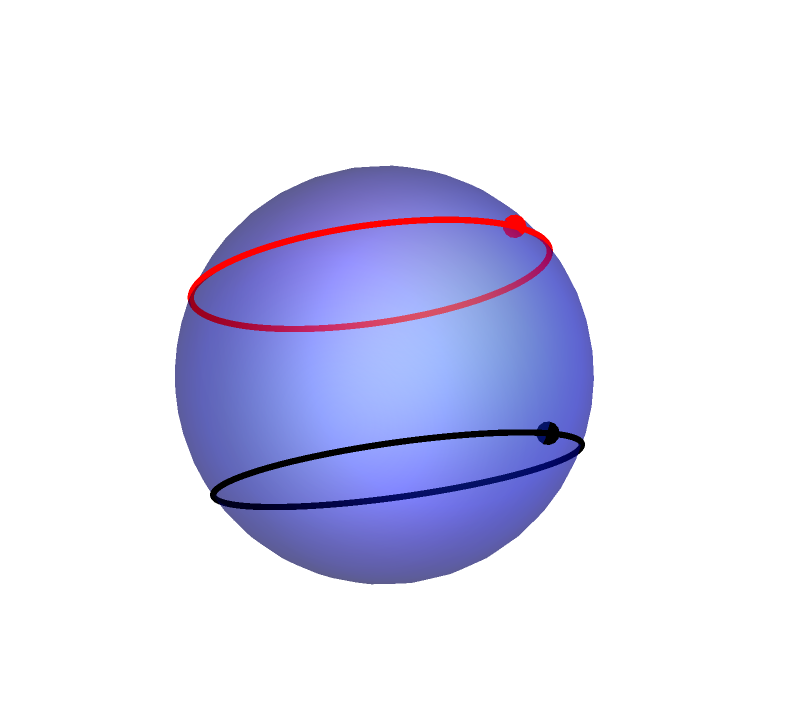}&
\includegraphics[width=5cm]{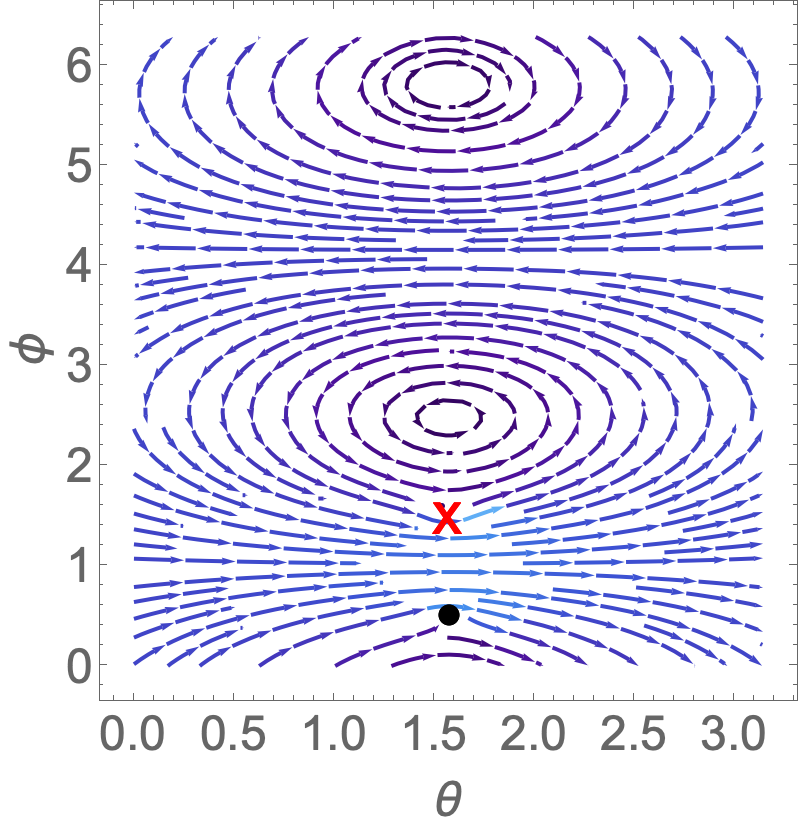}&
\includegraphics[width=5cm]{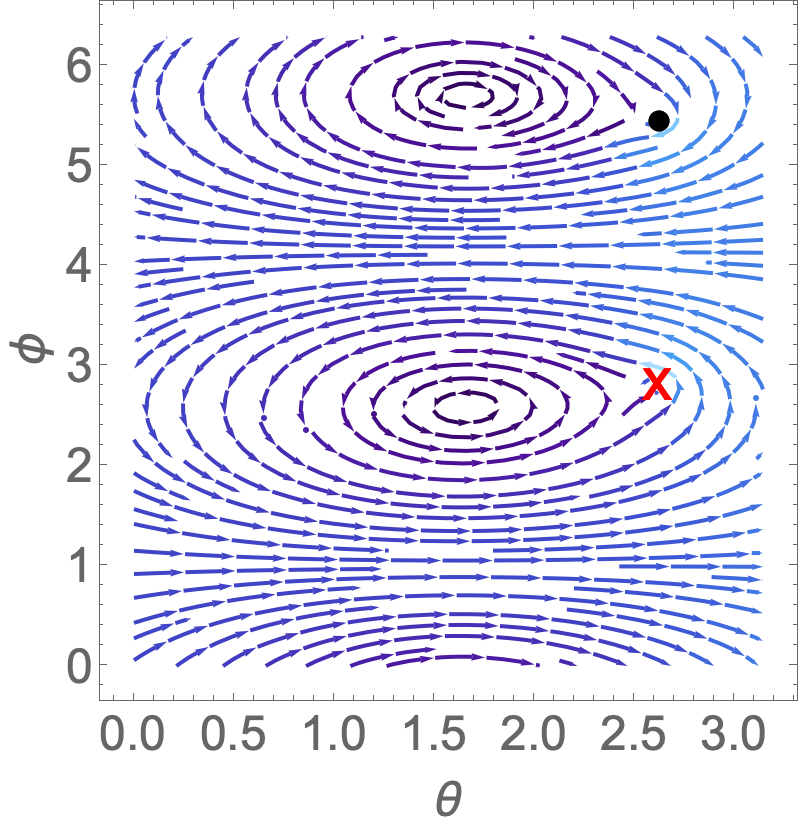}\\
\end{tabular}
    \caption{Streamline plot for two vortices, both at the equator in the high curvature regime, top row with same strength (positioned at $\phi =0.5$ and $\phi =3.3$) and bottom row with opposite strength (positioned at $\phi =0.5$ and $\phi =1.5$).}
     \label{fighc2} 
\end{figure}

 \begin{figure}[h]
   \centering
\begin{tabular}{lcc}
\includegraphics[width=5cm]{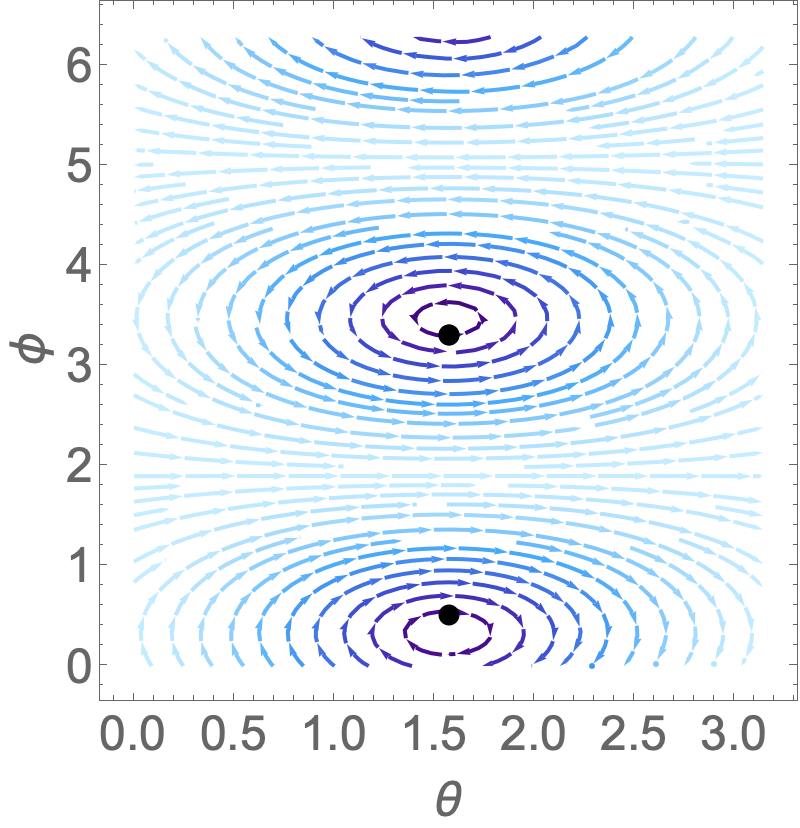}&
\includegraphics[width=5.6cm]{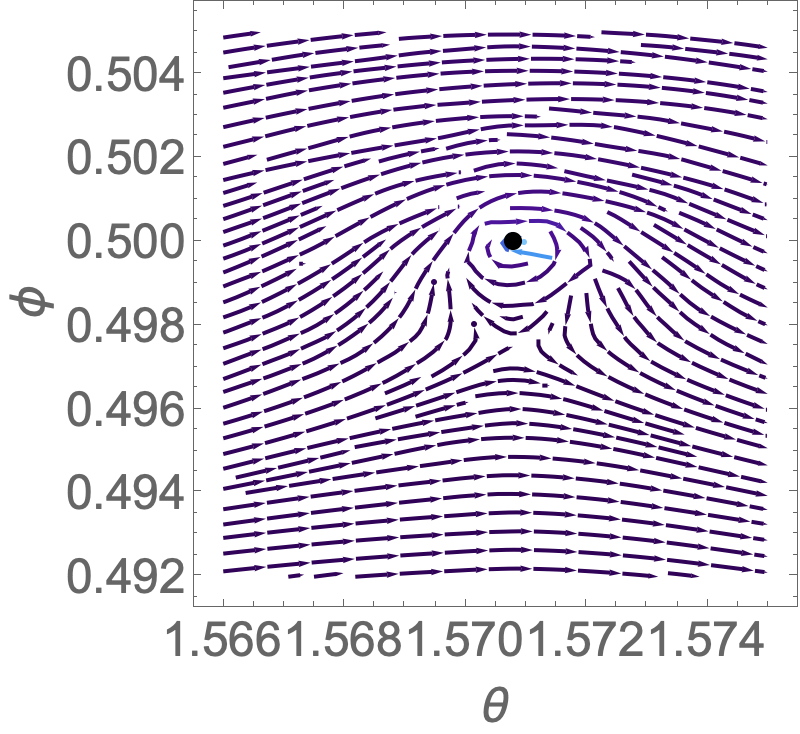}&
\includegraphics[width=5.6cm]{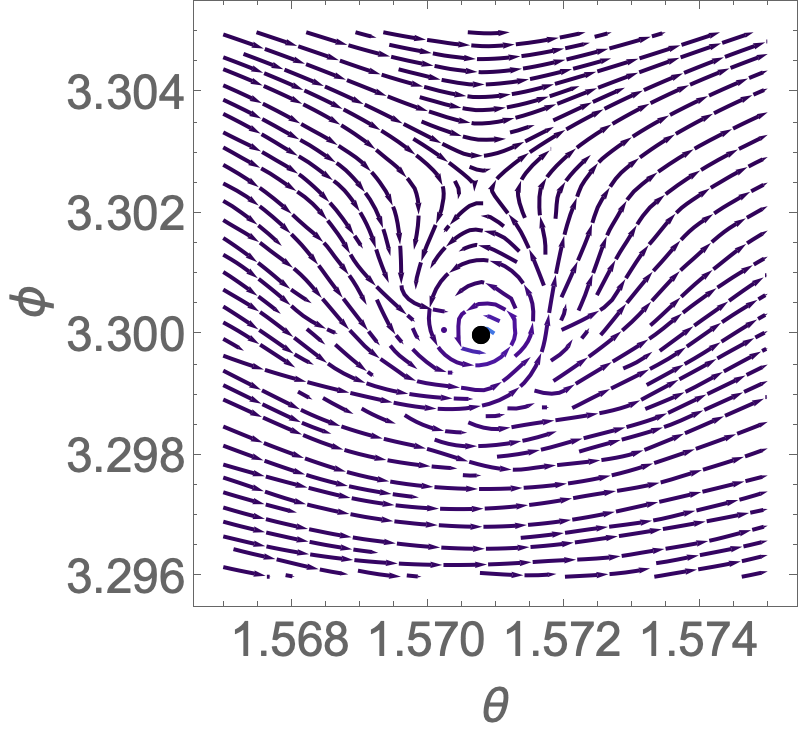}
\end{tabular}

    \caption{Streamline plot at $t=0$  for  two vortices, one at $ \theta=\pi/2, \phi =0.5$ and other at $ \theta=\pi/2, \phi =3.3$ in high curvature regime $R/\lambda \sim 10^{-3}$ for vortices of same circulation. On the left, we see the appearance of two global centers as expected (main text). The original vortices are  orbiting around the global center formed on the lower half of the leftmost figure. The newly formed saddles are tightly bound to the original vortices, as shown in the zoomed images on the right, showing each of the vortices separately near  $\phi =0.5$ and  $ \phi =3.3$ respectively.}
  \label{fighres} 
\end{figure}
\begin{figure}[h]
 \centering
\includegraphics[width=5cm]{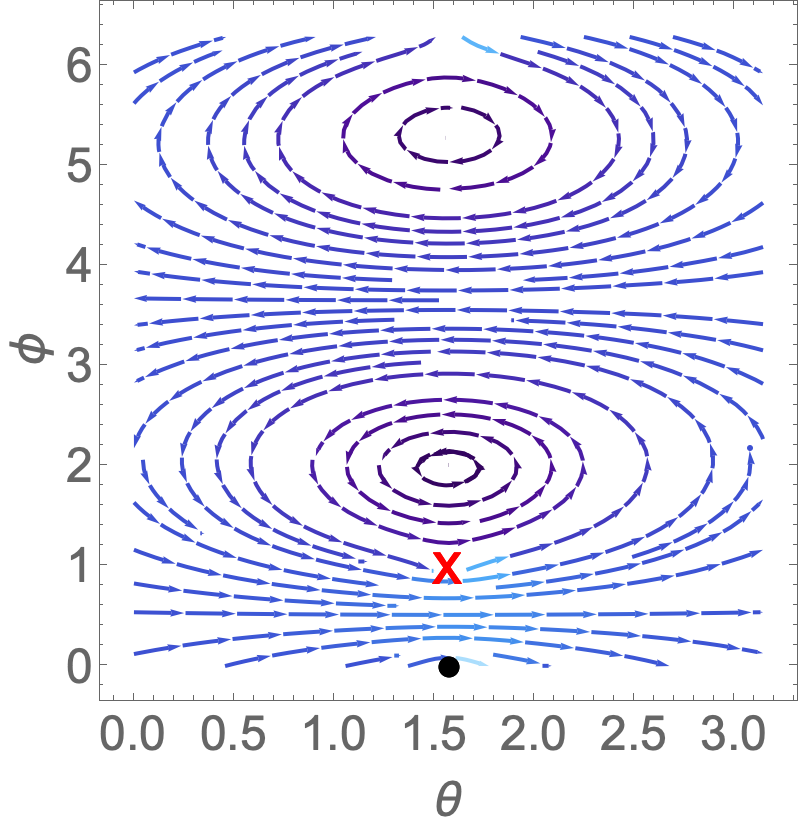}\\
      
    \caption{For vortex locations $\phi = 0$ and $\phi=1$ with opposite circulation, the global centers at the high curvature limit appear at $\theta_p = 2$ and $\theta_p = 5.2$ as predicted by Eq.~\ref{glbloc}.}
    \label{figroothc}
\end{figure}

The essential features for two vortices of same (opposite) circulation are detailed below.\\
\textbf{Same Circulation}: The dynamics of two vortices is similar to the low curvature case, with the rate given by Eq.~\ref{omlc}, but with the appropriate stream function for the high curvature regime, Eq.~\ref{reptq2}. In the limit of high curvature, this rate can be approximated by our estimate (Eq.~\ref{glbomega}).  In addition, the flow field develops two new centers due to the global rotation. The two vortices orbit around one of the global centers. Thus, compared to the low curvature regime, there is now an extra center and saddle appearing in the high curvature regime. The location of these new global centers is universal and will be calculated soon.  As curvature is increased, the saddles (anti-vortex) move towards the original vortices. This is reminiscent of a binding event of activity driven defects in nematic fluids \cite{srcb2018}. For membrane vortices, the binding is mediated by curvature versus the nematic case where it is driven by elasticity. Figure~ \ref{fighres}) shows the new stagnation points. Zoomed images show that curvature drives the newly formed saddles (of index -1) closer to the original vortices (index +1). \\
\textbf{Opposite Circulation}: Here as well the global rotation creates two new centers. Unlike the case of same circulation, the two vortices now orbit around different global centers.  Note that the new saddles are formed between the original vortices and the new global centers, as required by continuity of vector fields.  The location of these new global defects is again universal and independent of the details of the model (see below).  Here curvature also drives the newly formed saddles towards the original vortices, each of which now orbits a different global center.\\
\textbf{Location of global centers}: In the regime of high curvature, the stagnation points are still given by  Eq.~\ref{stgeq_txt}, and Eq.~\ref {fgeq_txt} as in the low-curvature regime, only the streamfunction used is the one appropriate for high curvature, see Eq.~\ref{reptq2} in Appendix \ref{aptq}. However, unlike the low curvature regime, there is no simple choice of parameters which simplifies the functional form of the stream function. One can still determine the location of the global defects, since in the limit where the stream function is dominated purely by the global term, one can approximate $F$ in Eq.~\ref{fgeq_txt} as follows :
\beqa
F=\frac{d \bm{\psi}}{d (\cos \gamma) } \sim -\frac{\eta_{2D}}{4 \pi R \eta_+}, 
\label{ff}
\eeqa
 while $G$ is a purely geometric factor, same as the low curvature regime (see Appendix \ref{apstg} ).
\beqa
G[z_p,z_j]= \frac{(1-|z_j|^2) (-2 \bar{z}_p)+ 4 \left((1+|z_p|^2) \frac{\bar{z}_j}{2} - Re[z_p \bar{z}_j] \bar{z}_p\right)}{(1+|z_j|^2) (1+|z_p|^2)^2}.
\label{Gdef}
\eeqa
Using these $F$ and $G$, the equation for stagnation points $z_p$, given by Eq.~\ref {fgeq_txt} simplifies considerably and is purely determined by the geometric function $G$.
\beqa
G(z_p,1) + \tau~ G(z_p, e^{i \phi}) \sim 0\nn\\
\Rightarrow \frac{1}{4} (1 - \bar{z}_p^2) + \frac{\tau}{2} \left(e^{- i \phi} - Re~[z_p e^{- i \phi}]\bar{z}_p\right) \sim 0
\eeqa
Substituting $z_p =e^{i \theta_p}$ in the above we get
\beqa
\frac{1}{4} (1 - e^{-2 i \theta}) + \frac{\tau}{2} \left(e^{- i \phi} - \cos( \theta - \phi) e^{- i \theta}\right)=0,
\label{stop_hc}
\eeqa
which has the following solutions
\beqa
&\theta_p = \pm \arccos \left[ \pm \frac{1+ \tau \cos \phi}{\sqrt{1+ \tau^2 + 2 \tau \cos \phi}}\right].
\label{glbloc}
\eeqa
For example, for $\tau=-1$ and $\phi=1$,  this yields $\theta_p =2.0708,1.0708,-2.0708, -1.0708 $. The corresponding streamlines in Fig.~\ref{figroothc} shows the global centers to be located at $\theta_p =2.0708 $ and $\theta_p =-1.0708 ~(= +5.21239)$. This coincides with two of the four solutions Eq.~\ref{glbloc} . The other two solutions are spurious because of our approximation (Eq.~\ref{ff}) and will be removed once the local corrections to Eq.~\ref{ff} are incorporated, similar to our low curvature computation. As expected, Eq.~\ref{glbloc} is independent of many details of the model and is only controlled by the vortex circulations and location.

\section { Many vortices }
\label{scmany}
\begin{figure}[tbh]
\begin{tabular}{ccc}
\includegraphics[width=4.5cm]{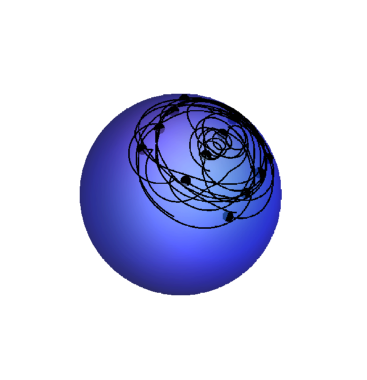}&
\includegraphics[width=3.8cm]{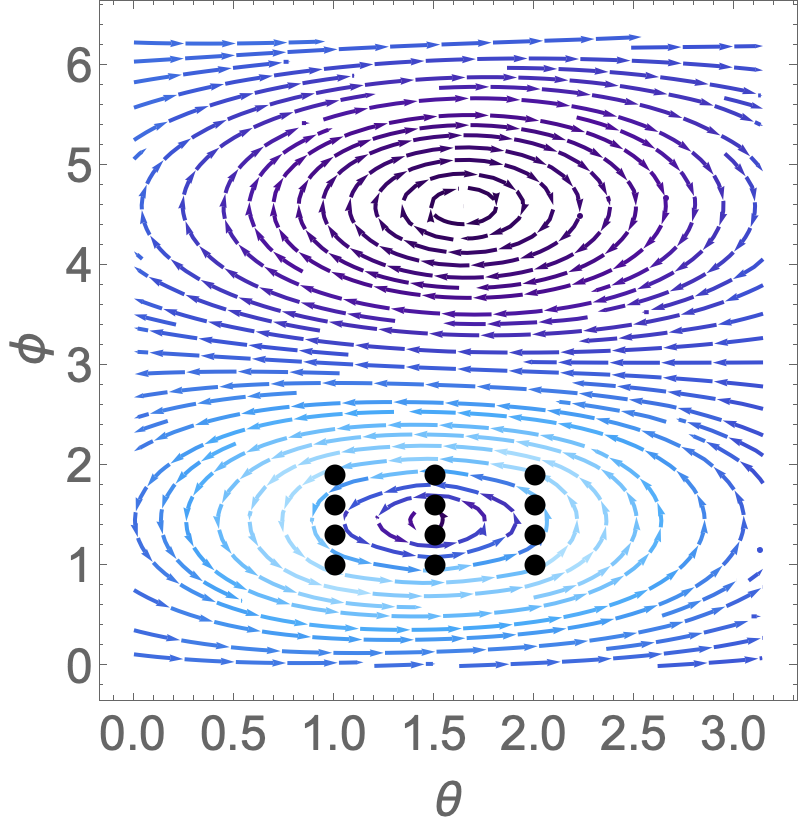}&
\includegraphics[width=3.8cm]{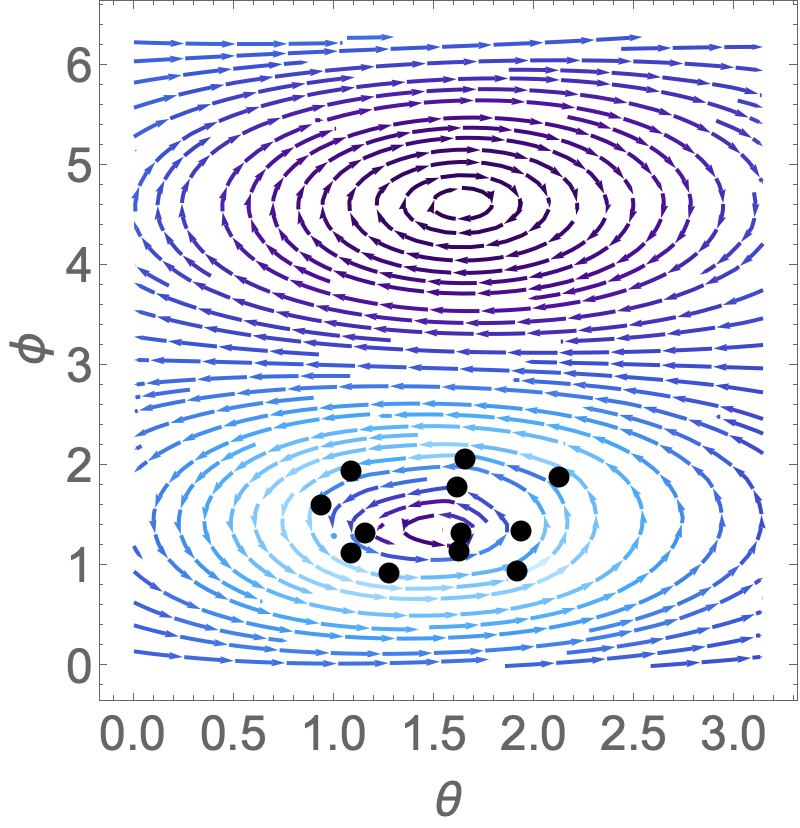}\\
\includegraphics[width=4cm]{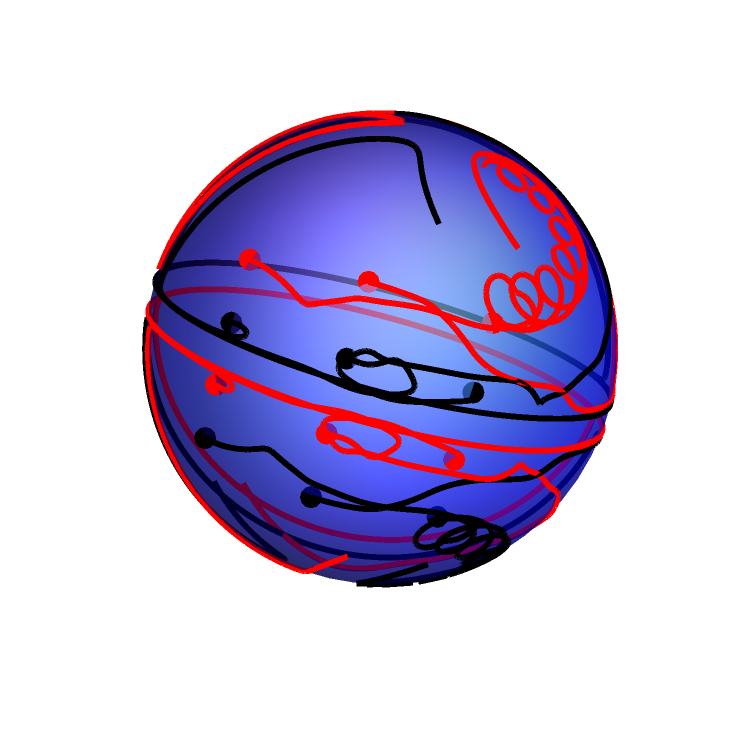}&
\includegraphics[width=3.8cm]{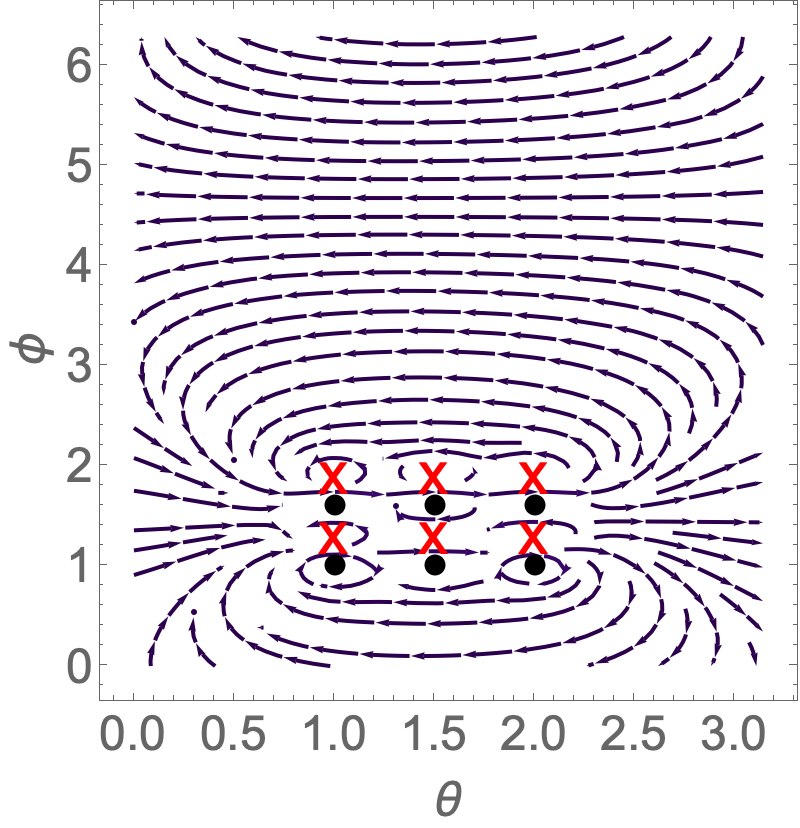}&
\includegraphics[width=3.8cm]{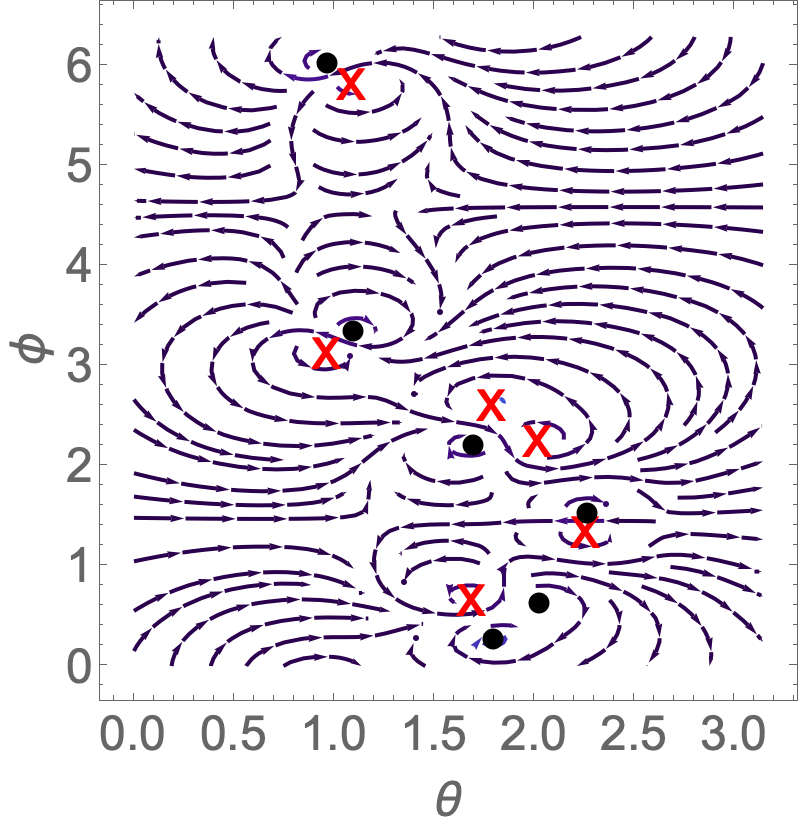}\\
\raisebox{.2\height}{\includegraphics[width=3cm]{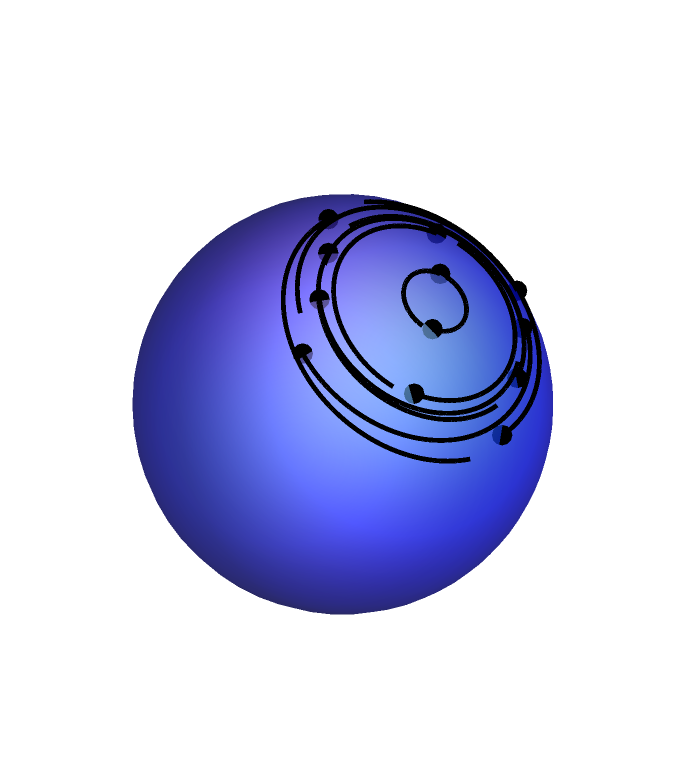}}&
\includegraphics[width=3.8cm]{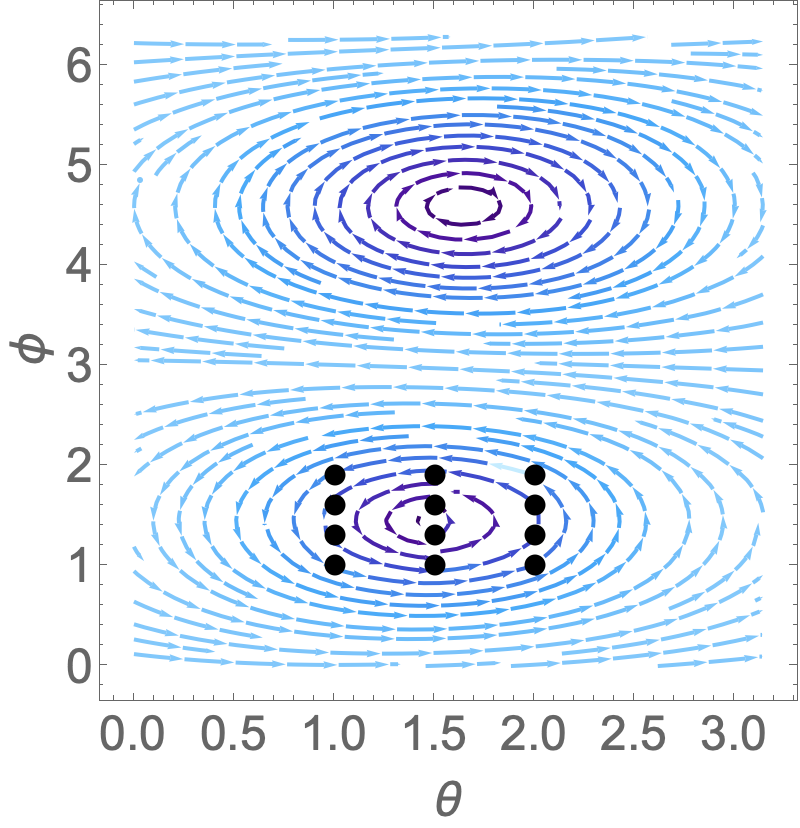}&
\includegraphics[width=3.8cm]{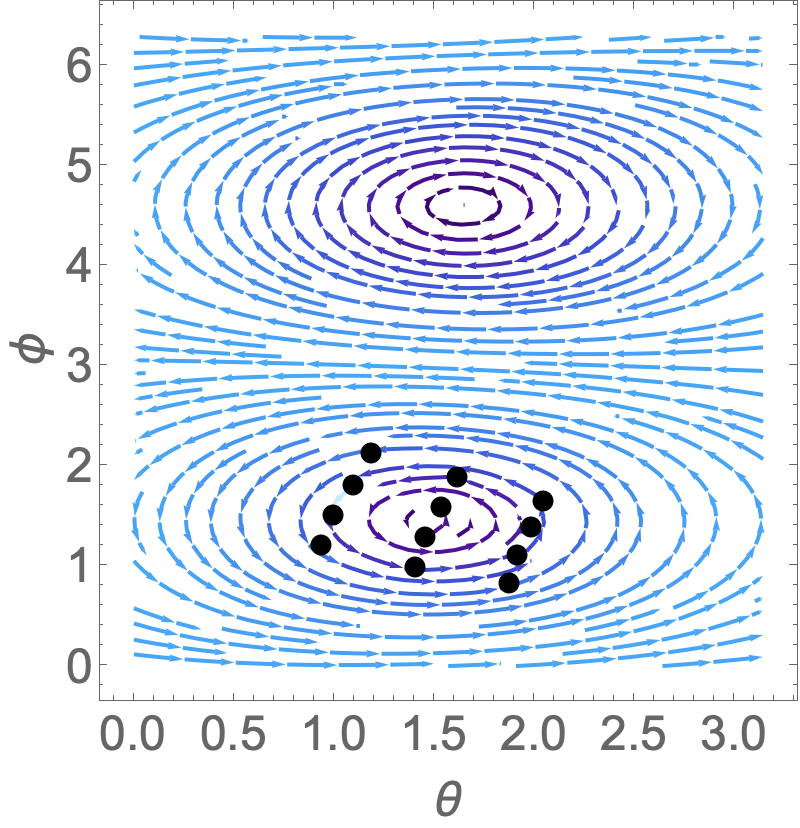}\\
\raisebox{.5\height}{\includegraphics[width=2.5cm]{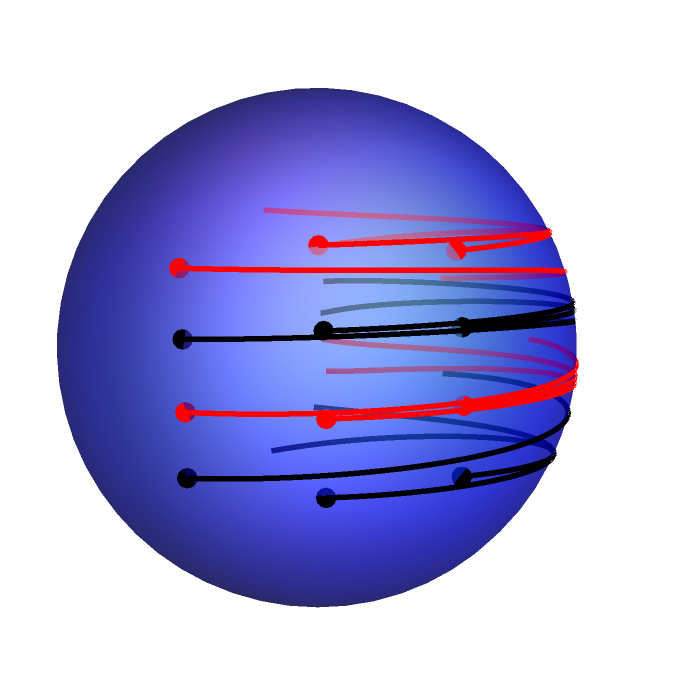}}&
\includegraphics[width=3.8cm]{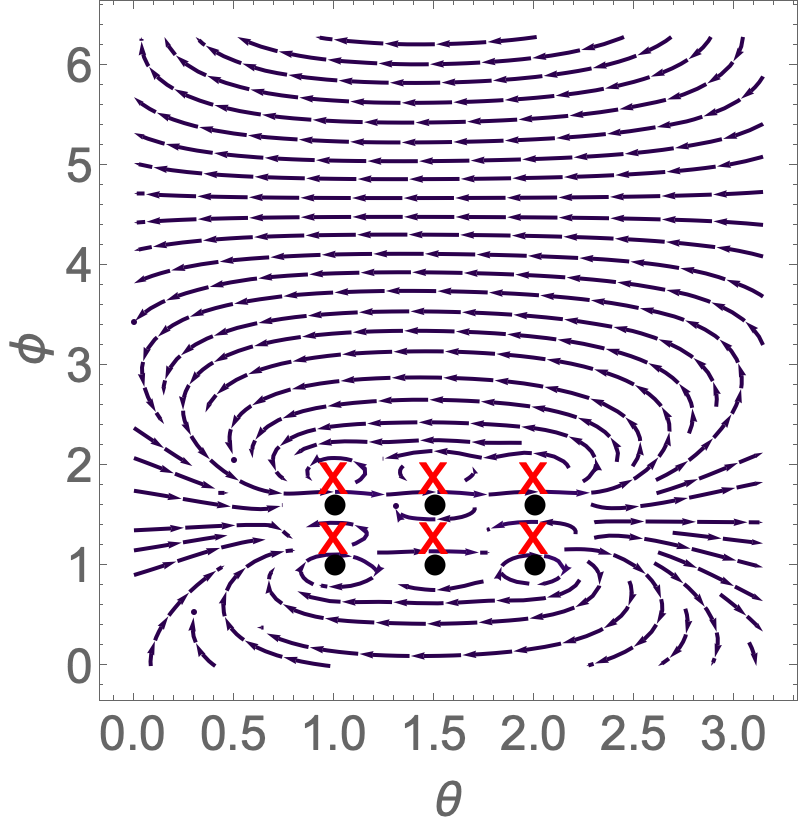}&
\includegraphics[width=3.8cm]{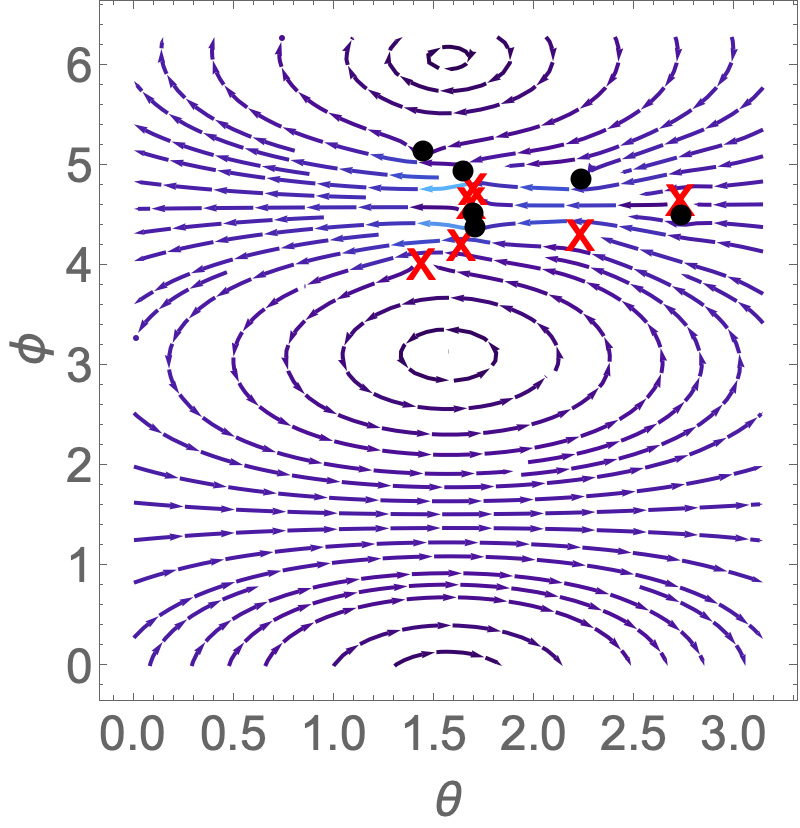}
\end{tabular}

    \caption{Dynamics and flows for 12 vortices. The first two rows are at low curvature (large radius): with same (row 1) and alternating (row 2) circulation strengths. Crosses (points) mark clockwise (counter-clockwise) circulation. The last two rows are at high curvature (small radius): with same (row 3) and alternating (row 4) circulation strengths. In all the rows, the vortices have the same initial locations centred in a small square. The left column shows the trajectories traced with time, the mid and right column show the streamlines at initial and final time respectively. Note how in the case of alternating circulations at the low curvature regime the dynamics span the entire spherical domain, whereas for same circulation vortices the dynamics stay bounded as resulting from the conservation of the second moment. Eq.~\ref{secondMoment}.}
        \label{figmany} 
\end{figure}

In this section we briefly discuss the situation of many rotating inclusions with varied circulations, with the dynamics and flow fields described by Eq.~\ref{dynmeq} and Eq.~\ref{field} respectively. For many vortices\footnote {Integrability is lost beyond $N=3$ vortices. For $N=4$ vortices, the system is still integrable if the sum of the circulations vanish.}, the dynamics is ergodic and it is in general difficult to perform an analytic investigation. However, one can still simulate the dynamics, numerically solving Eq.~\ref{dynmeq} and using the solution to keep track of spatio-temporal evolution of vortical defects via Eq.~\ref{field}. One can build some general intuition in specific situations, as explained in Fig.~\ref{figmany}.  For example, as shown in top row of Fig.~\ref{figmany}, in the low curvature regime, 12 closely spaced vortices of the same circulation tend to rotate together as a single effective center. Together with the creation of an isolated center,  the flow fields furnish a coarse grained version of Poincare Index Theorem. This also follows from the fact that the symmetries of the Hamiltonian we constructed (Eq.~\ref{hm}) preserve the second moment 
\begin{equation}
M = \sum_{i \neq j} \tau_i \tau_j C_{ij}^2, 
\label{secondMoment}
\end{equation}
where $C_{ij}$ denotes the chord distance between the vortices. In this situation since all circulations are the same, this implies that the vortices will remain geometrically confined within a region of the membrane.  \\

With alternating circulations (second row), this is no longer the case, with the 12 centers breaking into smaller groups and spreading across the whole membrane, while still conserving $M$. In the third and fourth rows of Fig.~\ref{figmany} we consider the same initial conditions, but with a high curvature. The high curvature leads to the creation of 2 global centers as expected, with vortices of the same circulation orbiting around one of the global centers, while vortices with alternating circulations get distributed among both the global centers. As expected from the conservation laws, the dynamics remains confined in the case of the same circulation vortices (third row) and unconfined for alternating circulations (fourth row). \\

\begin{figure}[h]
\includegraphics[width=13cm]{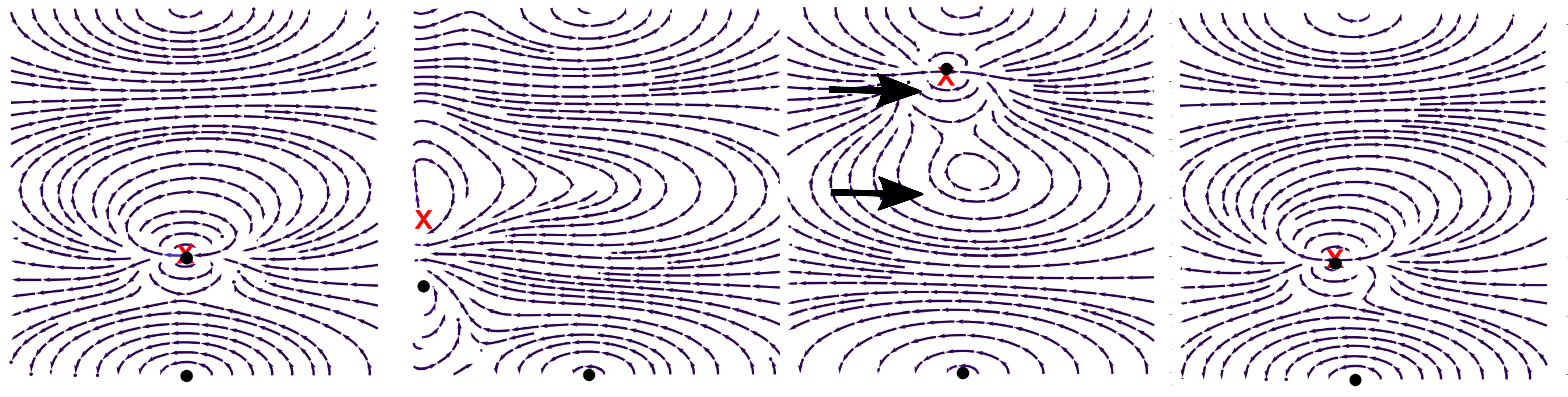}

   \caption{Left to right : an example of dynamical bifurcation for vortex strengths (1,1,-1) where a defect pair is created spontaneously in the third figure and then disappears as the dipole returns to its original position. Flow fields are shown in $(\theta, \phi)$ chart.}
        \label{figdp} 
\end{figure}

\textbf { Pair creation and fusion } Typically, collapse of vortices happens under very special initial conditions \cite{newton2000}. As we have seen, unlike the situation on the plane, the topology of the spherical membrane generically leads to the creation of new vortical defects in the flow fields. This creates the possibility for a spontaneous creation of vortical defect pairs as well as fusion events on the spherical membrane. We observe that this is indeed the case. The dynamics of vortices drives the system from one configuration of defects to another, with a different number of defects (still satisfying the Poincare Index theorem before and after the bifurcation). We are able to demonstrate these effects with a small number of vortices. Fig.~\ref{figdp} shows a temporal evolution, exhibiting  spontaneous creation and subsequent disappearance of a pair of vortical defects of opposite index (vortex anti-vortex pair).\\

\begin{figure}[h]
\includegraphics[width=10.2cm]{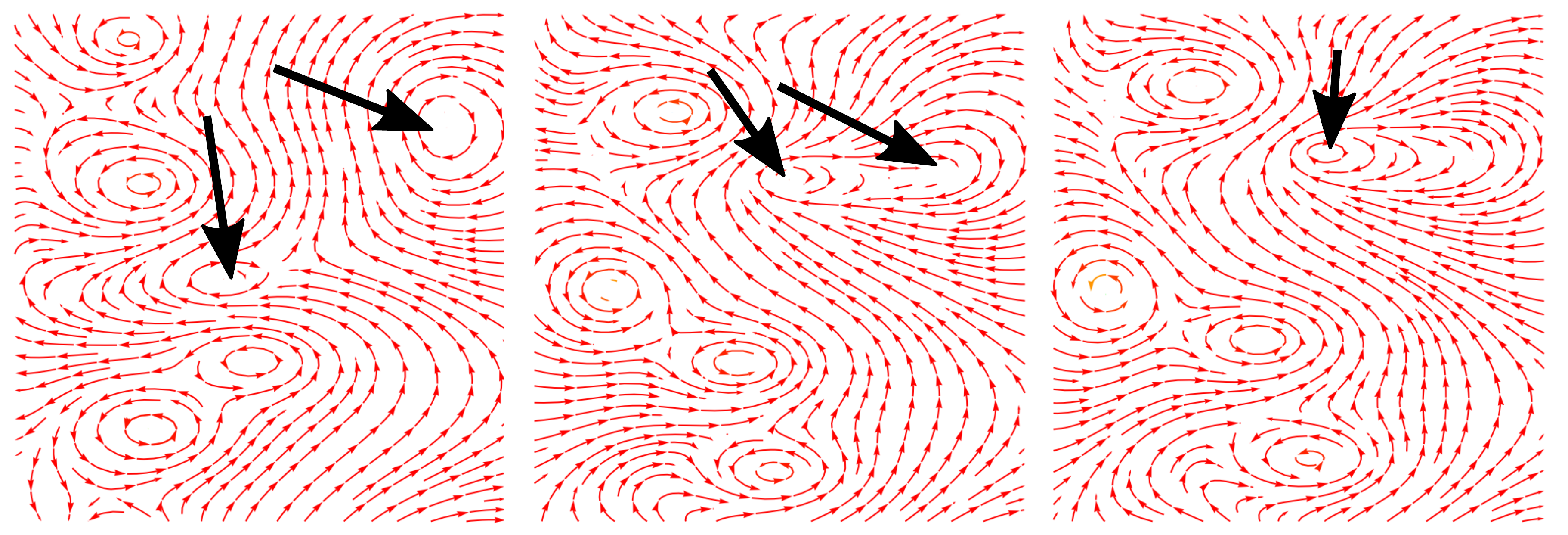}
 
    \caption{Dynamical fusion with five rotors. Black arrows mark the two centers which in the last frame combine into one. Flow fields are shown in $(\theta,  \phi)$ chart.}
        \label{figdf}
  \end{figure}

 Within the biological context, one may also incorporate the finite size of the rotating inclusions by introducing a soft repulsion between vortices, in addition to hydrodynamic interactions that we have considered so far. This makes the dynamics and flow fields very interesting, see Fig.~\ref{figdf} where we demonstrate a fusion event between an original membrane rotor and a newly created defect arising from the spherical topology. The fusion happens via a bridging saddle of negative index. We expect the number of such events to rapidly proliferate in the situation of large number of inclusions. The many rotor system will be explored in more detail in upcoming works. \\
 

 \section{Membranes of general curvature and other possible extensions}
\label{scgz}
It is straightforward (although computationally intensive) to generalize the calculations presented here to arbitrarily curved {\it static} geometries. Let us denote the coordinates in 3D ambient space by $x^{\alpha}$ and arbitrary coordinates on the membrane by $y^{a}$. A natural choice for the section of the ambient frame bundle for arbitrarily curved geometries (embedded in $ \mathcal{R}^3$) is the set $\hat{n}, \vec{e}_1, \vec{e}_2$ where
\beqa
\hat{n} =\frac{ \nabla \phi}{|\nabla \phi|}, ~~ e^\alpha_a =\frac{\partial{x^\alpha}}{\partial{y^a}}
\label{frame}
\eeqa
where the surface is described by the equation $\phi=0$, $a$ runs over $1,2$ for coordinates on the hypersurface while $\alpha $ runs over all three coordinates on the ambient space. 
It is a simple matter to repeat the calculations with the normal vector $\hat{n}$ now playing the role of the radial vector in the boundary conditions. For a given embedding of the surface, one first decomposes Lamb's solution \cite{lamb} for the 3D fluid (which we constructed in the spherical co-ordinate basis, see Appendix \ref{apf}) in the above new basis constructed out of the hypersurface embedding. Then we apply the stick and stress boundary conditions {\it in this basis}. The in-plane membrane velocity field can be extracted by solving the spectral eigenfunctions  $\psi(y,s)$ and eigenvalues $\lambda_s$  from the corresponding equations of membrane hydrodynamics (see Appendix \ref{apf} for a derivation) written in the basis $\vec{e}_1 $,  $\vec{e}_2  $ (defined in Eq.~\ref{frame}):
\beqa
 \eta_{2D} \left( 2 K(y) D^{a} \psi(y,s)  + D^{a} \Delta \psi(y,s)\right)  = \lambda_s  D^{a} \psi(y,s),
\eeqa
for an arbitrary local curvature $K(y)$ and where $D$ is associated with the pullback metric on the hypersurface
\beqa
h_{ab}= g_{\alpha \beta} e^{\alpha}_a  e^{\beta}_b.
\eeqa

For simple geometries, the spectrum is known and one can derive analytic expressions for the stream functions. It will be interesting to see, for example, how the anisotropies of an ellipsoid or of the negative curvature of hyperbolic spaces affect the streamline topologies.  We also plan to investigate the dynamics and streamline topology of flow fields in fluctuating biological membranes with bending rigidity and surface tension, see for example \cite{atzbergershape} for an experimental perspective. One can also use the flows we constructed in this paper to study the optimal time for navigation of microswimmers in spherical membranes, see \cite{optnvg}. We plan to report on these investigations in  subsequent works.\\

\section{Conclusion}
\label{sccl}
To summarize, in this work we explored in detail the topological aspects of 2D flows resulting from the dynamics of inclusions embedded in curved biological membranes. To get a first estimate of such flows, we considered a well known model \cite{saff1,saff2} adapted to spherical geometry, in the presence of external solvents.  In the examples of flows due to point sources (point force, torque and torque-dipole), we presented new closed form expressions for the respective Green's function in real space, using Appell Hypergeometric Functions. We investigated the topological features of the flow fields in some detail.  Such solutions can be used to model hydrodynamic interactions of proteins embedded in positively curved surfaces. The topology of the membrane creates many additional vortical defects. For the  simpler situation of two vortices, we were able to analytically predict the location of such defects with good precision. The point rotor solutions allowed us to construct a many-body geometric Hamiltonian that generates the dynamics of the vortices on the spherical membrane.  We studied the spatio-temporal evolution of defect mediated 2D flows in the spherical membrane. In particular, we found that at low curvature, the flows generated by the rotating inclusions are similar to flows generated by vortices in an ideal fluid . High curvature imparts a global rotation to the many-body system, with the individual vortices interacting locally.   Already in this simple model at low Reynolds, we saw surprisingly rich dynamics and flows mediated by the curvature and topology of the spherical membrane. The spatio-temporal evolution of streamlines revealed spontaneous creation and fusion of vortical defects, not present in the planar versions of the model. Some of our key formulas in this biological model are the dynamical equations Eq.~\ref{dynmeq}, \ref{field} and the rotation rates Eq.~\ref{omlc} and Eq.~\ref{glbomega}, which motivates experiments along the lines of Refs.~\cite{grzy},  \cite{terr} and \cite{vlvx}. This work also forms an essential building block to analyze a wide class of active inclusions, colloids, and 'swimmers' \cite{mik,mnk,chis,matsui,mh,adhar} on curved surfaces.\\

From an experimental point of view,\footnote { We thank the anonymous Referee for suggesting us to highlight this point. }one may achieve the transition from low to high curvature regime in a more controlled fashion by tuning the solvent viscosities $\eta_\pm$, keeping the radius R of the membrane fixed. With the advent of many advanced imaging techniques and fluorescent rotor probes \cite{rotorimg,mrotor}, it is now possible to perform accurate local transport measurements in curved biological membranes. Moreover, accurate particle tracking and velocity field imaging are also being conducted in many artificial setups \cite{atzbergershape,woodhouse,manp}. Such  investigations help us understand and characterize the biophysical properties of membranes and their impact on various cellular processes. Although such simple models overlook many complex details of a living cell, they are essential to get a first estimate of biophysical transport processes that routinely take place in living cells \cite{bm1,bm2}. A good understanding of flows happening in a membrane also has great potential to aid in efficient drug delivery \cite{optnvg}. We hope our results in this work will motivate more studies and experiments in these exciting directions.

\section{Acknowledgements}
R.S. acknowledges support from Department of Science and Technology, India (Grant No. IFA19-PH231). We would like to thank Haim Diamant for many insightful discussions. This research was also supported by the 
Israel Science Foundation (grant No. 1752/20).

\appendix
\section{More details on the Point Force}
\label{apf}
In this appendix section, we briefly review the computation of in-plane velocity-force response function, closely following \cite{henlev2008,henlev2010}.
We approximate the membrane as a two-dimensional  viscous incompressible fluid surrounded above and below by three-dimensional viscous fluids, with same notations for parameters as in the main text.  Greek indices are used to denote in-plane 2D objects while latin indices will be used for objects living in 3D ambient space. The  incompressibility and Stokes equations for the curved membrane thus have the general form :
\beqa
D^\alpha v_\alpha =0, ~~~ D^\beta \Pi_{\alpha \beta}=0
\label{curvedstokes}
\eeqa
where 
\beqa
 \Pi_{\alpha \beta}=p g_{\alpha \beta} - \eta _{\alpha \beta \mu \gamma} D^{\mu} v^{\gamma}
 \eeqa
 and 
 \beqa
 \eta_{\alpha \beta \mu \gamma } = \eta_{2D} \left( g_{\alpha \mu} g_{\beta \gamma } + g_{\alpha \gamma } g _{\mu \beta }\right) + ( \xi - \eta_{2d} ) g_{\alpha \beta} g_{\mu \gamma}
 \eeqa
where v denotes the in-plane 2D fluid velocity, $D$ is the 2D covariant derivative which generalizes the partial derivative of flat space, p is the local pressure and $g_{\mu \nu}$ is the 2D metric. $\eta_{2D}$ is the shear viscosity of the 2D membrane fluid while  $\xi$ is the bulk viscosity.\\\\
We now simplify the momentum equation of Eq.~\ref{curvedstokes} using the incompressibility property $D^\alpha v_\alpha =0$ and the metrinilic properties of the surface covariant derivative ie. $D_{\alpha} g_{\mu \nu}=0$, so that the metric $g_{\mu \nu}$ can freely pass in and out of the covariant derivative. The surface metric $g_{\mu \nu}$ is also used to raise/ lower appropriate indices.\\
\beqa
&D^{\beta} \Pi_{\alpha \beta} =0\nn\\
&\Rightarrow D^{\beta}\left( p g_{\alpha \beta} - \left( \eta_{2D} (g _{\alpha \mu} g_{\beta \gamma} + g_{\alpha \gamma} g _{\mu \beta}) +(\xi -\eta_{2D}) g_{\alpha \beta} g_{\mu \gamma}\right) D^{\mu} v^{\gamma}\right)=0\nn\\
&\Rightarrow D_{\alpha} p - \eta_{2D} \left(D_{\gamma} D_{\alpha} v^{\gamma} +D_\mu D^\mu v_\alpha\right)+(\xi - \eta_{2D}) D_{\alpha} \cancelto{0}{(D_\gamma v^\gamma)} =0\nn\\
&\Rightarrow D_{\alpha} p - \eta_{2D} \left(\underbrace{D_{\gamma} D_{\alpha} v^{\gamma}} +D_\mu D^\mu v_\alpha\right)=0\nn\\
\label{stk1}
\eeqa
Let us consider the term shown in braces in the above equation. In flat space , the derivatives $D_\gamma$ and $D_{\alpha}$ are just partial derivatives and they may be interchanged (flat space derivatives commute), then using incompressibility $\partial_\gamma v^\gamma=0$ the term vanishes. However they no longer commute in curved surfaces and the commutator is proportional to the local Gaussian curvature $K(x)$. Thus,
\beqa
D_{\gamma} D_{\alpha} v^{\gamma}= D^{\gamma} D_{\alpha} v_{\gamma}=\commutator{ D^{\gamma}}{ D_{\alpha}} v_\gamma + D_\alpha \cancelto{0}{D^\gamma v_\gamma }=K(x)~ v_\alpha
\eeqa
Plugging this back to Eq.~\ref{stk1} we get
\beqa
& D_{\alpha} p - \eta_{2D} \left(K(x) ~ v_\alpha +D_\mu D^\mu v_\alpha\right)=0\nn\\
\label{stk2}
\eeqa
Abbreviating the curved laplacian $D_\mu D^\mu$ by $\Delta$ , we have the final form of the Stokes equation written in terms of the Gaussian curvature.
\beqa
\boxed{\eta_{2D} \left( K(x)  + \Delta\right)  v_\alpha = D_\alpha p}
\label{stkf}
\eeqa
In the limit of zero curvature, one thus recovers the usual 2D Stokes equations.\\
We now turn to the analysis of the spectrum of the operator $\eta_{2D} \left( K(x)  + \Delta\right)$.
\beqa
\eta_{2D} \left( K(x)  + \Delta\right) v_\alpha (\vec{x},s) = \lambda_s v_\alpha(\vec{x},s)
\label{egveq}
\eeqa
where we use $s$ to label the eigenvalues $\lambda_s$ and eigenfunctions $v_\alpha(\vec{x},s)$.
Any arbitrary velocity field on the curved surface can be decomposed in terms of the eigenfunctions $v_\alpha(\vec{x},s)$ as follows:\\
\beqa
 v_\alpha(x) =\sum_s A_s v_\alpha(\vec{x},s)
 \label{vdcmp}
\eeqa
It helps to write the velocity eigenfunctions $v_\alpha(\vec{x},s)$ in terms of a stream function eigenmodes $\phi(\vec{x},s)$, satisfying the incompressibility requirement $D^\alpha v_\alpha =0$,   as follows:
\beqa
v_\alpha(\vec{x},s) =\epsilon_{\alpha \gamma} D^{\gamma} \phi(\vec{x},s)
\label{vstrm}
\eeqa
where $\epsilon$ is proportional to the totally antisymmetric permutation symbol $e$.
\beqa
\epsilon_{\alpha \gamma} =\sqrt{g} ~ e_{\alpha \gamma}
\eeqa
and $\sqrt{g}$ denotes the determinant of the surface metric $g_{\mu \nu}$.\\
Plugging Eq.~\ref{vstrm} in Eq.~\ref{vdcmp} we get the total stream function $\phi(\vec{x})$ ie.
\beqa
 v_\alpha(x) =\sum_s A_s \epsilon_{\alpha \gamma} D^{\gamma} \phi(\vec{x},s):= \epsilon_{\alpha \gamma} D^{\gamma}   \phi(\vec{x})
\eeqa
where the total stream function is given by a decomposition in the eigenmodes
\beqa
 \phi(\vec{x})= \sum_s A_s \phi(\vec{x},s)
 \label{vstrmtotal}
 \eeqa
 We now rewrite the eigenvalue equation Eq.~\ref{egveq} in terms of the stream function $\phi$.

 \beqa
 \eta_{2D} \left( 2 K(x) D^{\gamma} \phi(\vec{x},s)  + D^{\gamma} \Delta \phi(\vec{x},s)\right)  = \lambda_s  D^{\gamma} \phi(\vec{x},s)\nn\\
\label{finalegvl}
 \eeqa
For a general local Gaussian curvature, one can solve this equation numerically , however for surfaces of constant curvature, there is an additional simplification :
\beqa
  \Delta~ \phi(\vec{x},s)  =  \frac{\lambda_s  - 2 K \eta_{2D}}{\eta_{2D}} ~~ \phi(\vec{x},s)\nn\\
\label{constk}
 \eeqa
 We now turn to a discussion of the external solvents and boundary conditions.\\
 \textbf { External fluids } : We have ignored the role of external fluids in the fairly generic treatment above. We now incorporate the effects of the external solvent. The external fluids are described by the usual 3D Stokes equations. Let us denote the velocity, pressure, viscosity for $r>R$ by $+$ and same quantities for $r<R$ by $-$.
 \beqa
 \vec{\nabla} \cdot \vec{v}_\pm=0, ~~ \eta_\pm \nabla^2 v_\pm = \vec{\nabla} p_\pm
 \eeqa
We have two boundary conditions : \\
 a)  \textbf{Stick boundary condition } : velocities must coincide on the membrane surface.
 \beqa
 \boxed {v_\pm | _{r=R}=v}
 \label{stick}
 \eeqa
b) \textbf {Stress Balance condition on the membrane}
 \beqa
 \sigma^{ext}_\alpha= D_\alpha p -\eta_{2D} \left(K(x) +\Delta\right) v_\alpha + T_\alpha
 \label{stressbalance}
 \eeqa
 where $\sigma_\alpha^{ext}$ is the external point force or point torque applied to the membrane  and $T_\alpha$ is the traction due to coupling of the membrane fluid with the external solvents.
 \beqa
 T_\alpha= \sigma_{\alpha r}^{-} -\sigma_{\alpha r}^+ |_ {r=R}, ~~ \sigma_{ij}^\pm =\eta_\pm \left(D_i v_j^\pm +D_j v_i^\pm\right)- g_{ij} p_{\pm}
 \label{trc}
 \eeqa
Rewriting the stress balance equation Eq.~\ref{stressbalance} in terms of stream function using Eq.~\ref{vstrm} and eliminating 
 the membrane pressure by taking the antisymmetric derivative one arrives at,
  \beqa
\epsilon^{\alpha \beta} D_{\beta} \sigma^{ext}_\alpha=  - \sum_s A_s \lambda_s  \Delta \phi_s +  \epsilon^{\alpha \beta} D_{\beta} T_\alpha\nn\\
 \label{stressbalancemain}
 \eeqa
 where the traction $T$ is given by Eq.~\ref{trc}. So far our discussions are applicable to any curved static membrane geometry. We now specialize to spherical membrane.\\
 \textbf{Spherical membrane} The 2D sphere metric and Levi civita are listed below along with their inverses :
\begin{equation}
  g_{\alpha \beta}=\begin{pmatrix}
R^2 & 0\\\\
    0 & R^2 \sin^2 \theta \\\\
     \end{pmatrix}
     \end{equation}
     
     \begin{equation}
  g^{\alpha \beta}=\begin{pmatrix}
\frac{1}{R^2} & 0\\\\
    0 & \frac{1}{R^2 \sin^{2} \theta} \\\\
     \end{pmatrix}
     \end{equation}
     
       \begin{equation}
  \epsilon_{\alpha \beta}=\begin{pmatrix}
0 & R^2 \sin \theta\\\\
    -R^2 \sin \theta & 0 \\\\
     \end{pmatrix}
     \end{equation}
     
     \begin{equation}
  \epsilon^{\alpha \beta}=\begin{pmatrix}
0 & \frac{\csc \theta} {R^2}\\\\
    - \frac{\csc \theta} {R^2} & 0 \\\\
     \end{pmatrix}
     \end{equation}

The spectrum of the Laplace Beltrami operator on a sphere  is known, Eq.~\ref{constk} immediately yields the eigenvalues $\lambda_s$ in terms of  the constant curvature K and $\eta_{2D}$. For the sphere of radius R , the eigenvalue label s is given by the spherical harmonic mode index $(l,m)$ . Further, $ K =1/R^2$ and the known spectrum for spherical laplacian  is given by
\beqa
\Delta \phi_{lm} = - \frac{l(l+1)}{R^2} \phi_{lm}.
\label{knownsp}
\eeqa
Comparing Eq.~\ref{constk} and Eq.~\ref{knownsp}  gives 
\beqa
\lambda_l = \frac{2 -l(l+1)}{R^2} ~ \eta_{2D}
\label{sphereev}
\eeqa
and the eigenfunctions are given by
\beqa
\phi_{lm} = Y_{lm}(\theta, \phi)
\label{sphereef}
\eeqa
The mode decomposition for the velocity field on the sphere is given by Eq.~\ref{vdcmp} with the role of s played by $(l,m)$.
\beqa
v_\alpha =\sum_{lm} A_{lm} \epsilon_{\alpha \gamma} D^{\gamma} Y_{lm}
\label{vdcmps2}
\eeqa
where we used Eq.~\ref{sphereef}.  
The stress balance condition Eq.~\ref{stressbalance}  will help us determine the unknown coefficients $A_{lm}$  in Eq.~\ref{vdcmps2} for the membrane velocity on the sphere in terms of the applied force. For this, we need to compute the traction $T_\alpha$ appearing in the stress balance condition using the knowledge of the known Lamb's solution \cite{lamb} for the external solvent Eq.~\ref{vext}.  We carry out the steps below :\\\\
\textbf{External solvent in spherical co-ordinate- Lamb's solution} : Let us denote the velocities for $r>R$ by $v_+$ and for $r<R$ by $v_-$. The solution of 3D Stokes equations  \cite{lamb} is given by
\beqa
\vec{v}_- = \sum_{l=1}^{\infty} v^-_{l}, v^-_{l} = \vec{\nabla} \times ( \vec{r} q_l^-) + \vec{\nabla} w_l^{-} + \frac{1}{\eta_-(l+1) (2l+3)}\left(\frac{1}{2} (l+3) r^2 \vec{\nabla} p_l^- - l \vec{r} p_l^-\right)
\label{vext}
\eeqa
where $q_l^-,w_l^-, p_l^-$ are harmonic functions of $(r,\theta,\phi)$ ie. $\nabla^2 q_l^- =0,\nabla^2 w_l^-=0,\nabla^2p_l^-=0$. 
\beqa
q_l^- = \sum_{m=-l}^{m=l} q_{l,m}^-~ r^l Y_{lm} (\theta, \phi)\nn\\
w_l^- = \sum_{m=-l}^{m=l} w_{l,m}^-~ r^l Y_{lm} (\theta, \phi)\nn\\
p_l^{-} =\sum_{m=-l}^{m=l} p_{l,m}^-~ r^l Y_{lm} (\theta, \phi)\nn\\
\label{sphm}
\eeqa
Similarly for $r>R$ the solution is obtained by the replacement $l \rightarrow -l-1$. However, the stick boundary conditions demand
\beqa
w_l^- =0,~~ p_l^-=0 ,w_l^+=0,~~p_l^+ =0
\label{psol1}
\eeqa
and 
\beqa
q_{lm}^- = \frac{A_{lm}}{R^{l+1}}, ~~ q_{lm}^+ = R^l A_{lm}
\label{psol2}
\eeqa

Meanwhile, the traction can be computes using the definition Eq.~\ref{trc}.
\beqa
T_\alpha =\sum_{lm} \left( \frac{\eta_-}{R} (l-1) + \frac{\eta_+}{R}(l+2)\right) A_{lm} \epsilon_{\alpha \beta} D^{\beta} Y_{lm}(\theta, \phi) 
\label{trcfinal}
\eeqa
Finally using the stress balance condition Eq.~\ref{stressbalancemain} and decomposing the point force localized at $(\theta_0,\phi_0)$ via
\beqa
\boxed {\sigma^{ext}_\alpha=\frac{F_{0_\alpha}}{R^2}\underbrace{\sum_{l=0}^{\infty} \sum_{m=-l}^{l} Y_{lm}(\theta, \phi) Y_{lm}^*( \theta_0,\phi_0)}_{\frac{1}{\sin \theta_0} \delta(\theta -\theta_0) \delta (\phi -\phi_0)}},
\eeqa
we can solve for the unknown membrane velocity coefficients $A_{lm}$ in terms of force components $ F_{0_\alpha}$

\beqa
A_{lm} 
=\frac{\csc \theta_0}{\eta_{2D} s_l l (l+1)} 
\left(F_{\theta_0} \partial_{\phi_0} Y_{lm}^* (\theta_0 , \phi_0) - F_{\phi_0} \partial_{\theta_0} Y_{lm}^* (\theta_0 , \phi_0)\right)
\eeqa
where $s_l =l(l+1) -2 +\frac{R}{\lambda_-}(l-1)+\frac{R}{\lambda_+}(l+2)$ and $\lambda_{\pm}= \frac{\eta_{2D}}{\eta_{\pm}}$. Let us note that the traction contribution kills the zero mode $l=1$.
Plugging the $A_{lm}$ into the expression for the velocity field , we finally arrive at the stream function corresponding to the velocity field on the membrane surface for the point force:
\beqa
\bm{\psi}= \sum_{lm} A_{lm} Y_{lm}( \theta, \phi) = \sum_{lm}\frac{\csc \theta_0}{\eta_{2D} s_l l (l+1)} \left(F_{\theta_0} \partial_{\phi_0} Y_{lm}^{*} [\theta_0, \phi_0] - F_{\phi_0} \partial_{\theta_0} Y_{lm}^{*} [\theta_0, \phi_0]\right) Y_{lm} ( \theta, \phi)
\label{strmpforce}
\eeqa
where $s_l =l(l+1) -2 +\frac{R}{\lambda_-}(l-1)+\frac{R}{\lambda_+}(l+2)$ and $\lambda_{\pm}= \frac{\eta_{2D}}{\eta_{\pm}}$. \\
Performing the sum over m, this yields
\beqa
\bm{\psi} = \frac{ \csc \theta_0}{4 \pi \eta_{2D}}  \left( F_{\theta_0} \partial_{\phi_0} S -  F_{\phi_0} \partial_{\theta_0}S\right) 
\eeqa
where
\beqa
S= \sum_l \frac{2l+1}{s_l l (l+1)} P_l[\cos \gamma]
\label{Sdef}
\eeqa
where  $\cos \gamma = \cos \angle \left((\theta_0,\phi_0), (\theta, \phi) \right)= \sin \theta \sin \theta_0 \cos(\phi -\phi_0) + \cos \theta \cos \theta_0$ .\\
The physical velocity field on the membrane surface is summarized by an Oseen tensor on ${\mathbb S^2}$ given by
$v_\theta = G_{\theta \theta} F_\theta + G_{\theta \phi} F_\phi\nn\\
v_\phi = G_{\phi \theta} F_\theta + G_{\phi \phi} F_\phi$
These are physical components of velocity and force as opposed to components in a covariant basis.
\beqa
&G_{\theta \theta} = \frac{\csc \theta \csc \theta_0}{ 4 \pi \eta_{2D}} \partial_\phi \partial_{\phi_0} S\nn\\
&G_{\theta \phi} = -\frac{\csc \theta }{ 4 \pi \eta_{2D}} \partial_\phi \partial_{\theta_0} S\nn\\
&G_{\phi \theta} =- \frac{\csc \theta_0}{ 4 \pi \eta_{2D}} \partial_\theta \partial_{\phi_0} S\nn\\
&G_{\phi \phi} = \frac{1}{ 4 \pi \eta_{2D}} \partial_\theta \partial_{\theta_0} S
\eeqa
We now perform the sum over Legendre Polynomials to obtain a closed form expression for S, Eq.~\ref{Sdef}. We proceed first by noting that the roots of $s_l=0$ are given by
\beqa
l_p= \frac{ -(\eta_{2d} + R \eta_- + R \eta_+) + \sqrt{9 \eta_{2d}^2 + 6R \eta_{2d} (\eta_- -\eta_+)+ R^2 (\eta_- + \eta_+)^2}}{2 \eta_{2d}}\nn\\
l_m= \frac{ -(\eta_{2d} + R \eta_- + R \eta_+) - \sqrt{9 \eta_{2d}^2 + 6R \eta_{2d} (\eta_- -\eta_+)+ R^2 (\eta_- + \eta_+)^2}}{2 \eta_{2d}}\nn\\
\label{roots}
\eeqa
We analyze the structure of the roots in detail in Appendix \ref{aproot}. Here we just import those results. Depending on the parameters of the model, $l_p$ lies in the range $-2 < l_p \leq 1$ . On the other hand, $l_m$ is always negative. Breaking S (Eq.~\ref{Sdef}) into partial fractions and summing the individual parts , one gets two different real space representations of S depending on the sign of $l_p$, as we show in Eq.~\ref{repf1} and Eq.~\ref{repf2} below.\\
  
Further, in the situation when the internal and external Saffman lengths are same ie. $\lambda_- = \lambda_+:= \lambda$, the analysis in Appendix \ref{aproot} shows that for large radius $R \gg  \lambda$, the root $l_p$ is negative.  In the opposite situation of high curvature (small radius) ie.  $R \ll  \lambda$, the root $l_p$ is positive. Thus one has different representations of S depending on the sign of the root $l_p$. We use the appropriate one for our simulations. We list them below :

\textbf{Case 1 } : $-2 < l_p < 0$  ( Low curvature regime)
\beqa
S_{l_p<0}= \frac{1}{l_m l_p} \left(\log[2] -\log(-\cos \gamma + \sqrt{2- 2 \cos \gamma}+1)\right)+ \frac{1}{(1+l_m)(1+l_p)} \log[\frac{\cos \gamma -  \sqrt{2- 2 \cos \gamma}-1}{\cos \gamma-1}]+  \nn\\
\frac{1+2 l_m}{l_m (1+l_m)(l_m-l_p)} A[l_m] +
\frac{1+2 l_p}{l_p (1+l_p)(l_p-l_m)} A[l_p]
\label{repf1}
\eeqa
where the function $A[l_m]$ is defined by a combination of Appell Hypergeometric function.
\beqa
&A[l_m]=\nn\\
&\frac{(-1+l_m) l_m ~\mathcal{A}[2-l_m,\frac{1}{2},\frac{1}{2}, 3-l_m, e^{i \gamma}, e^{-i \gamma}] -(-2+l_m)\left((-1+l_m) ~\mathcal{A}[-l_m,-\frac{1}{2},-\frac{1}{2}, 1-l_m, e^{i \gamma} , e^{-i \gamma}]+ 2 l_m~ \mathcal{A}[1-l_m, \frac{1}{2},\frac{1}{2}, 2 -l_m, e^{i \gamma}, e^{- i \gamma}]~ Cos ~\gamma \right)}{(-2+l_m)(-1+l_m) l_m}\nn\\
\label{Adef}
\eeqa
and a similar relation for  $A[l_p]$.\\\\
The function $A$  becomes simpler for integer values of the negative root. We list some of them below.
\beqa
A|_{l_m=0} =\log 2 - \log \left[-x+\sqrt{2-2x}+1\right]|_{x= \cos \gamma}\nn\\
A|_{l_m=-1} =\log \left[\frac{x-\sqrt{2-2x}-1}{x-1}\right]|_{x= \cos \gamma}\nn\\
A|_{l_m=-2} =\sqrt{2-2x} +2  x \coth^{-1} ( \sqrt{2-2x}+1)|_{x= \cos \gamma}
\label{ints}
\eeqa
\textbf{Case 2} : $0< l_p <1$ (High curvature regime).\\
In this situation,
\beqa
S_{l_p>0}= \frac{1}{l_m l_p} \left(\log[2] -\log(-\cos \gamma + \sqrt{2- 2 \cos \gamma}+1)\right)+ \frac{1}{(1+l_m)(1+l_p)} \log[\frac{\cos \gamma -  \sqrt{2- 2 \cos \gamma}-1}{\cos \gamma-1}]+  \nn\\
\frac{1+2 l_m}{l_m (1+l_m)(l_m-l_p)} A[l_m] +
\frac{1+2 l_p}{l_p (1+l_p)(l_p-l_m)} B[l_p]
\label{repf2}
\eeqa 
where
\beqa
B[l_p]= -\frac{1}{l_p}+ \frac{1- \mathcal{A}~[-l_p,\frac{1}{2},\frac{1}{2},1-l_p,e^{i \gamma}, e^{-i \gamma}]}{l_p}
\label{Bdef}
\eeqa
There are some special points in the parameter space where the above representations need to be supplemented by the following:

\textbf{Case 3} : one of the roots is zero\\
This enforces the other root , let us call it $\tilde{l} =\frac{\eta_-+4 \eta_+}{\eta_--2 \eta_+}$ and $\eta_{2d} =\frac{R}{2}(2 \eta_+ - \eta_-)$. Note that for $\eta_{2d}>0$ we need $2 \eta_+ > \eta_-$, which implies $\tilde{l}$ has to be negative.
\beqa
S=  \sum_l \frac{2l+1}{(l-\tilde{l}) l^2 (l+1)} P_l[\cos \gamma]
=\frac{-1- \tilde{l}}{\tilde{l}^2} \left(\log[2] -\log(-\cos \gamma + \sqrt{2- 2 \cos \gamma}+1)\right) \nn\\
-\frac{1}{\tilde{l}} S_0 + \frac{1}{1+\tilde{l}} A[-1] +\frac{1+2 \tilde{l}}{\tilde{l}^2 (1+\tilde{l})} A[\tilde{l}]
\eeqa
where
\beqa
S_0=\sum_{l=1}^\infty \frac{P_l[\cos \gamma]}{l^2}
\eeqa
 is convergent and can be evaluated numerically.\\
 \textbf{Case 4} : one of the roots is -1\\
 This enforces the other root , let us call it $\tilde{l} =\frac{2( \eta_-+ \eta_+)}{2\eta_-- \eta_+}$ and $\eta_{2d} =\frac{R}{2}( \eta_+ -2 \eta_-)$. Note that for $\eta_{2d}>0$ we need $ \eta_+ > 2\eta_-$, which implies $\tilde{l}$ has to be negative.
\beqa
S=  \sum_l \frac{2l+1}{(l-l_p) l (l+1)^2} P_l[\cos \gamma]
=\frac{-1}{\tilde{l}} \left(\log[2] -\log(-\cos \gamma + \sqrt{2- 2 \cos \gamma}+1)\right) + \frac{1}{-\tilde{l}-1} \tilde{S}_0 \nn\\
+  \frac{\tilde{l}}{(\tilde{l}+1)^2} A[-1] +\frac{1+2 \tilde{l}}{\tilde{l} (1+\tilde{l})^2} A[\tilde{l}] 
\eeqa
where
\beqa
\tilde{S}_0=\sum_{l=1}^\infty \frac{P_l[\cos \gamma]}{(l+1)^2}
\eeqa
 is convergent and can be evaluated numerically.\\

\section{ More details on the Point Torque}
\label{aptq}

Keeping the same notations as the force-velocity response calculation, the equation for stress balance Eq.~\ref{stressbalancemain}  in the situation of a rotor  embedded in the spherical membrane takes the following form in the basis of spherical harmonics:
\beqa
&\boxed {-\epsilon^{\alpha \gamma} D_{\gamma} [\tau~ \epsilon_{\alpha \beta} D^\beta ]|_{\theta_0,\phi_0} \frac{1}{R^2}\left(\underbrace{ \sum _{l,m} Y_{lm} (\theta , \phi) Y_ {lm} ^{*} ( \theta_0, \phi_0)}_{\frac{1}{\sin \theta} \delta( \theta -\theta_0) \delta(\phi -\phi_0)}\right)=   \sum_{lm} \frac{\eta_{2D} l(l+1)}{R^4} \underbrace{s_l}_{\text{membrane stress } + \text{traction}} A_{lm}  Y_{lm}(\theta, \phi) },\nn\\
& s_l =l(l+1) -2 +\frac{R}{l_-}(l-1)+\frac{R}{l_+}(l+2). 
\eeqa
where the in-plane membrane velocity field is decomposed as Eq.~\ref{vdcmps2}, ie.  $v_\alpha =\sum_{lm} A_{lm} \epsilon_{\alpha \gamma} D^{\gamma} Y_{lm}$ and $tau$ denotes the rotor circulation. Solving for the unknown coefficients $A_{lm}$ from the above equation yields
\beqa
\boxed{A_{lm}= \frac{\tau  Y_ {lm} ^{*} ( \theta_0, \phi_0)}{\eta_{2D}~  s_l}}
\label{almsol}
\eeqa
Plugging this into the mode expansion for the velocity field Eq.~\ref{vdcmps2}, we get
\beqa
v_\alpha = \sum_{lm} \frac{\tau  Y_ {lm} ^{*} ( \theta_0, \phi_0)}{\eta_{2D}~  s_l} \epsilon_{\alpha \gamma} D^\gamma|_{\theta,\phi} Y_{lm}(\theta, \phi)
\eeqa
Performing the sum over m, $\sum_{m=-l}^{m=l}  Y_{lm}(\theta_1,\phi_1)  Y_{lm}^*(\theta_2,\phi_2) =\frac{2l+1}{4 \pi} P_l(\cos \gamma)$ we finally have
\beqa
v_\alpha = \sum_{l} \frac{\tau (2l+1)}{4 \pi \eta_{2D}~  s_l} \epsilon_{\alpha \gamma} D^\gamma|_{\theta,\phi} P_l (\cos \gamma)
\eeqa
where $\gamma$ is the geodesic angle between $(\theta,\phi)$ and $(\theta_0,\phi_0)$.

Introducing the operator  $[ \nabla_\perp ^{\mathbb S^2}] =- \left(\hat {\theta} \frac{1}{R \sin \theta}\partial_\phi - \hat{\phi} \frac{1}{R} \partial_\theta\right)$ we find that the physical velocity field can be expressed as 
\beqa 
\boxed{v =\frac{\tau}{\eta_{2D}} [\nabla_\perp ^{\mathbb S^2}] ~\bm{\psi}}
\eeqa
where the dimensionless stream function $\bm{\psi}$ is given by
\beqa
\boxed {\bm{\psi}~ [\theta, \phi, \theta_0, \phi_0] =  \sum_l \frac{ (2l+1)}{4 \pi ~  s_l}  P_l (\cos \gamma) }
\label{strmptrq}
\eeqa
where  $s_l =l(l+1) -2 +\frac{R}{l_-}(l-1)+\frac{R}{l_+}(l+2)$ and $\lambda_{\pm}= \frac{\eta_{2D}}{\eta_{\pm}}$ and $\cos \gamma = \sin \theta \sin \theta_0 \cos(\phi -\phi_0) + \cos \theta \cos \theta_0$ .\\\\
\textbf{Vanishing of membrane pressure in rotors} : Taking the symmetric derivative $D^\alpha$ of the stress balance condition $\sigma_\alpha^{ext} = D_\alpha p - \sum_{lm} A_{lm} \lambda_l \epsilon_{\alpha \gamma} D^\gamma \phi_{lm} + T_\alpha $ we get 
\beqa
\Delta p =0  
\eeqa
for the case of a rotor ie. $\sigma_\alpha^{ext} =\tau \epsilon_{\alpha \beta} D^\beta \delta(\theta -\theta_0, \phi -\phi_0)$ and traction $T_\alpha$ given by Eq.~\ref{trcfinal}.
Since $\Delta$ is a Laplace Operator on a compact manifold $S^2$ (see \cite {heiko} for details), this means p can only be a harmonic function with eigenvalue zero which forces it to be a constant and drops out of the hydrodynamic equation because it appears as a gradient. Let us note that this does not happen for the external point force where we will get a non zero membrane pressure.\\\\
\textbf{Flat membrane limit} : The planar limit of Eq.~\ref{strmptrq} can be understood by introducing a momentum variable $q = \frac{l}{R}$ and converting the sum into an integral in the limit of large radius.
\beqa
&\bm{\psi}~ [\theta, \phi, \theta_0, \phi_0] =  \sum_{l=1}^{l_{max}} \underbrace{\frac{1}{R}}_{dq = \frac{d^2 q}{ 2 \pi q}} ~\underbrace{\frac{ (2l+1) R}{4 \pi ~  s_l}}_{\frac{1}{2 \pi (q+\lambda^{-1})}} \underbrace{ P_l (\cos \gamma)}_{e^{iq.r}} \nn\\
&\rightarrow  \int \frac{d^2 q }{q(q+\lambda^{-1})}e^{iq.r}
\eeqa
where we used
\beqa
&\frac{ (2l+1) R}{4 \pi ~  s_l}=\frac{(2qR+1)R}{4 \pi \left(qR(qR+1) -2 +\frac{R}{\lambda_-}(qR-1)+\frac{R}{\lambda_+}(qR+2)\right)  }\nn\\
&\sim \frac{2qR^2}{4 \pi \left(q^2 R^2  +\frac{R}{\lambda_-}(qR)+\frac{R}{\lambda_+}(qR)\right)  }\nn\\
& \sim \frac{1}{ \left(q  +\frac{1}{\lambda_-}+\frac{1}{\lambda_+}\right)  }\nn\\
\eeqa
\textbf { Performing the sum over Legendre Polynomials}  
In order to find closed form expression for $\bm{\psi}$ given by Eq.~\ref{strmptrq}, we proceed exactly as the situation of the point force.
Depending on the nature of the roots of the equation  $s_l=0$, we again have different representations of $\bm{\psi}$. We list them below :\\
\textbf{Case 1 } :  $-2 < l_p < 0$  (low curvature regime).
\beqa
\bm{\psi}_{l_p<0}=\frac{1}{4 \pi} \left(\frac{2 l_m+1}{l_m -l_p} A[l_m]-\frac{2 l_p+1}{l_m - l_p} A [l_p]\right) 
\label{reptq1}
\eeqa
with A  by Eq.~\ref{Adef}.\\
\textbf{Case 2} :  $0< l_p <1$ (high curvature regime).\\
\beqa
\bm{\psi}_{l_p>0}=\frac{1}{4 \pi} \left(\frac{2 l_m+1}{l_m -l_p} A[l_m]-\frac{2 l_p+1}{l_m - l_p} B[l_p]\right) 
\label{reptq2}
\eeqa
with A and B given by Eq.~\ref{Adef} and \ref{Bdef}.\\
\textbf{Case 3 } : $l_p =0$ \\
In this situation,
\beqa
&\bm{\psi}_{l_p=0}=\sum_l\frac{2l+1}{l(l-l_m)} P_l [\cos\gamma]\nn\\
& = -\frac{1}{l_m} \log[\frac{2}{-\cos \gamma +\sqrt{2-2 \cos \gamma} +1}] +\frac{1+ 2l_m}{l_m}(A[l_m] + \frac{1}{l_m})
\eeqa
where A is defined in Eq.~\ref{Adef}.
\begin{figure}
   \centering
\begin{tabular}{lcc}
\includegraphics[width=4.7cm]{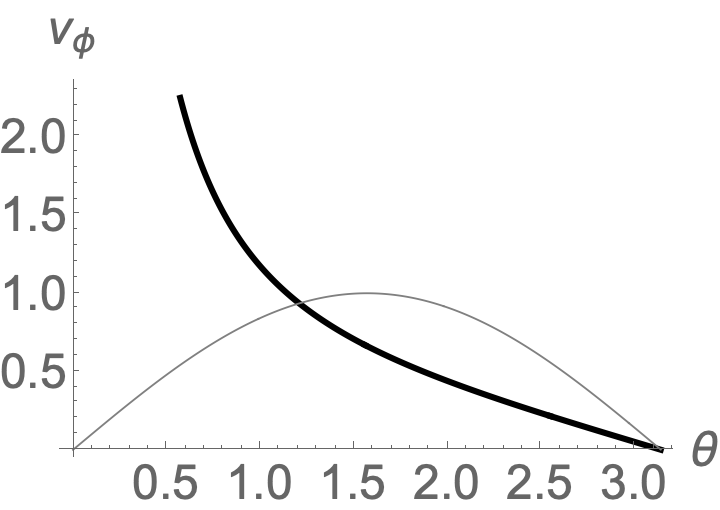}&
\includegraphics[width=4.7cm]{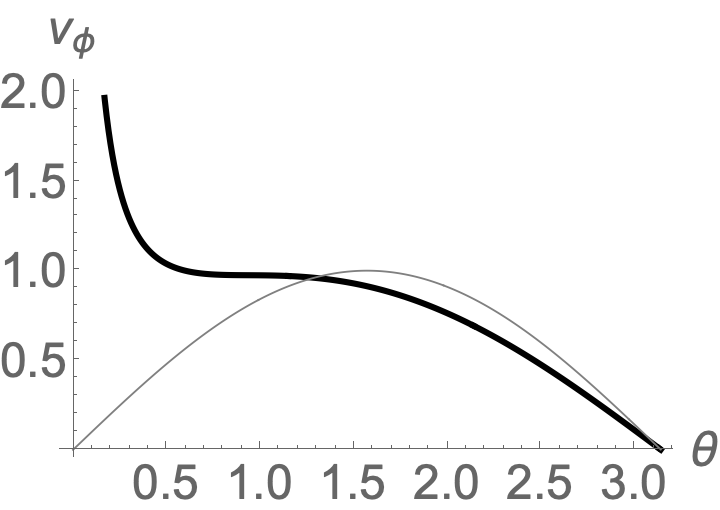} \\
\includegraphics[width=4.7cm]{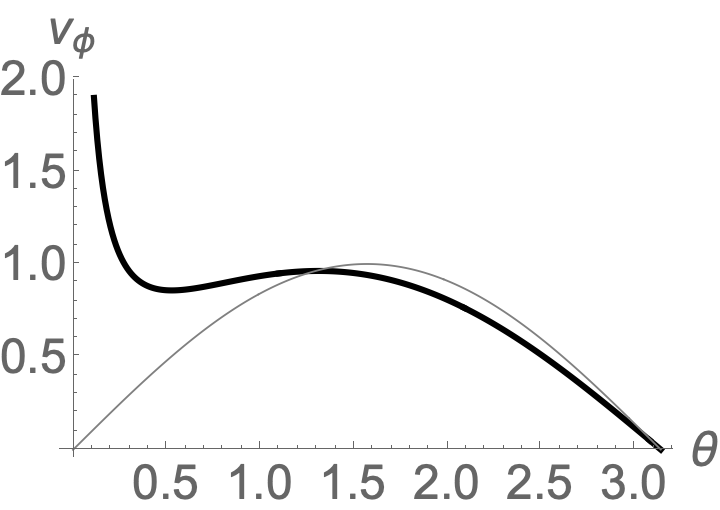}&
\includegraphics[width=4.7cm]{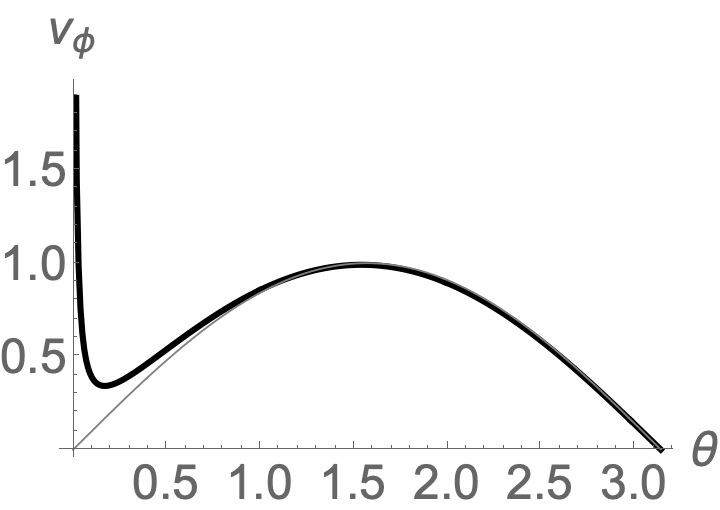}\\
\end{tabular}

    \caption{Comparison of the velocity component $v_\phi$ vs $\theta$ (using  our analytic expression Eq.~\ref{reptq2})  with the contribution to velocity purely from global rotation term (Eq.~\ref{vph} of main text) shown in gray, for the point torque positioned at the north pole. Top left to bottom right, the curvature is progressively increased. The non-monotonicity (which appears first on the second plot from the left) is due to the dominance of the global rotation term as curvature is increased. }
 \label{figvphi} 
\end{figure}
\section{ More details on the Counter Rotating Torque}
\label{apctq}
We now consider the situation of  counter rotating torque dipole, one situated at $r=R$ with torque $\tau$ and the other positioned at $r=R+d$ with torque $-\tau$. This is a useful first approximation of rotor proteins in a membrane \cite{oppshelley}. Following Lamb's solution \cite{lamb} in spherical co-ordinates, we first consider the solutions of the Stokes equations in 3D , as given in Eq.~\ref{vext} with Eq.~\ref{psol1} \footnote{  In our notation the physical components of the 3D  velocity field are given by $\vec{v} =v_r \hat{r} + v_\theta \hat{\theta} + v_{\phi} \hat{\phi}$ and similarly for the membrane $\vec{v}= v_\theta \hat{\theta} + v_{\phi} \hat{\phi}$}. We list the profiles in the appropriate domains ( $v^-, v_{int}, v^+$ denote the innermost, intermediate and outermost  velocity fields respectively) :

%
%

For $r<R$ we have 
\beqa
&{{[v]}_r^-} =0,\nn\\
& {{[v]}_\theta^-} =\sum_{lm} \frac{1}{\sin \theta} ~  q_{lm}^- r^l \partial_{\phi} Y_{lm} (\theta, \phi)\nn\\
&{{[v]}_\phi^-}=-\sum_{lm}   q_{lm}^- r^l \partial_\theta Y_{lm} (\theta, \phi)
\label{v3minustd}
\eeqa
and for $r>R+d$ are given by
\beqa
&{{[v]}_r^+} =0,\nn\\
& {{[v]}_\theta^+} =\sum_{lm} \frac{1}{\sin \theta} ~  q_{lm}^+ r^{-l-1} \partial_{\phi} Y_{lm} (\theta, \phi)\nn\\
&{{[v]}_\phi^+}=-\sum_{lm}   q_{lm}^+ r^{-l-1} \partial_\theta Y_{lm} (\theta, \phi)
\label{v3plustd}
\eeqa
For $ R< r < R+d$ we have  both the rising and falling solutions in the intermediate velocity field $v_{int}$.
\beqa
&{{[v_{int}]}_r^-} =0,\nn\\
& {{[v_{int}]}_\theta^-} =\sum_{lm} \left(\frac{1}{\sin \theta} ~ \tilde{ q}_{lm}^- r^l \partial_{\phi} Y_{lm} (\theta, \phi) + \frac{1}{\sin \theta} ~  \tilde{q}_{lm}^+ r^{-l-1} \partial_{\phi} Y_{lm} (\theta, \phi) \right)\nn\\
&{{[v_{int}]}_\phi^-}=-\sum_{lm}  \left( \tilde{q}_{lm}^- r^l \partial_\theta Y_{lm} (\theta, \phi) +\sum_{lm}   \tilde{q}_{lm}^+ r^{-l-1} \partial_\theta Y_{lm} (\theta, \phi)\right)
\label{vint}
\eeqa
The physical components of the membrane velocity field are as follows :
\beqa
&v_\theta =  = \frac{1}{R \sin \theta} \sum_{lm} A_{lm} \partial_\phi Y_{lm}(\theta, \phi)\nn\\
&v_\phi  = -\frac{1}{R } \sum_{lm} A_{lm} \partial_\theta Y_{lm}(\theta, \phi).\nn\\
\label{mbvtd}
\eeqa
Thus altogether we have 5 undetermined coefficients $A_{lm},q_{lm}^-,q_{lm}^+,  \tilde{q}_{lm}^+, \tilde{q}_{lm}^-$ which are to be determined via 5 equations : 3 equations from velocity matching and 2 stress balance equations at $r=R$ and $r=R+d$.
Velocity matching at $r=R$ gives 2 equations since the membrane velocity field has to match with $v_{int}$ and $v_-$ simultaneously at $r=R$. This gives
\beqa
&\frac{A_{lm}}{R} = q_{lm}^- R^l = \tilde{q}_{lm}^- R^l + \tilde{q}_{lm}^+ R^{-l-1}\nn\\
\label{velmatchR}
\eeqa
Further $v_{int}$ has to match with $v^+$ at $r=R+d$. This gives us 
\beqa
&\tilde{q}_{lm}^- (R+d)^l +  \tilde{q}_{lm}^+ \frac{1}{(R+d)^{l+1}} = \frac{q_{lm}^+ }{(R+d)^{l+1}}\nn\\
\eeqa
\textbf{ Traction at $r=R+d$ : } 
\beqa
\tilde{T}_\alpha = \eta^+ \left[\sum_{lm} \tilde{q}_{lm}^- (R+d)^l (l-1) + (\tilde{q}_{lm}^+ - q_{lm}^+) (R+d) ^{-l-1} (-l-2)\right]  \epsilon_{\alpha \beta } D^{\beta} Y_{lm}
\label{tildet}
\eeqa

Stress balance at $r=R+d$ : The equation for stress balance in this surface(plugged $\tilde{T}$ from Eq.~\ref{tildet}) gives

\beqa
\frac{\tau}{(R+d)^2} Y_{lm}^* (\theta_0,\phi_0) =\eta^+ \left(\tilde{q}_{lm}^-(R+d)^l (l-1) +( \tilde{q}_{lm}^+-q_{lm}^+) (R+d)^{-l-1} (-l-2)  \right)\nn\\
\eeqa

\textbf{Traction at $r=R$} :

\beqa
T_\alpha = \left[\eta_- q_{lm}^{-} (l-1) R^l - \eta_+ \tilde{q}_{lm}^- (l-1) R^l -\eta_+ \tilde{q}_{lm}^+ (-l-2) R^{-l-1}\right] \epsilon_{\alpha \beta} D^{\beta} Y_{lm}
\eeqa

The stress balance at $r=R$ now becomes
\beqa
&-\frac{\tau}{R^2} Y_{lm}^* (\theta_0,\phi_0)=- \frac{A_{lm}(2- l(l+1))}{R^2} \eta_{2D} + \left(\eta_-  q_{lm}^- (l-1)R^l - \eta_+ \tilde{q}_{lm}^- (l-1)R^l -\eta_+ \tilde{q}_{lm}^+ (-l-2) R^{-l-1} \right)
\eeqa
Thus overall we have the following set of 5 equations :
\beqa
&\frac{A_{lm}}{R} = q_{lm}^- R^l = \tilde{q}_{lm}^- R^l + \tilde{q}_{lm}^+ R^{-l-1}\nn\\
&\tilde{q}_{lm}^- (R+d)^l +  \tilde{q}_{lm}^+ \frac{1}{(R+d)^{l+1}} = \frac{q_{lm}^+ }{(R+d)^{l+1}}\nn\\
&\frac{\tau}{(R+d)^2} Y_{lm}^* (\theta_0,\phi_0) =\eta^+ \left(\tilde{q}_{lm}^-(R+d)^l (l-1) +( \tilde{q}_{lm}^+-q_{lm}^+) (R+d)^{-l-1} (-l-2)  \right)\nn\\
&-\frac{\tau}{R^2} Y_{lm}^* (\theta_0,\phi_0)=- \frac{A_{lm}(2- l(l+1))}{R^2} \eta_{2D} + \left(\eta_-  q_{lm}^- (l-1)R^l - \eta_+ \tilde{q}_{lm}^- (l-1)R^l -\eta_+ \tilde{q}_{lm}^+ (-l-2) R^{-l-1} \right)
\eeqa
 
The above system of five equations can be solved for the five unknowns $A_{lm},q_{lm}^-,q_{lm}^+,  \tilde{q}_{lm}^+, \tilde{q}_{lm}^-$.\\

The solution is given by
\beqa
&A_{lm} = \tau \frac{(R+d)^ {-2-l} \left(R^{2+l} - d^2 (d+R)^l - 2 d R (d+R)^l - R^2 (d+R)^l\right)}{(-2+l+l^2)~ \eta_{2D} + R\left[\eta_- (l-1) + \eta_+ (l+2) \right]}Y_{lm}^* (\theta_0,\phi_0) \nn\\
& = \frac{\tau}{\eta_{2D}} \frac{(R+d)^ {-2-l} \left(R^{2+l} - d^2 (d+R)^l - 2 d R (d+R)^l - R^2 (d+R)^l\right)}{s_l}~Y_{lm}^* (\theta_0,\phi_0) \nn\\
\eeqa
\beqa
& q_{lm}^-= \frac{\tau}{\eta_{2D}}\frac{ R^{-l-1} (d+R)-2-l(R^{2+l} - d^2 (d+R)^l -2 d R(d+R)^l -R^2 (d+R)^l)~Y_{lm}^* (\theta_0,\phi_0) }{s_l}\nn\\
&q_{lm}^+= \frac{\tau}{\eta_{2D}}\frac{(d+R)^{-2-l}\left((1-l) R^{1+2l} ( (2+l) \eta_{2d} +R(\eta_- - \eta_+)) -(1+2l)R^l (d +R)^{2+l} \eta_p +(d+R)^{1+2l}((-1+l)((2+l) \eta_{2d} +R \eta_-) +(2+l) R \eta_+)\right)}{(1+2l)\eta_+ s_l}~Y_{lm}^* (\theta_0,\phi_0)\nn\\
& \tilde{q}_{lm}^+ = \frac{\tau}{\eta_{2d}} \frac{R^l ((1-l) R^{1+l} (d+R)^{-2-l} ((2+l) \eta_{2d} +R(\eta_- -\eta_+))- (1+2l)\eta_+}{(1+2l)\eta_+ s_l}~Y_{lm}^* (\theta_0,\phi_0)\nn\\
& \tilde{q}_{lm}^- = \frac{(d+R)^{-2-l} \tau Y_{lm}^* (\theta_0,\phi_0) }{(2l+1) \eta_+}
\eeqa

\beqa
&v_\alpha = \sum_{lm} A_{lm}\epsilon_{\alpha \gamma} D^{\gamma} Y_{lm} \nn\\
&=\sum_{lm} \frac{\tau}{\eta_{2D}} \frac{(R+d)^ {-2-l} \left(R^{2+l} - d^2 (d+R)^l - 2 d R (d+R)^l - R^2 (d+R)^l\right)}{s_l}~Y_{lm}^* (\theta_0,\phi_0) \epsilon_{\alpha \gamma} D^{\gamma} Y_{lm}(\theta ,\phi)\nn\\
\eeqa

Performing the sum over m, we get
\beqa
\boxed {v_\alpha =\sum_l \frac{\tau}{\eta_{2D}} \frac{C_l}{s_l}~ \frac{2l+1}{4 \pi}~ \epsilon_{\alpha \gamma} D^{\gamma} P_l (\cos \gamma)}
\eeqa
where 
\beqa
&\boxed {C_l := (R+d)^ {-2-l} \left(R^{2+l} - d^2 (d+R)^l - 2 d R (d+R)^l - R^2 (d+R)^l\right)}\nn\\
&=\frac{d(-2-l)}{R} + \mathcal{O} (d^2)
\label{Cl}
\eeqa
Introducing the operator ~ $[ \nabla_\perp ^{\mathbb S^2}] =- \left(\hat {\theta} \frac{1}{R \sin \theta}\partial_\phi - \hat{\phi} \frac{1}{R} \partial_\theta\right)$ we find that the physical velocity field can be expressed as 
\beqa 
\boxed{v =\frac{\tau}{\eta_{2D}} [\nabla_\perp ^{\mathbb S^2}] ~\bm{\psi}}
\eeqa
where the dimensionless stream function $\bm{\psi}$ is given by
\beqa
\boxed {\bm{\psi}~ [\theta, \phi, \theta_0, \phi_0] =  \sum_l \frac{ (2l+1)C_l}{4 \pi ~  s_l}  P_l (\cos \gamma) }
\label{strmctrq}
\eeqa
where  $s_l =l(l+1) -2 +\frac{R}{l_-}(l-1)+\frac{R}{l_+}(l+2)$ and $l_{\pm}= \frac{\eta_{2D}}{\eta_{\pm}}$ and $\cos \gamma = \sin \theta \sin \theta_0 \cos(\phi -\phi_0) + \cos \theta \cos \theta_0$  and $C_l$ is defined in Eq.~\ref{Cl}.Let us also note from Eq.~\ref{Cl} we note that the velocity field vanishes when the distance d between the counter rotating torques go to zero.\\\\
\textbf{Flat Membrane limit} : The planar limit of Eq.~\ref{strmctrq} can be understood by again introducing a momentum variable $q = \frac{l}{R}$ and converting the sum into an integral in the limit of large radius.
\beqa
&\bm{\psi}~ [\theta, \phi, \theta_0, \phi_0] =  \sum_{l=1}^{l_{max}} \underbrace{\frac{1}{R}}_{dq = \frac{d^2 q}{ 2 \pi q}} ~\underbrace{\frac{ (2l+1)C_l R}{4 \pi ~  s_l}}_{\frac{q d}{2 \pi (q+\lambda^{-1})}+\mathcal{O} (d^2)} \underbrace{ P_l (\cos \gamma)}_{e^{iq.r}} \nn\\
&\rightarrow d \times \int \frac{d^2 q }{(q+\lambda^{-1})}e^{iq.r}
\eeqa
where we used
\beqa
&\frac{ (2l+1)C_l R}{4 \pi ~  s_l}=\frac{(2qR+1)\frac{d}{R} (-2- qR)R}{4 \pi \left(qR(qR+1) -2 +\frac{R}{\lambda_-}(qR-1)+\frac{R}{\lambda_+}(qR+2)\right)  }\nn\\
&\sim \frac{2d q^2R^2}{4 \pi \left(q^2 R^2  +\frac{R}{\lambda_-}(qR)+\frac{R}{\lambda_+}(qR)\right)  }\nn\\
& \sim \frac{q}{ \left(q  +\frac{1}{\lambda_-}+\frac{1}{\lambda_+}\right)  }\nn\\
\eeqa
Thus the velocity has dimensions $ \frac{\tau d}{\eta_{2D}} \times \frac{1}{\lambda}$ .\\
We need to perform the sum ( to lowest order in d ). Once again, the roots of the equation  $s_l=0$ are given by Eq.~\ref{roots}.\\\\
\textbf{Case 1}: $-2<l_p <0$ ( low curvature)
 \beqa
  &\bm{\psi}_{l_p<0}=  -\frac{d}{4 \pi R} \left [\frac{2}{\sqrt{2 -2\cos \gamma}} + \frac{2 +5 l_m +2 l_m^{2}}{l_m-l_p} A[l_m] + \frac{-2 -5 l_p -2 l_p^{2}}{l_m-l_p} A[l_p] \right]
  \label{repctq1}
  \eeqa
  where A is defined by Eq.~\ref{Adef}.\\
  \textbf{ Case 2} : If $0<l_p <1$  (high curvature)
 \beqa
  &\bm{\psi}_{l_p>0}=  -\frac{d}{4 \pi R} \left [\frac{2}{\sqrt{2 -2\cos \gamma}} + \frac{2 +5 l_m +2 l_m^{2}}{l_m-l_p} A[l_m] + \frac{-2 -5 l_p -2 l_p^{2}}{l_m-l_p} B[l_p] \right] \nn\\
  \label{repctq2}
 \eeqa
  where A and B are defined by Eq.~\ref{Adef} and Eq.~\ref{Bdef}.\\
\textbf{ Case 3} : $l_p =0$ \\
In this situation,
\beqa
&\bm {\psi}_{l_p=0}=\sum_l\frac{(2l+1) (-l-2) }{l(l-l_m)} P_l [\cos\gamma]\nn\\
& = -\frac{2}{\sqrt{2-2 \cos \gamma}}+\frac{2}{l_m} \log[\frac{2}{-\cos \gamma +\sqrt{2-2 \cos \gamma} +1}] +\frac{-2 -5 l_m - 2 l_m^2}{l_m}(A[l_m] + \frac{1}{l_m})
\eeqa
where A is defined in Eq.~\ref{Adef}.\\

\section{ Analytical Investigations of Streamline Topologies}
\label{apstg}
In this section, we provide details of the derivation of Eq.~\ref{rootlc_txt} of main text. We initially consider $N $ rotors and later specialize to two rotors.
One can project the dynamical equations Eq.~\ref{dynmeq} of main text via stereographic projection on the plane. If we denote the plane polar co-ordinates by ( r, $\tilde{\theta}$), then the stereographic map relates ( r, $\tilde{\theta}$) to coordinates ($\theta, \phi$) on the sphere via the relations
\beqa
&\tilde{\theta} = \phi\nn\\
&r= \tan\frac{\theta}{2}
\eeqa
Using this mapping, the hydrodynamic evolution equations take the form
\beqa
&\frac{d}{dt} r_i^2 =\frac{1}{\eta_{2D} R^2} \sum_{j \neq i}^N \frac{\tau_j (1+r_i^2)^2}{2}~ \partial_{\tilde{\theta}_i} \bm{\psi} [\gamma _{ij}]\nn\\
&\frac{d}{dt}\tilde{\theta_i} =\frac{1}{\eta_{2D} R^2} \sum_{j \neq i}^N \frac{-\tau_j (1+r_i^2)^2}{2}~ \partial_{r_i^2} \bm{\psi} [\gamma _{ij}]
\eeqa
where 
\beqa
\gamma_{ij}= \arccos \left(\frac{(1-r_i^2)(1-r_j^2)+4 r_i r_j \cos(\tilde \theta_i - \tilde \theta_j)}{(1+r_i^2)(1+r_j^2)}\right)
\eeqa
Similarly, the equation of a tracer particle (denoted by suffix p) moving in the presence of N rotors can be written in terms of  $H_p$  as follows :
\beqa
&\frac{d}{dt} r_p^2 =\frac{1}{\eta_{2D} R^2}  \frac{ (1+r_p^2)^2}{2}~ \partial_{\tilde{\theta}_p} H_p\nn\\
&\frac{d}{dt}\tilde{\theta_p} =\frac{1}{\eta_{2D} R^2} \frac{- (1+r_p^2)^2}{2}~ \partial_{r_p^2} H_p
\label{eqtracer}
\eeqa

\beqa
H_p= \sum_{j}^N \tau_j ~\bm{\psi}[\gamma_{pj}]
\label{hmp}
\eeqa
and
\beqa
\gamma_{pj}= \arccos \left(\frac{(1-r_p^2)(1-r_j^2)+4 r_p r_j \cos(\tilde \theta_p - \tilde \theta_j)}{(1+r_p^2)(1+r_j^2)}\right)
\eeqa
  Introducing complex coordinates on the plane $z = r e^{i \tilde{\theta}}$ we can write Eq.~\ref{eqtracer} in complex notation
  \beqa
 \frac{d}{dt} z_p^* = \frac{i}{\eta_{2D} R^2} \frac{(1+ |z_p|^2)^2}{2} ~\partial_{z_p} H_p
 \label{hmcp}
 \eeqa
 where  $H_p$  is the same as defined in Eq.~\ref{hm} \footnote{Let us note that upon substituting the stream function for ideal vortices given by $ \bm{\psi}[\gamma_{pj}]= \log(1-\cos \gamma_{pj})$ into Eq.~\ref{hm}, one gets the standard Hamiltonian for ideal vortices on the sphere given by 
\beqa
 H_p^{ideal}=\sum_{j}^N \tau_j\log\left(\frac{ |z_p -z_j|^2}{(1+ |z_p|^2)(1+|z_j|^2)}\right).
 \eeqa } with the geodesic distance in complex notation given by
 \beqa
\gamma_{pj}= \arccos \left(\frac{(1-|z_p|^2)(1-|z_j|^2)+4 ~ Re[z_p \bar{z}_j ]}{(1+|z_p|^2)(1+|z_j|^2)}\right)
\eeqa
In this section, we perform an analytical treatment of the location of stagnation points on the spherical membrane in the regimes of low and high curvature separately.
\subsection { Regime of low curvature }
In general, it follows from Eq.~\ref{hmcp} that solving for stagnation points amounts to solving for solutions to
  \beqa
  \frac{d}{dt} \bar{z}_p= \frac{i}{\eta_{2D} R^2} \frac{(1+ |z_p|^2)^2}{2} ~\partial_{z_p}\left( \sum_{j}^N \tau_j ~\bm{\psi}[\gamma_{pj}]\right) =0 
  \label{stgeq}
 \eeqa
with the geodesic distance in complex notation given by
 \beqa
\gamma_{pj}= \arccos \left(\frac{(1-|z_p|^2)(1-|z_j|^2)+4 ~ Re[z_p \bar{z}_j ]}{(1+|z_p|^2)(1+|z_j|^2)}\right)
\eeqa
 and the stream function $\bm{\psi}$ given by Eq.~\ref{strmptrq}. Because the stream function is complicated in structure after performing the Legendre sum, we propose here to choose a set of parameters that enables us to simplify the stream function and subsequent analysis of stagnation points.\\\\
To be concrete, let us choose $\eta_{2D} = 3/2, \eta_- =1, \eta_+=2, R=1$ for which $\lambda/R =1/2$. This yields the two roots of $s_l=0$ to be $l_m=-3, l_p=0$ \\
Using Case 3 of summed up versions of Eq.~\ref{strmptrq}, we get
 \beqa
\bm{\psi}[\gamma]= \frac{1}{12 \pi} \left[ \frac{5}{2} \big{[} (6 \cos^2 \gamma -2) \arccoth (\sqrt{2-2 \cos \gamma} +1) + 3 \cos \gamma (\sqrt{2-2 \cos \gamma} -1)+\sqrt{2-2 \cos \gamma} \right) \nn\\
- \log(- \cos \gamma + \sqrt{2-2 \cos \gamma} +1)\big{]}
\label{strmsimple}
\eeqa
Plugging Eq.~\ref{strmsimple} into Eq.~\ref{stgeq}
\beqa
\frac{i}{\eta_{2D} R^2} \frac{(1+ |z_p|^2)^2}{2} ~\left( \sum_{j}^N \tau_j ~F[z_p,z_j]~ G [ z_p, z_j] \right) =0
\label{fgeq}
\eeqa
where the factors $F$ and $G$ arise from the derivative of the stream function ie. $\partial_{z_p} \bm{\psi} = \frac{\partial \bm{\psi}}{\partial \cos \gamma}~~ \partial_{z_p} \cos \gamma : = F \times G$. Let us note that although F is dependent on the choice of parameters, the factor G is essentially purely geometric. For our choice of parameters,
\beqa
&F[z_p, z_j] =\frac{10 - 8 \sqrt{2-2 \cos \gamma_{pj}} + \cos \gamma_{pj} \left( -1 +5 \sqrt{2-2 \cos \gamma_{pj}} +15 (-1+\sqrt{2-2 \cos \gamma_{pj}}) \cos \gamma_{pj} - 30 \cos \gamma_{pj} \sin^2 \gamma_{pj} \arccoth(1+\sqrt{2-2 \cos \gamma_{pj}})\right)}{-12 \pi \sin^2 \gamma_{pj}} \nn\\
&G[z_p,z_j] = \frac{(1-|z_j|^2) (-2 \bar{z}_p)+ 4 \left((1+|z_p|^2) \frac{\bar{z}_j}{2} - Re[z_p \bar{z}_j] \bar{z}_p\right)}{(1+|z_j|^2) (1+|z_p|^2)^2}
\label{fg}
\eeqa
where in the expression of F we have
\beqa
\cos \gamma_{pj}=  \left(\frac{(1-|z_p|^2)(1-|z_j|^2)+4 ~ Re[z_p \bar{z}_j ]}{(1+|z_p|^2)(1+|z_j|^2)}\right)
\eeqa
We now specialize to the case of two rotors on the spherical membrane. Let their positions in the complex plane be denoted by $z_1$ and $z_2$.  Since the stagnation points are always constrained to lie on the great circle joining the two locations, we can essentially map the dynamics to the unit circle on the complex plane. We choose coordinates such that the location of first rotor is at $\tilde{\theta}_1=0$ and the second rotor at $\tilde{\theta}_1=\phi$.  We further choose, without loss of generality, the strength of the first rotor to be 1 and relative strength between the rotors be denoted by $\tau$. Thus
\beqa
z_1 = 1, z_2 = e^{i \phi}, \tau_1=1, \tau_2 = \tau
\label{loc}
\eeqa
Plugging in Eq.~\ref{loc} into Eq.~\ref{fgeq} and Eq.~\ref{fg},we convert it to a effective two parameter problem where the stagnation point $z_p$ has to be solved as a function of the relative vortex strength $\tau$ and the location of the second vortex parametrized by $\phi$ , from the equation
\beqa
&F[z_p,1]~ G[z_p,1] + \tau_2~  F[z_p, e^{i \phi}]~ G[z_p, e^{i \phi}] =0\nn\\
&\Rightarrow f[Re(z_p)] \frac{1}{4} (1- \bar{z}_p^2) + \tau f [Re (z_p e^{-i \phi})] \frac{1}{2} \left(e^{-i \phi} - Re ( z_p e^{-i \phi}) \bar{z}_p \right)=0
\eeqa
where 
\beqa
f= \frac{10 - 8 \sqrt{2-2 x} + x \left( -1 +5 \sqrt{2-2 x} +15 (-1+\sqrt{2-2 x}) x + 30 x~ (x^2-1) \arccoth(1+\sqrt{2-2 x})\right)}{12 \pi (x^2-1)} 
\eeqa
Substituting ansatz $ z_p= e^{i \theta_p}$ into Eq.~\ref{rootlc}, we get
\beqa
f[\cos \theta_p] \frac{1}{4} (1- e^{-2 i \theta_p}) + \tau f [\cos(\theta_p -\phi)] \frac{1}{2} \left(e^{-i \phi} - \cos(\theta_p -\phi) e^{-i \theta_p} \right)=0
\label{rootlc}
\eeqa

\subsection { Regime of high curvature }
In the regime of high curvature,  ignoring all local corrections,
\beqa
F=\frac{d \bm{\psi}}{d (\cos \gamma) } = -\frac{\eta_{2D}}{4 \pi R \eta_+}
\eeqa
This follows directly from the fact that the global rotation term is 
\beqa
\frac{d}{d \gamma} \bm{\psi} =\frac{\eta_{2D}}{4 \pi R \eta_+} \sin \gamma
\eeqa
Using this F, the equation for stagnation points simplifies considerably.
\beqa
G(z_p,1) + \tau~ G(z_p, e^{i \phi}) \sim 0\nn\\
\Rightarrow \frac{1}{4} (1 - \bar{z}_p^2) + \frac{\tau}{2} \left(e^{- i \phi} - Re~[z_p e^{- i \phi}]\bar{z}_p\right)=0
\eeqa
Substituting $z_p =e^{i \theta_p}$ in the above we get
\beqa
\frac{1}{4} (1 - e^{-2 i \theta}) + \frac{\tau}{2} \left(e^{- i \phi} - \cos( \theta - \phi) e^{- i \theta}\right)=0
\eeqa
This is the equation discussed in main text Eq.~\ref{stop_hc}.
\section{Roots of $s_l$ : Poles of the stream function in Legendre basis on the sphere.}
\label{aproot}
In all the examples we studied in the main text, the dimensionless stream function on the spherical membrane has the following generic structure  in the basis of Legendre polynomials:
\beqa
\bm{\psi}  [\theta, \phi, \theta_0, \phi_0] =  \sum_l \frac{ f_l}{4 \pi ~  s_l~ g_l}  P_l (\cos \gamma) 
\label{apstgen}
\eeqa
where $f_l$ and $g_l$ are some polynomials in Legendre modes denoted by $l$ and $s_l=l(l+1) -2 +\frac{R}{\lambda_-}(l-1)+\frac{R}{\lambda_+}(l+2)$. The geodesic angle between the source and response locations is denoted by $\gamma$. In order to find the real space Greens function, one is thus left with the task of performing the sum Eq.~\ref{apstgen}. As mentioned in Appendix \ref{apf}, \ref{aptq}, \ref{apctq}, the real space representation of the stream function crucially depends  on the root structure of the equation $s_l =0$.\\
In this appendix we discuss the nature of the roots of the equation $s_l =0$.

\beqa
s_l =  l(l+1)-2 + R \frac{\eta_-}{\eta_{2D}} (l-1) + \frac{R \eta^+}{\eta_{2D}} (l+2)=0\nn\\
\Rightarrow l^2 + l \left( 1+\frac{R \eta_-}{\eta_{2d}} +\frac{R \eta_+}{\eta_{2d}} \right) + \left( -2 - \frac{R \eta_-}{\eta_{2d}
} + \frac{2 R \eta^+}{\eta_{2d}} \right)=0
\label{rooteq}
\eeqa
\beqa
l_p= \frac{ -(\eta_{2d} + R \eta_- + R \eta_+) + \sqrt{9 \eta_{2d}^2 + 6R \eta_{2d} (\eta_- -\eta_+)+ R^2 (\eta_- + \eta_+)^2}}{2 \eta_{2d}}\nn\\
l_m= \frac{ -(\eta_{2d} + R \eta_- + R \eta_+) - \sqrt{9 \eta_{2d}^2 + 6R \eta_{2d} (\eta_- -\eta_+)+ R^2 (\eta_- + \eta_+)^2}}{2 \eta_{2d}}\nn\\
\eeqa
Let us now discuss the nature of the roots in the space of parameters :\\
\textbf{ Nature of the root $l_m$} : always negative. \\
\textbf{ Nature of the root $l_p$} : The range of this root is $-2 < l_p \leq 1$ . Thus, this root changes sign as parameters are varied. As we saw in Appendix \ref{apf},\ref{aptq}, \ref{apctq} the stream function has two different representations in real space depending on the sign of $l_p$. \\ 
In order to understand this better, let us first consider the simpler situation  $\eta_+ =\eta_- := \eta_{3d}$. Defining the unique Saffman length as $ \lambda := \frac{ \eta_{2d}}{2 \eta_{3d}}$, we first note from Eq.~\ref{rooteq} that the product of the two roots is 
\beqa
 l_m l_p=  \frac{R}{2 \lambda} -2  .
\eeqa 
Since $l_m$ is always negative,  it is clear that for large radius $R> 4 \lambda$, the root $l_p$ is negative.  In the opposite situation of high curvature (small radius) ie.  $R<4 \lambda$, the root $l_p$ is positive. \\

To explore the more generic situation where $\eta_+ \neq \eta_-$, let us consider expansions of $l_p$ in terms of radius R.\\ For small radius,
\beqa
l_p = 1-\frac{R \eta_+}{\eta_{2d}} +\mathcal{O}(R^2)
\eeqa
Thus shows that $l_p<1$ for small radius (high curvature) and attains the limiting value one in the limit of vanishing external solvent or radius.\\
Let us also identify the regime where $l_p<0$. This demands 
  \beqa
  l_p<0,  ~~  l_m l_p=\left( -2 - \frac{R \eta_-}{\eta_{2d}
} + \frac{2 R \eta^+}{\eta_{2d}} \right) >0
\eeqa
which is satisfied when
\beqa
 2 \eta_+ -\eta_- >0 , ~~\eta_{2d} < \frac{R}{2} ( 2 \eta_+ -\eta_-)
\eeqa
For large radius ,
\beqa
l_p = -\frac{2 \eta_+ - \eta_-}{\eta_+ + \eta_-}  +\mathcal{O}(1/R)
\eeqa
One notes that now the sign of $l_p$ is more subtle, positive if $  \eta_- > 2 \eta_+$ and negative for $  \eta_- < 2 \eta_+$.

\section{Data Availability Statement}
The analytical data that supports the findings of this study is available within the article and its supplementary material. Numerical details and additional data are available from the authors upon reasonable request.

\begin{footnotesize}

\end{footnotesize}

\end{document}